\documentclass[%
superscriptaddress,
bibnotes,
 amsmath,amssymb,
 aps,
]{revtex4-2}

\usepackage{graphicx}
\usepackage{dcolumn}
\usepackage{bm}
\usepackage{hyperref}
\usepackage{subcaption}
\usepackage{xcolor}	
\usepackage{soul}
\usepackage{siunitx}
\usepackage[outdir=./]{epstopdf}
\usepackage[export]{adjustbox}

\begin{document}

\preprint{APS/123-QED}


\title{Axisymmetric Pushing of a Spherical Cargo using\\ an Active Spherical Janus Motor}

\author{Subramaniam \surname{Chembai Ganesh}} 
 \affiliation{Benjamin Levich Institute and Department of Chemical Engineering, City College of the City University \\ of New York, New York, New York 10031, USA }

\author{Jessica S. Rosenberg}
\affiliation{Benjamin Levich Institute and Department of Physics, City College of the City University \\ of New York, New York, New York 10031, USA}%

\author{Jeffrey F. Morris}
 \affiliation{Benjamin Levich Institute and Department of Chemical Engineering, City College of the City University \\ of New York, New York, New York 10031, USA }

\author{Joel Koplik}
\affiliation{Benjamin Levich Institute and Department of Physics, City College of the City University \\ of New York, New York, New York 10031, USA}%

\author{Charles Maldarelli}
 \affiliation{Benjamin Levich Institute and Department of Chemical Engineering, City College of the City University \\ of New York, New York, New York 10031, USA }

\date{\today}

\begin{abstract}
We analyze the interaction between a self-diffusiophoretic spherical Janus motor and an inert spherical cargo particle in an axisymmetric configuration in the Stokes regime.  To study the different configurations of the two spheres and their motions, we develop an analog to the twin multipole approach to numerically determine the axisymmetric stream function for the flow field.  We verify the validity and accuracy of this approach using existing literature and COMSOL Multiphysics. We study the effects of the size of the Janus cap, the relative ratio of sizes of the two spheres, and their separation distance on their interactions. For the case of a stationary cargo, we identify the existence of a distinct regime where the Janus motor hovers at a finite separation distance from the cargo and summarize the results using a phase diagram. In the presence of a freely moving cargo, we analyze the steady terminal velocities of the Janus motor and the cargo to identify distinct conditions at which the two spheres can translate with equal velocities while maintaining a finite separation distance. 

\end{abstract}

\maketitle


\section{\label{sec:intro}Introduction}

The study of microscopic self-propelled particles ('swimmers') is of interest due to their relevance in natural biological systems \cite{Lauga2009hydrodynamics}, targeted drug delivery \cite{su2019janus}, \cite{tran2014janus}, medical diagnostics \cite{ho2011monodisperse} and for their ability to self-assemble \cite{walther2013janus}. These swimmers typically operate in the low Reynolds number regime where the absence of inertia requires that the net propulsion of any particle at this scale with moving parts must employ a mechanism that breaks time symmetry \cite{purcell1977life}. We study Janus motors (JMs) which are Janus particles whose surfaces are functionalized to interact asymmetrically with the surrounding environment to generate the necessary propulsion forces for their motion. One such self propulsion mechanism is the well studied Self-diffusiophoresis. Here, the chemically active JM reacts enables the asymmetric efflux of some solute, creating a local concentration gradient that drives its motion \cite{anderson1984diffusiophoresis}, \cite{anderson1989colloid}, \cite{golestanian2005propulsion}, \cite{howse2007self}. Spherical JMs (typically in the range of 1-10 microns) have been studied both in an infinite medium both experimentally \cite{ebbens2011direct}, \cite{ebbens2012size}, \cite{zheng2013non}, \cite{paxton2004catalytic} and theoretically \cite{sharifi2013diffusiophoretic}, \cite{popescu2016self}, \cite{brady2011particle} (For review see \cite{ebbens2010pursuit}, \cite{wang2013small}, \cite{kapral2013perspective}, \cite{moran2017phoretic}). 

To understand the dynamics of JMs for applications such as colloidal assembly \cite{walther2013janus}, or cargo towing \cite{baraban2012catalytic},it is essential to understand the interaction of JMs among themselves or with other particles/obstacles. In an effort to simplify the analysis for the general interactions involving JMs, there have been many efforts in recent literature to understand the pair interactions of JMs and JMs with other particles \cite{baraban2012transport}, \cite{bayati2016dynamics}, \cite{nasouri2020exact}, \cite{rojas2021hydrochemical}, \cite{sharifi2016pair}.  Baraban et al. 2012 \cite{baraban2012transport} experimentally studied the interaction between JMs with varying activities on their surfaces and (inert) cargo particles. They used polystyrene particles coated with platinum placed in an aqueous solution of hydrogen peroxide. The platinum cap on the JM catalyses the decomposition of the hydrogen peroxide in the solution to create a concentration gradient of oxygen around the JM, thereby driving it's motion.  Baraban et al. found that when a single JM with a hemispherical active cap collides with the cargo, it can allow continued pushing; showing a kind of towing.  

While complete theoretical studies of cargo towing haven’t been undertaken, there have been studies of the interaction between JMs.  Sharifi-Mood et al. 2016 studied pairs of two JMs using the Reynolds Reciprocal Theorem \cite{happel2012low}, \cite{teubner1982motion} to solve for the terminal velocities of the particles. They studied different configurations of the two JMs; including how different orientations between the active faces, sizes of the active faces, and distances between the two JMs affected their motion.  Their research concluded with a phase diagram that makes explicit the effects of these variables, and whether they lead to approach, assembly or the two particles moving apart.  

Golestanian et al. studied axisymmetric JMs in pairs. They made use of bispherical coordinates \cite{lee1980motion} to fully solve for the forces, torques, as well as the self-diffusiophoretic slip velocity on the JM's surface.  They created generic geometric functions that can be summed to find the relative velocity of the two particles and exposes the effect that the JM’s activity has on the analytical solution. While Golestanian et al. and Sharifi-Mood both look at two JMs, the focus of our study is to develop a more complete description of the of the axisymmetric interaction between a JM and an inert cargo with emphasis on the cargo-towing application. In our paper we apply the twin multipole approach \cite{ross1968potential} to solve the hydrodynamic axisymmetric stream function.  The twin multipole approach has been popularized by Jeffrey \cite{jeffrey1973conduction} who solved the temperature field for the conduction of heat in the presence of two particles and later by \cite{jeffrey1984calculation} for the hydrodynamic interactions between two spheres, where the solutions could be described by spherical harmonics. The novelty of our approach comes from extending the technique to the axisymmetric stream function which does not satisfy the harmonic equation. Further, the twin multipole approach employs relatively straightforward spherical coordinates for solving the governing equations. This allows for simpler application of boundary conditions, and easier evaluation of quantities of interest (such as stress fields and net forces) compared to more complex solution techniques such as using a bispherical basis etc.

\section{\label{sec:Problem}Problem Statement}

We consider an active spherical JM of radius $R_1$  and an inert spherical cargo of radius $R_2$  immersed in a liquid of density $\rho (\frac{kg}{m^3})$ and viscosity $\mu (\frac{kg}{m s})$. We limit our discussions to the axisymmetric configuration where the vector pointing to the cargo’s center from the center of the JM is parallel to $\bm{\hat{e}_z}$ . 

\begin{figure}
\includegraphics[width = 0.5\textwidth]{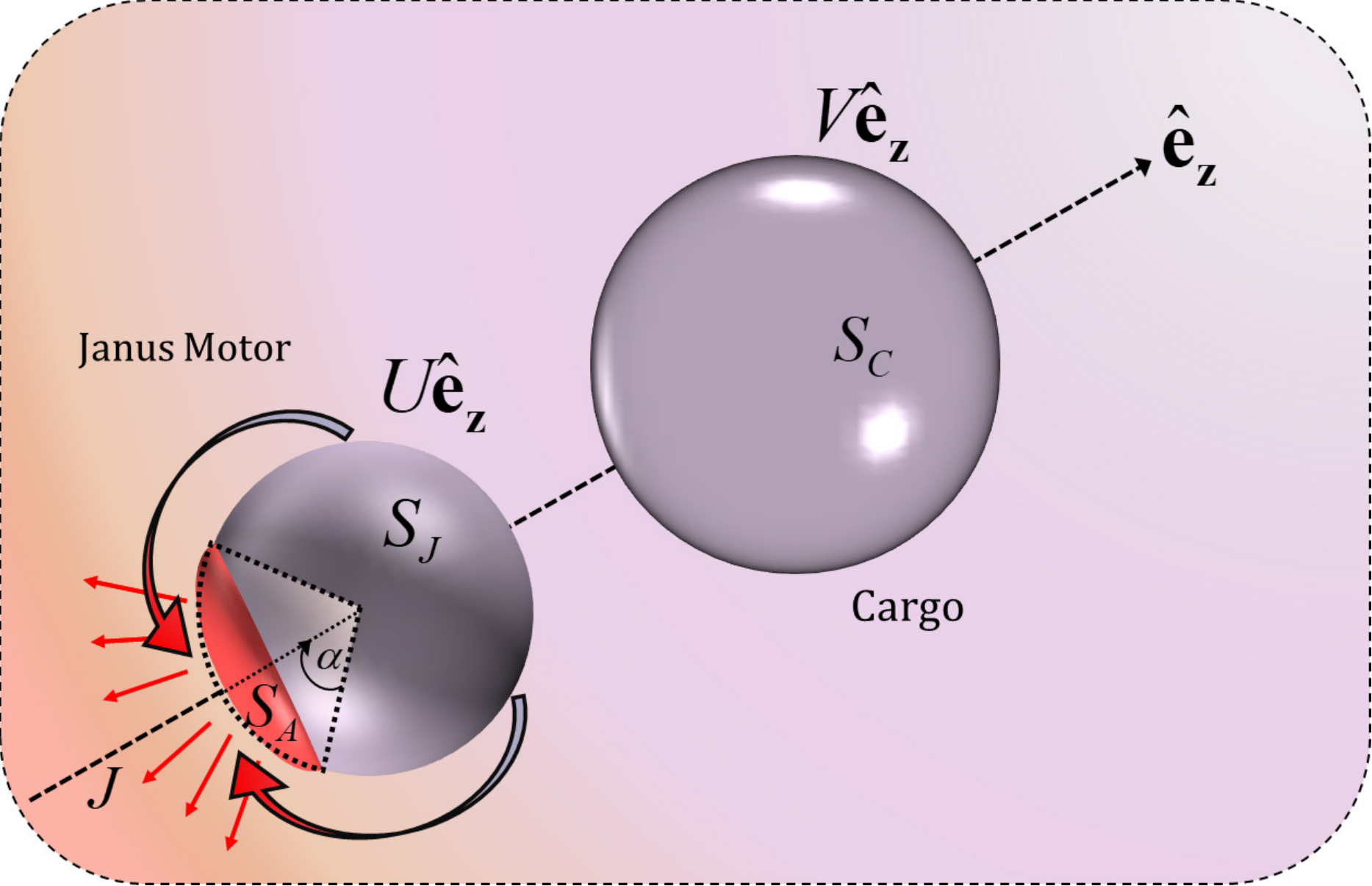}
\caption{A JM and cargo in an axisymmetric configuration moving with velocities $U {\bm{\hat{e}_z}} $ and $V {\bm{\hat{e}_z}}$ respectively. The flux of the solute is indicated by $J$, the cap angle by $\alpha$ along a meridional plane.} 
\label{fig:JMCargoScheme}
\end{figure}

The size of the active cap $\bm{S_A}$ of the JM is represented by the angle defined with respect to $\bm{\hat{e}_z}$ along any meridional plane as shown in Figure \ref{fig:JMCargoScheme} and indicated by $\alpha$.  We assume the active cap produces a constant flux of magnitude, $J (\frac{mol}{m^2s})$  of some solute in the normal direction to its surface.  The remaining portion of the JM, $\bm{S_J}$ and the entire cargo, $\bm{S_C}$ are both inert (no flux).  A local gradient of solute concentration $C (\frac{mol}{m^3})$ created by the difference in chemical properties on the JM’s face induces an osmotic flow along the surface of the JM \cite{anderson1984diffusiophoresis} which propels the motor and the cargo with terminal velocities $U \bm{\hat{e}_z}$ and $V \bm{\hat{e}_z} (\frac{m}{s})$ respectively. 

We assume the solute has a diffusion coefficient $D (\frac{m^2}{s})$ in the surrounding fluid, so we can define the dimensionless P\'eclet number $Pe = UR_1/D$ and the Reynolds number $Re = \rho UR_1/\mu$ for the problem, using the radius $(R_1)$ and velocity $(U)$ of the JM as our characteristic length and velocity scales respectively.

\section{\label{sec:C Field}Concentration Field}

We assume our concentration field of the solute is purely diffusive ($Pe \rightarrow 0$).  Therefore the field is governed by the Laplace equation, 
\begin{equation} \label{eq:laplace}
\nabla^2C = 0. 
\end{equation}

Since the presence of any constant ambient concentration of the solutes far away does not affect the gradients in the concentration field, We can also assume that the concentration field decays to zero far field, giving us the corresponding boundary conditions,

\begin{equation} \label{eq:BC1}
C \rightarrow 0 \quad\{\bm{x} \rightarrow \infty\},
\end{equation}
\begin{equation} \label{eq:BC2}
{\bm{\hat{n}}} \cdot \nabla C = 0 \quad \{\bm{x} \in \bm{S_C} , \bm{S_J}  \},
 \end{equation}
\begin{equation} \label{eq:BC3}
{\bm{\hat{n}}} \cdot \nabla C = \frac{\textrm{-} J}{D} \quad \{\bm{x} \in \bm{S_A} \}.
\end{equation}

We nondimensionalize equations \ref{eq:laplace}, \ref{eq:BC1}, \ref{eq:BC2}, \ref{eq:BC3} by length $R_1$, implying a natural scale of $JR_1/D$ for the concentration field.  

\begin{figure}
\includegraphics[width=0.2\textwidth]{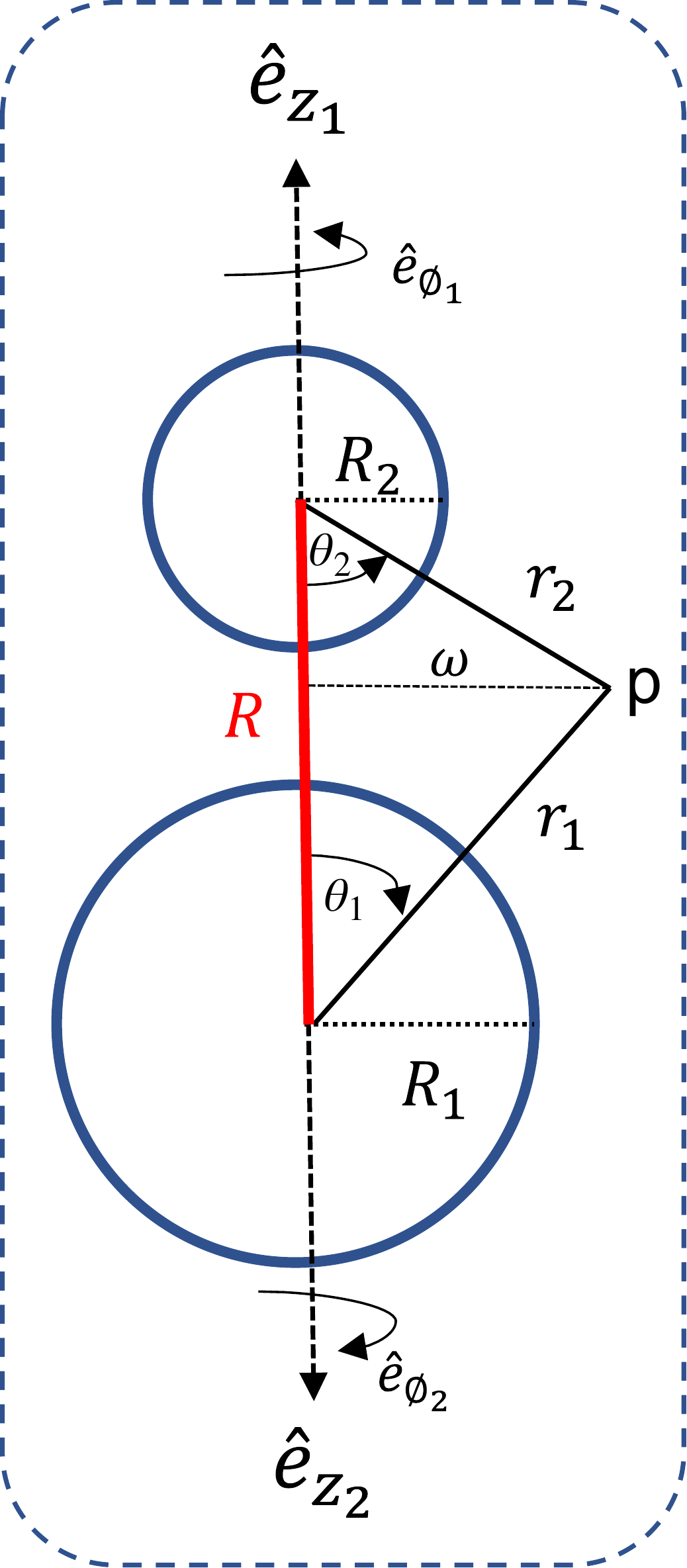}
\caption{Description of the two basis systems used for the twin multipole approach.}
\label{fig:spherical}
\end{figure}

We implement the twin multipole approach to solve for the concentration field. Here, we write the solutions to the concentration field as the sum of two concentration fields $C = C_1 + C_2$ using two spherical coordinate systems $(r_1,\theta_1,\phi_1)$ $(r_2,\theta_2, \phi_2)$ with their origins at the center of the JM and cargo respectively as shown in Figure \ref{fig:spherical}.  The two fields satisfy the governing equation \ref{eq:laplace}.  We can then write the general solutions as an expansion in solid harmonics that decay to zero at infinity (from \ref{eq:BC1}).

\begin{equation} \label{eq:conc1}
C = \sum_{i=1}^2 C_i = \sum_{i=1}^2 \sum_{n=0}^\infty \Bigl[ \big( B_i \big)_{n} r_i^{\textrm{-}(n+1)} \Bigr] p_n(\gamma_i) \qquad \forall \{\gamma_i = cos\theta_i  \},
\end{equation}

where $p_n$ is the $n^{th}$ order Legendre polynomial function. We then apply boundary conditions \ref{eq:BC1}, \ref{eq:BC2}, \ref{eq:BC3} using the translation theorem, 

\begin{equation} \label{eq:translation}
\frac{P_n(cos\theta_2)}{r_{_2}^{^{n+1}}} = \sum_{m=0}^{\infty} {m + n \choose n} \biggl( \frac{r_{1}^{m}}{R^{^{m+n+1}}}  \biggr) P_m(cos\theta_1).
\end{equation}

Equation \ref{eq:translation} is useful because the boundary conditions can be applied on the JM and by interchanging the indices $1, 2$, the boundary conditions can be applied on the cargo. The unknown coefficients $(B_1)_n$, $(B_2)_n$ can be solved for simultaneously by numerically inverting the resulting system of equations.  

\section{\label{sec:V Field}Velocity Field}
In problems involving neutral diffusiophoresis, the flow field is typically assumed to be inertia-less $(Re \rightarrow 0)$ \cite{,sharifi2016pair,nasouri2020exact} and the velocity, $\bm{u}$ and pressure $p$ fields are governed by the incompressible Stokes equations

\begin{equation}\label{eq:stokes1}
\textrm{-}\nabla p + \mu \nabla^2 {\bm{u}} = 0,
\end{equation}
\begin{equation}\label{eq:stokes2}
\nabla \cdot {\bm{u}} = 0.
\end{equation}
The corresponding boundary conditions are 
\begin{equation}\label{eq:BCv1}
{\bm{u}} =0 \quad\{x \rightarrow \infty\},
\end{equation}

\begin{equation}\label{eq:BCv2}
{\bm{u}}= U{\bm{\hat{e}}_z}+ \beta \nabla_sC \quad \{x \in S_C , S_J  \},
\end{equation}

\begin{equation}\label{eq:BCv3}
{\bm{u}} = V{\bm{\hat{e}}_z}  \quad \{x \in S_C  \}.
\end{equation}

Where $\beta$ is the diffusiophoretic slip coefficient which has positive values when the interaction between the JM and the solute is repulsive, ${\bm{\hat{n}}}$ is the normal unit vector pointing into the fluid from the solid surfaces and $\nabla_s = (\boldsymbol{\delta} -\bm{\hat{n}\hat{n} }) \cdot \nabla$.  Since there are no external forces, the net force on the cargo $(F_{_C}^{^{net}} {\bm{\hat{e}}_z})$ and the JM $(F_{_J}^{^{net}} {\bm{\hat{e}}_z})$ are equal to zero.  

\begin{equation}\label{eq:FJnet}
F_{_J}^{^{net}} {\bm{\hat{e}_z}} = \iint\limits_{S_J,S_A} ( {\bm{\hat{e}_z}} \cdot  \boldsymbol{\sigma} \cdot {\bm{\hat{n}}})ds = 0
\end{equation}

\begin{equation}\label{eq:FCnet}
F_{_C}^{^{net}} {\bm{\hat{e}_z}} = \iint\limits_{S_C} ( {\bm{\hat{e}_z}} \cdot  \boldsymbol{\sigma} \cdot {\bm{\hat{n}}})ds = 0
\end{equation}

where $\boldsymbol{\sigma}$ is the Cauchy stress tensor.  We can use the Stokes stream function to simplify the calculations given the axisymmetric nature of the problem.  We define the velocity field in terms of the stream function,

\begin{equation} \label{eq:velstream}
{\bm{u}} = \nabla \times \biggl( \frac{\boldsymbol{\Psi}}{\omega} \biggr).
\end{equation}

The stream function satisfies the linear equation $\nabla \times \nabla \times \nabla \times \nabla \times  \biggl( {\bm{\hat{e}}}_{_{\boldsymbol{\phi}}} \frac{\psi}{\omega} \biggr) = E^4 \psi = 0$.  We use a twin multipole approach to solve for the stream function. The eigenfunction expansion for the stream function for an unbounded fluid can be written using the two spherical coordinates defined in Figure \ref{fig:spherical} as 
\begin{equation} \label{eq:eigenexpansion}
{\boldsymbol{\Psi}} = \sum_{i=1}^2 \psi_i {\bm{\hat{e}}}_{\boldsymbol{\phi}i}  = \sum_{i=1}^2 \sum_{n=1}^\infty \Bigl[ \big( X_i \big)_n r_{_i}^{^{2-n}} \big(Y_i \big)_n r_{_i}^{^{-n}}  \Bigr]g_n(cos\theta_i) {\bm{\hat{e}}}_{\boldsymbol{\phi}i}. 
\end{equation}
Here, the eigenfunctions are defined in terms of the modified Gegenbauer polynomials $g_n(\gamma) = -\mathbb{C}_{n+1}^{-1/2}(\gamma) $ \cite{leal2007advanced}.  Note, the vector sum of the stream function gives us $\boldsymbol{\Psi} = (\psi_1 - \psi_2) {\bm{\hat{e}}}_{\boldsymbol{\phi}_1} =  (\psi_2 - \psi_1) {\bm{\hat{e}}}_{\boldsymbol{\phi}_2}$.  We now rewrite the boundary conditions \ref{eq:BCv1}, \ref{eq:BCv2}, \ref{eq:BCv3} in terms of the stream functions and satisfy these boundary conditions on the surface of the spheres using the following translational theorems
\begin{equation}\label{eq:transtheorem1}
\frac{g_n(\gamma_2)}{r_2^n} =   \sum _{m=0}^\infty \left(\frac{R_{1}^{m}}{2n+1}\right)\Bigl[  \Omega_{1}p_{m}(\gamma_1) +  \Omega_{2}p_{m-1}(\gamma_1)  + \Omega_{3}p_{m+1}(\gamma_1)  \Bigr], 
\end{equation} 

\begin{equation}\label{eq:transtheorem2}
\frac{g_n(\gamma_2)}{r_2^{n-2}} =\sum _{m=0}^\infty  \left(\frac{R_{1}^{m}}{2n+1}\right)  \Bigl[ \Omega_{4}p_{m}(\gamma_1) +  \Omega_{5}p_{m-1}(\gamma_1) +  \Omega_{6}p_{m+1}(\gamma_1) +  \Omega_{7}p_{m-2}(\gamma_1) +  \Omega_{8}p_{m+2}(\gamma_1)        \Bigr]. 
\end{equation}

We define the functions $\Omega_1(r_1, r_2, R)$ through  $\Omega_8(r_1, r_2, R)$ in Appendix \ref{sec:apA}.  The following identities are useful in applying the boundary conditions.  

\begin{equation}\label{eq:bonnet}
g_n(x) = \frac{p_{n+1}(x) - p_{n-1}(x)}{2n+1} \quad \text{(Bonnet's recursion theorem)}, 
\end{equation}

\begin{equation}\label{eq:bonnet2}
\frac{d}{dx}g_n(x) = p_n(x).
\end{equation}

\begin{figure} [htb!]
  \centering
  \begin{subfigure}{.3\linewidth}
    \centering
    \includegraphics[width = \linewidth]{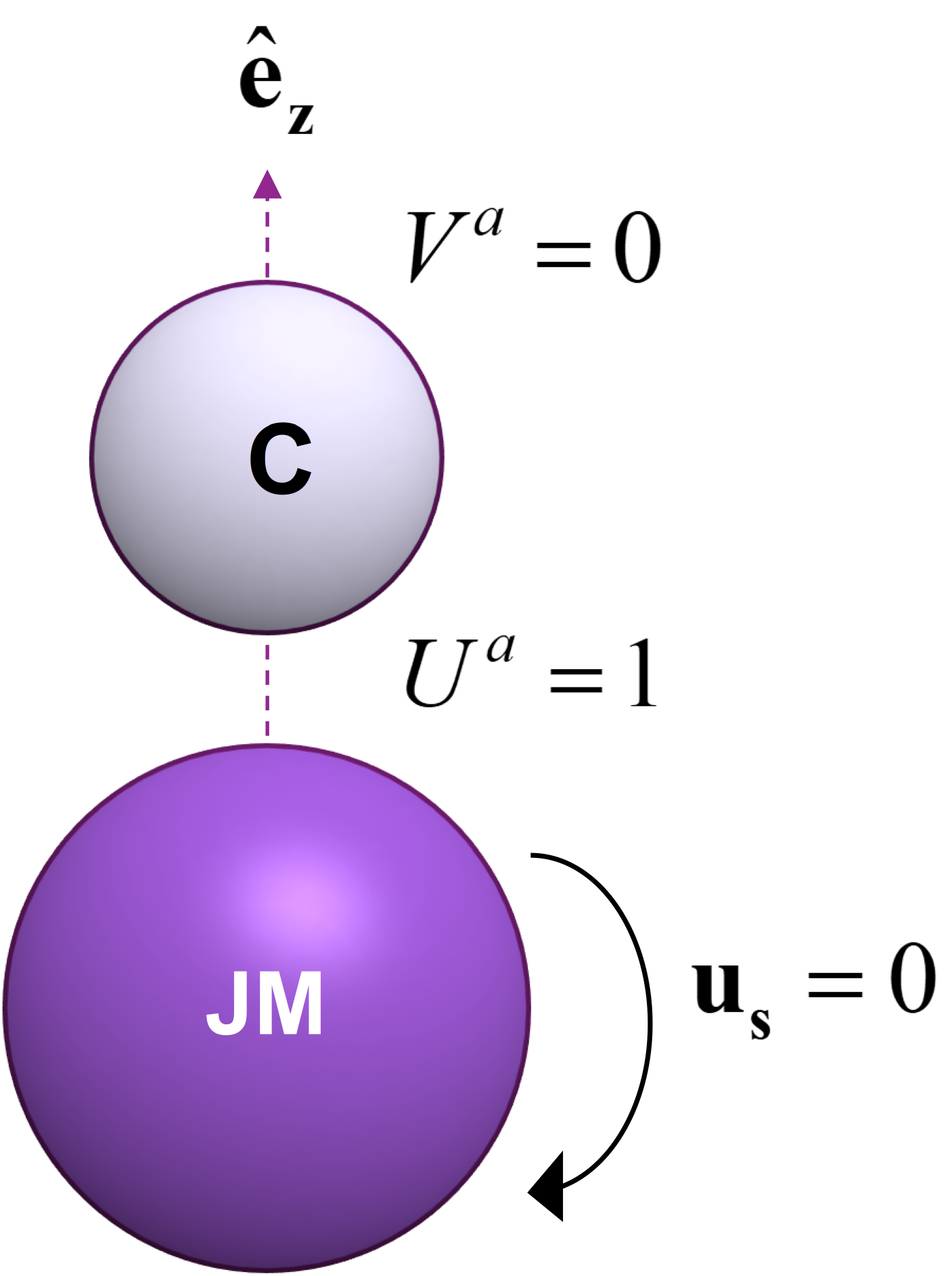}
    \caption{}
    \label{fig:sub1}
  \end{subfigure}%
  \hspace{1.56em}
  \begin{subfigure}{.29\linewidth}
    \centering
    \includegraphics[width = \linewidth]{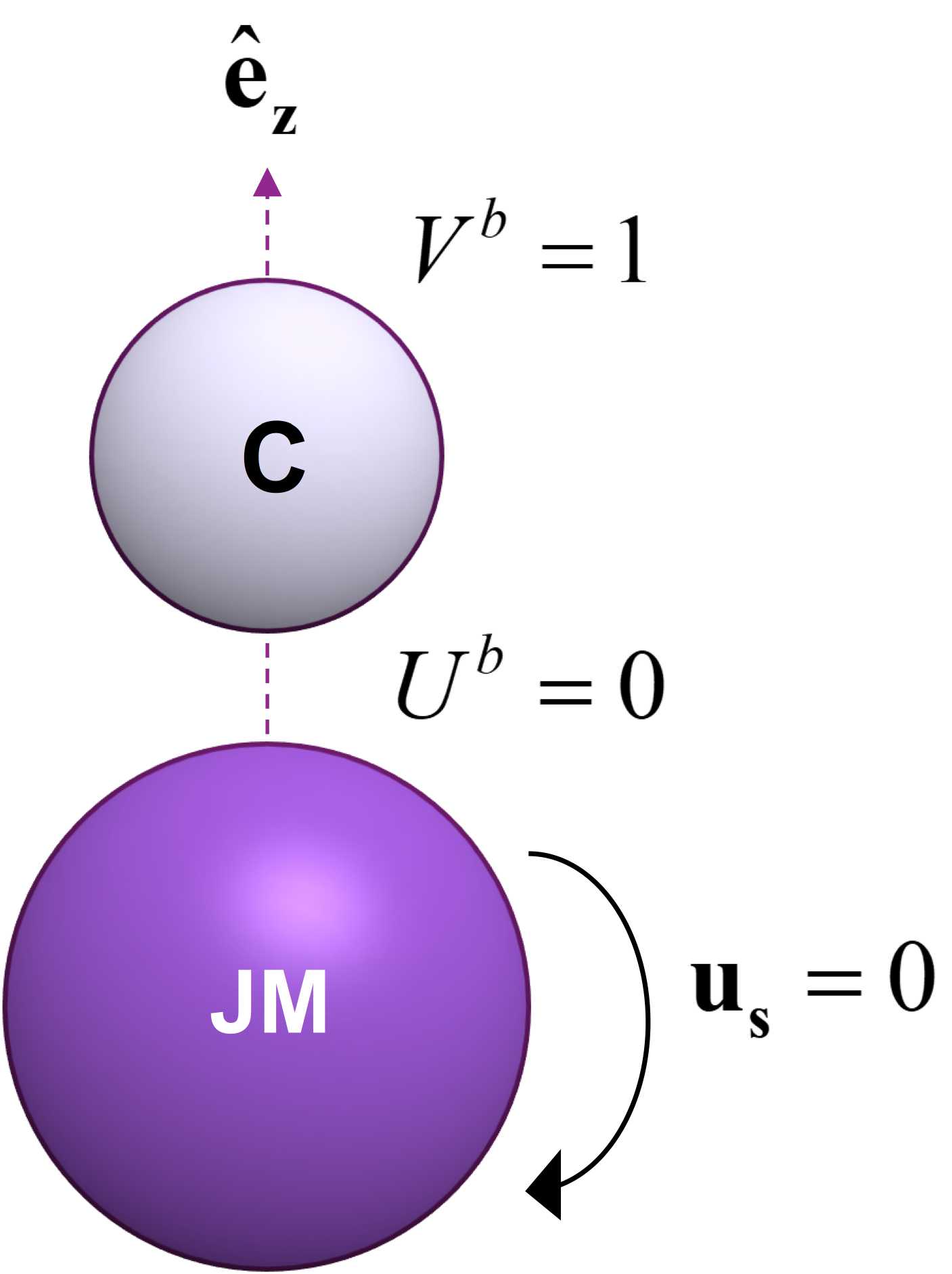}
    \caption{}
    \label{fig:sub2}
  \end{subfigure}%
  \hspace{1.4em}
  \begin{subfigure}{.35\linewidth}
    \centering
    \includegraphics[width = \linewidth]{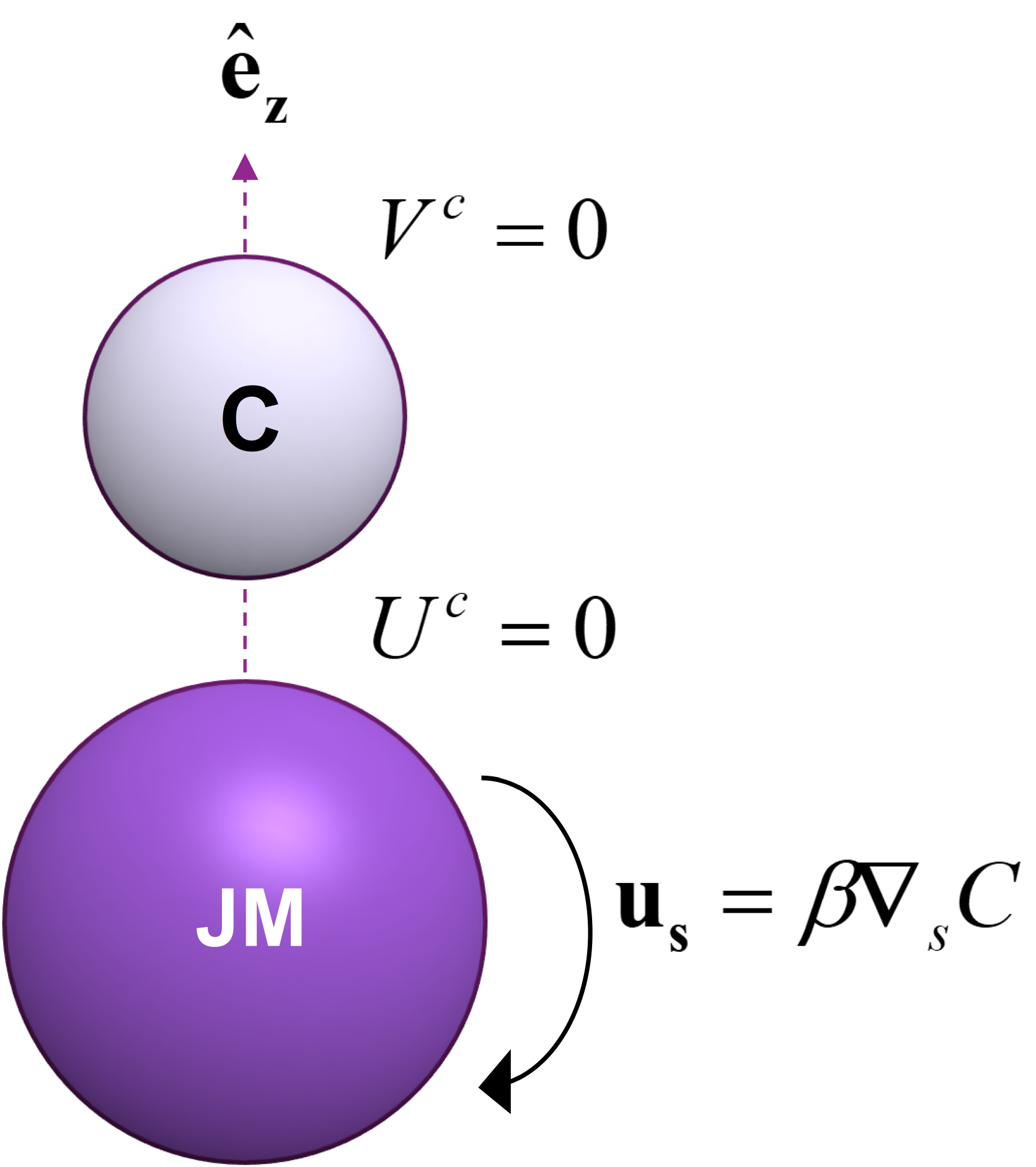}
    \caption{}
    \label{fig:sub3}
  \end{subfigure}
  \caption{\ref{fig:sub1} represents the sub-problem in which only the JM has translational velocity.  \ref{fig:sub2} shows the case in which only the cargo has translational velcoity.  Finally, \ref{fig:sub3} shows the problem where neither particle has any translational velocity and there is only a slip velocity, $\bm{u_s}$, prescribed on the JM's surface due to diffusiophoresis.}
  \label{fig:threesubprobs}
\end{figure}

Since Stokes flow is linear, we split the net force on each particle into three sub-forces respectively; \\ $F^{net} = (UF^a + VF^b + F^c){\bm{\hat{e}_z}}  $ each of which corresponds to the three sub-problems described in Figure \ref{fig:threesubprobs}.  We use superscripts to differentiate the sub-problems.  We will refer to the force due to sub-problem 3 (Figure \ref{fig:sub3}) with the slip velocity as the propulsion force since this force drives the JM and the cargo. The other two forces correspond to the velocity dependent hydrodynamic drags that balance the propulsion force. 

The boundary conditions \ref{eq:BCv1}, \ref{eq:BCv2}, \ref{eq:BCv3} can be rewritten in terms of the stream functions, and like the concentration field, can be split into the three sub-problems.  These boundary conditions can then be applied to the general solution \ref{eq:eigenexpansion} to obtain a coupled system of equations which we can then numerically invert for the coefficients \\ $X_{1n}^j,X_{2n}^j, Y_{1n}^j, Y_{2n}^j , j \in \{a,b,c \} $ for each sub-problem.  

Finally, we can calculate the propulsion forces for both spheres either by integrating the stress tensor corresponding to sub-problem 3 (Figure \ref{fig:sub3}) about the surface of the two spheres as defined in Equations \ref{eq:FJnet} and \ref{eq:FCnet} or by using the reciprocal theorem \cite{happel2012low} which relates the forces $F^c$ to the stress fields $\bm{\sigma}^a, \bm{\sigma}^b$ 

\begin{equation}\label{eq:FJRecip}
F_{_J}^{^{c}} {\bm{\hat{e}_z}} = \iint\limits_{S_J,S_A} ( {\bm{\hat{e}_z}} \cdot  \boldsymbol{\sigma}^a \cdot {\bm{\hat{n}}})ds,
\end{equation}

\begin{equation}\label{eq:FCRecip}
F_{_C}^{^{c}} {\bm{\hat{e}_z}} = \iint\limits_{S_J,S_A} ( {\bm{\hat{e}_z}} \cdot  \boldsymbol{\sigma}^b \cdot {\bm{\hat{n}}})ds.  
\end{equation}

The advantage of using the latter approach is that we can directly calculate the propulsion forces from the stress tensors of sub-problems 1 and 2 (Figures \ref{fig:sub1} and \ref{fig:sub2}) which have already been well studied in literature \cite{kim2013microhydrodynamics,jeffrey1984calculation}. Note that the integrals in \ref{eq:FJRecip} and \ref{eq:FCRecip} are only over the surface of the JM. Finally, we reduce the force free conditions in Equations \ref{eq:FJnet} and \ref{eq:FCnet} to Equation \ref{eq:matrix}

\begin{equation}\label{eq:matrix}
\begin{bmatrix}
F_{_J}^{^a} \quad & F_{_J}^{^b} \\[6pt]
F_{_C}^{^b} & F_{_C}^{^a}             
\end{bmatrix}
\begin{bmatrix}
U  \\[6pt]
V            
\end{bmatrix}
=
\begin{bmatrix}
-F_{_J}^{^c} \\[6pt]
-F_{_C}^{^c}             
\end{bmatrix}. 
\end{equation}

\section{\label{sec:Results}Results and Discussion}
We define the size of the cargo relative to the JM $(R_1)$ as  $\lambda = R_2/R_1$.  We also define a scaled edge-to-edge separation as $e = R/R_1 - (1+ \lambda)$. The results are evaluated by truncating the concentration field \ref{eq:conc1} and stream functions \ref{eq:eigenexpansion} to 100 terms in and using the first 25 terms of the stream function expansion to evaluate the forces on the JM and the cargo by integrating the stress on each sphere for each sub-problem. The forces presented in this section are scaled by a factor of $\frac{6\pi\mu\beta JR_1}{D}$ and the terminal velocities are presented scaled by a factor of $\frac{\beta J}{D}$. The twin multiple approach is singular at the limit of $e\to\infty$ \cite{jeffrey1984calculation} so we limit the closest approach in our results to $e = 0.1$.

\begin{figure}
\includegraphics[width=0.2\textwidth]{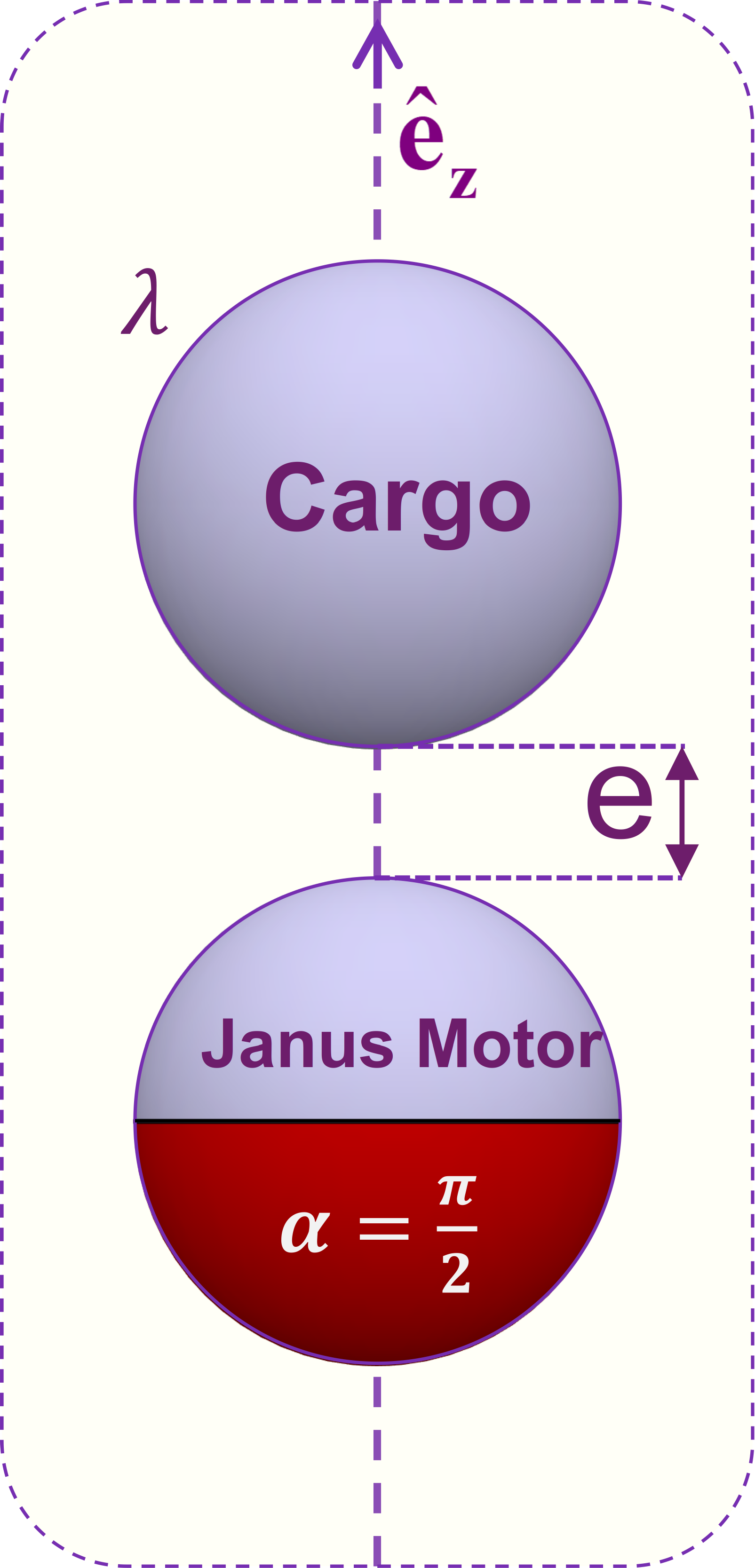}
\caption{An illustration of a JM and a cargo with $\alpha = \pi/2$. }
\label{fig:axi}
\end{figure}

\subsection{\label{sec:stationarycargo}Stationary Cargo}

\begin{figure} 
  \centering
  \begin{subfigure}{.3\linewidth}
    \centering
    \includegraphics[width = \linewidth]{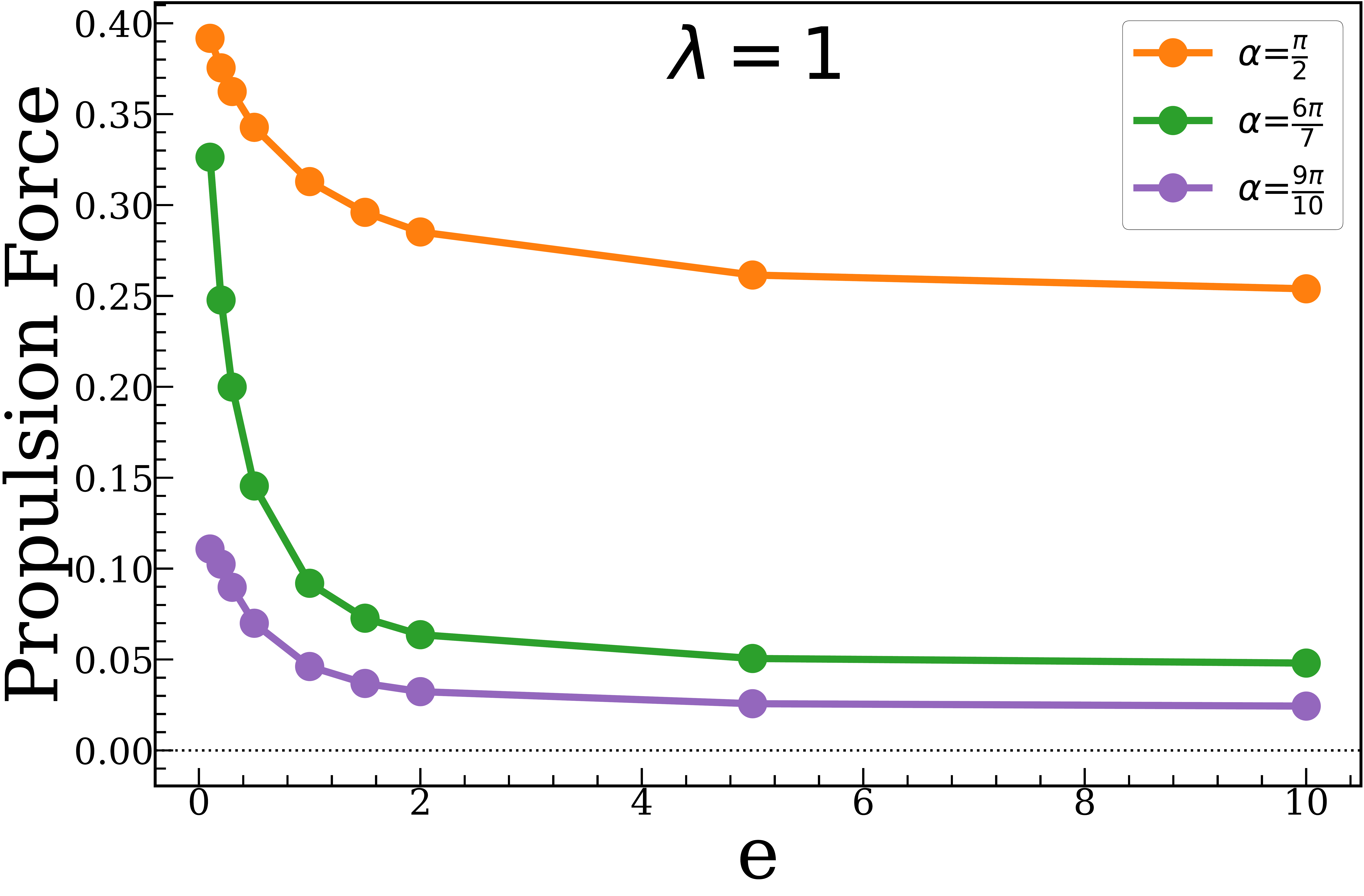}
    \caption{}
    \label{fig:plot_9_19_14_JM}
  \end{subfigure}%
  \hspace{2em}
   \begin{subfigure}{.3\linewidth}
    \centering
    \includegraphics[width = \linewidth]{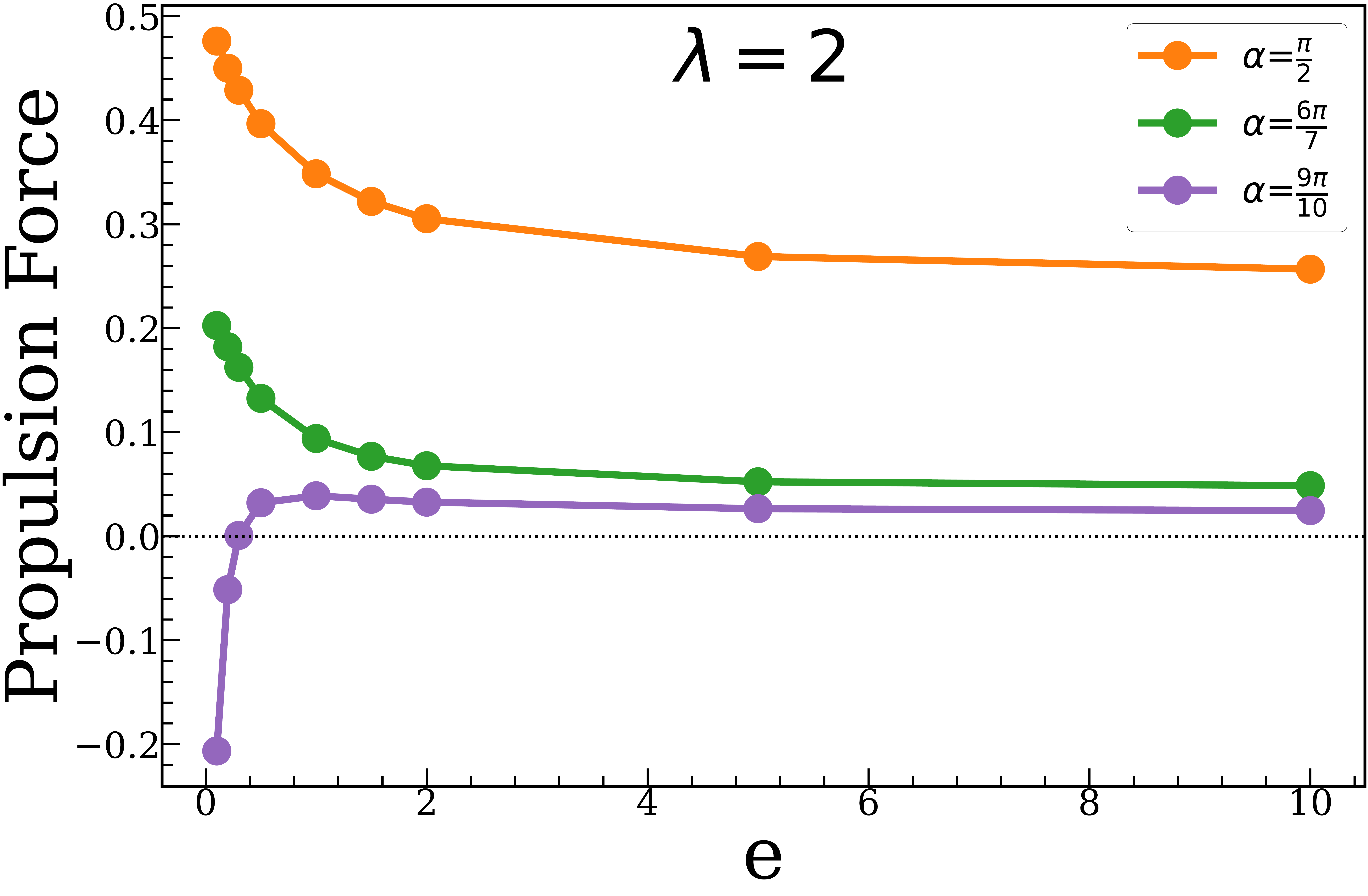}
    \caption{}
    \label{fig:plot_17_18_13_JM}
  \end{subfigure}%
  \hspace{2em}
   \begin{subfigure}{.3\linewidth}
    \centering
    \includegraphics[width = \linewidth]{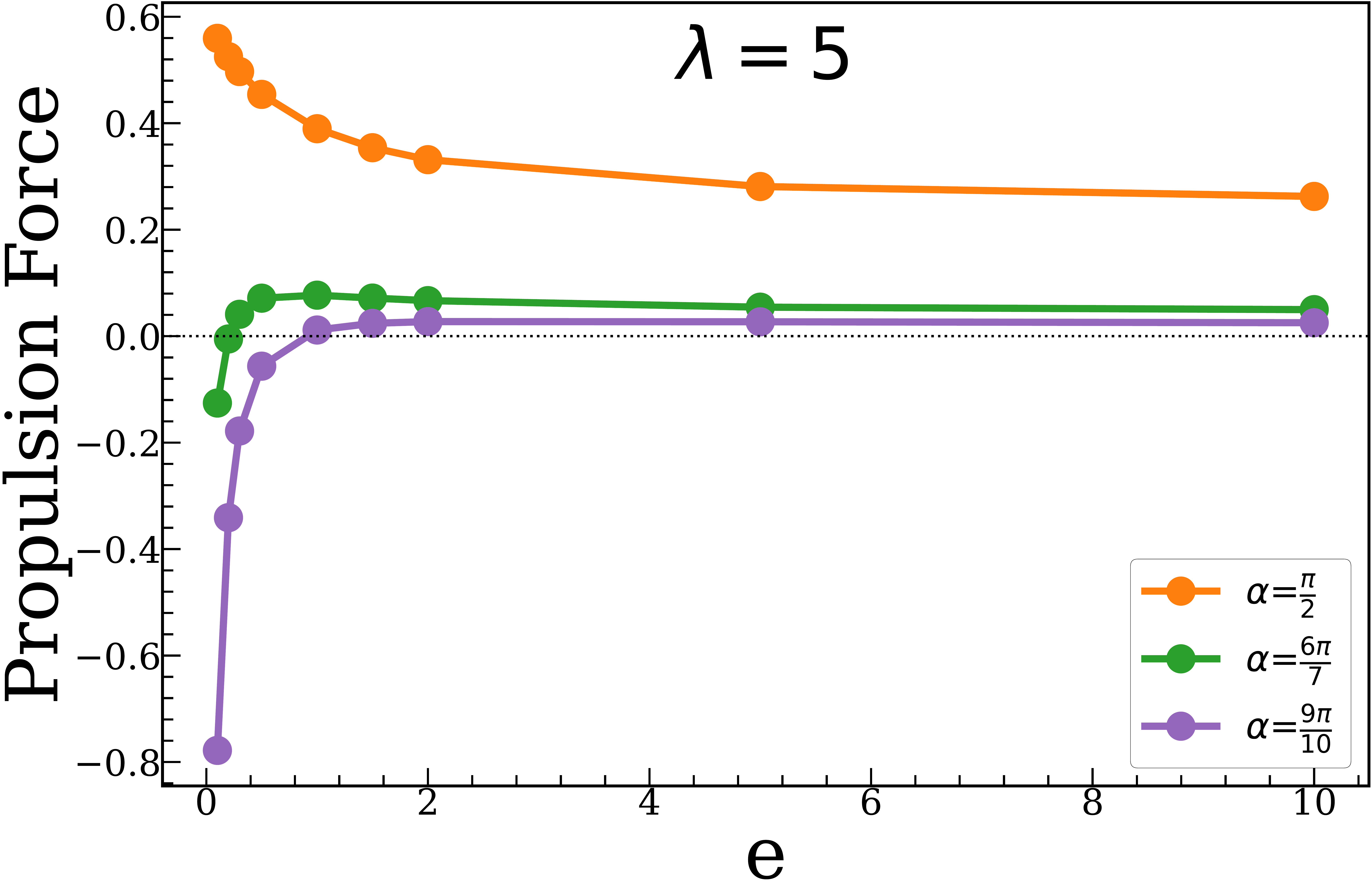}
    \caption{}
    \label{fig:plot_15_1_4_JM}
  \end{subfigure}%
  \hspace{2em}

 \centering
  \begin{subfigure}{.3\linewidth}
    \centering
    \includegraphics[width = \linewidth]{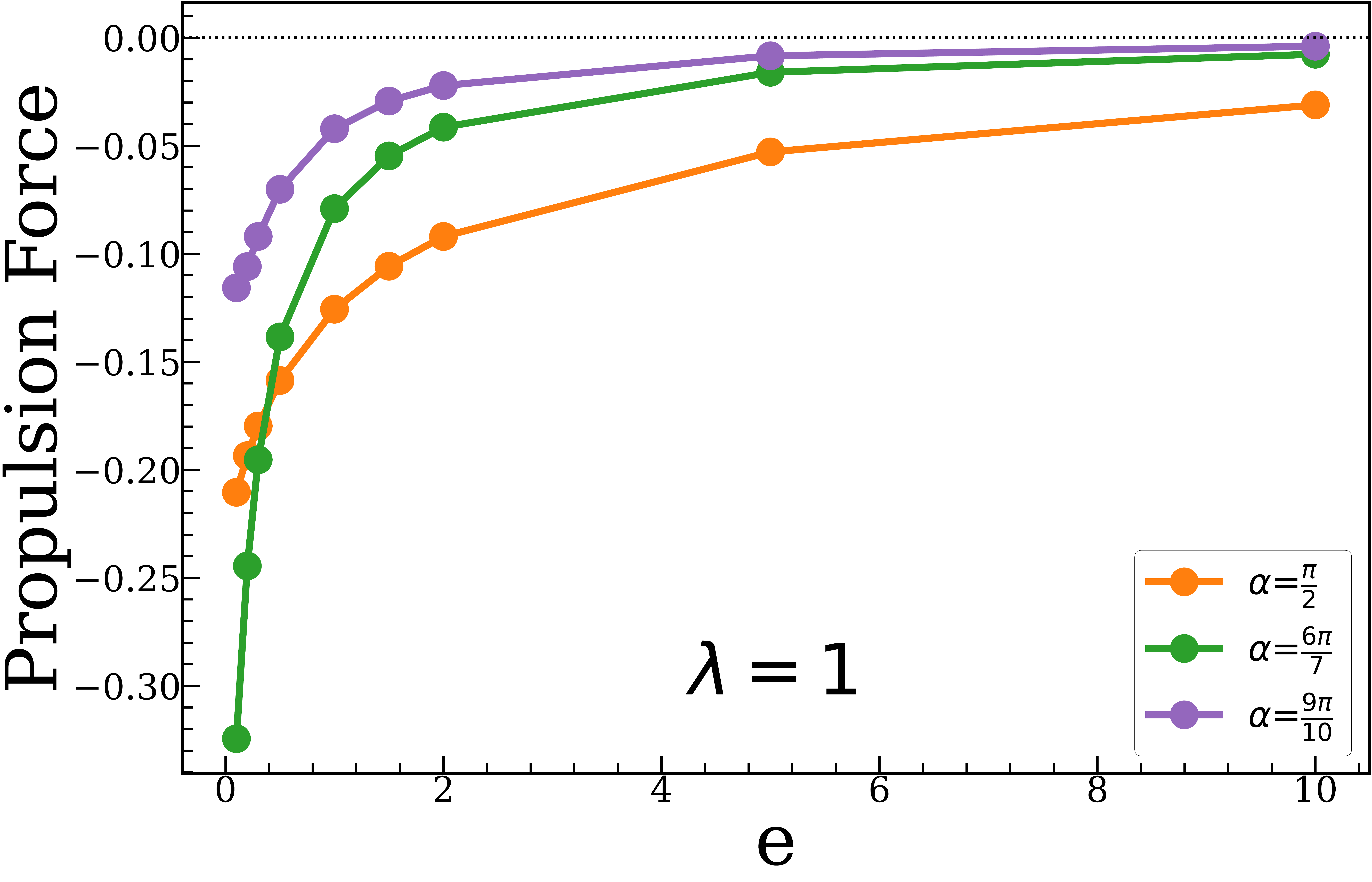}
    \caption{}
    \label{fig:plot_9_19_14_Cargo}
  \end{subfigure}%
  \hspace{2em}
     \begin{subfigure}{.3\linewidth}
    \centering
    \includegraphics[width = \linewidth]{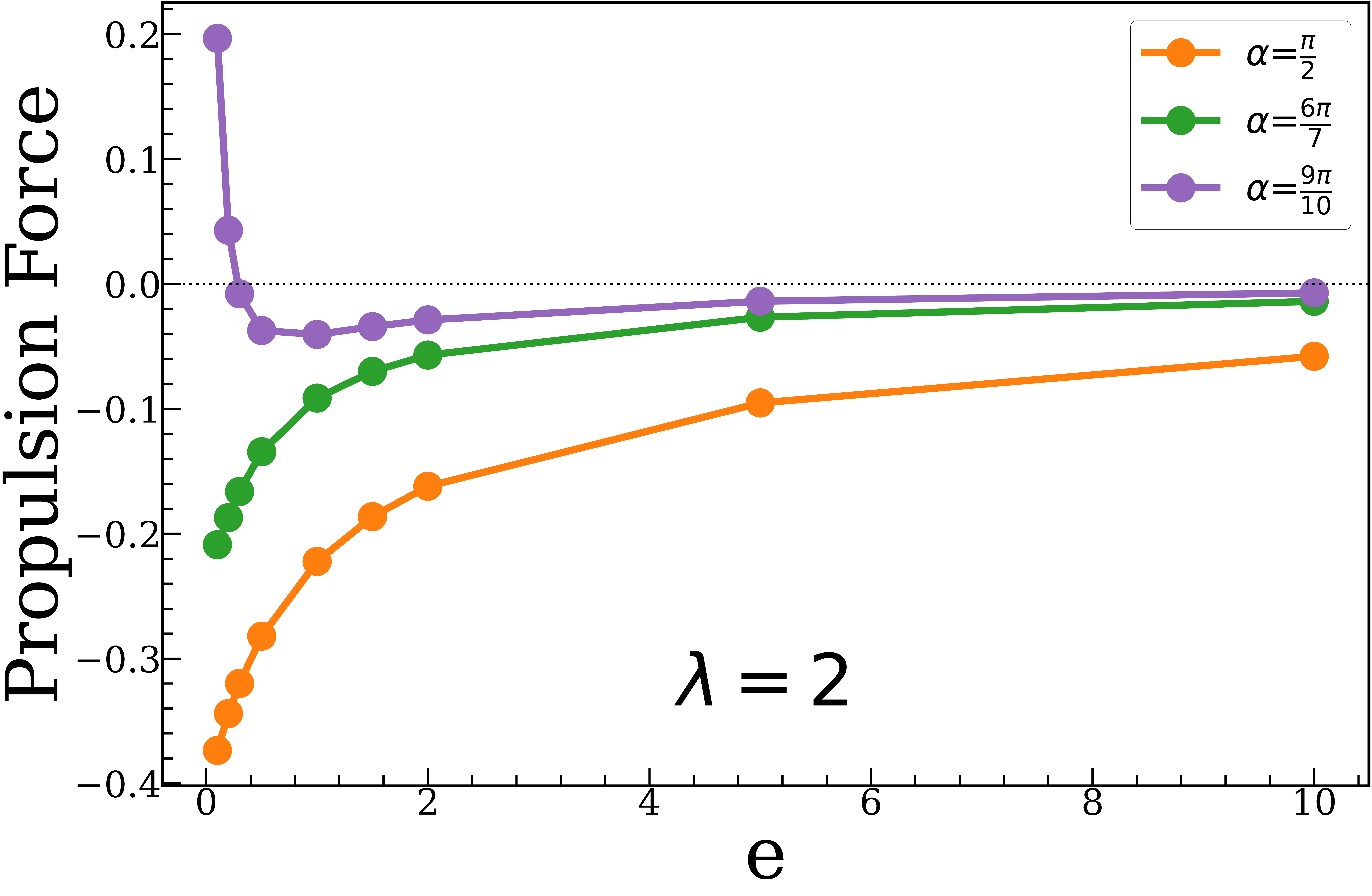}
    \caption{}
    \label{fig:plot_17_18_13_Cargo}
  \end{subfigure}%
  \hspace{2em}
  \begin{subfigure}{.3\linewidth}
    \centering
    \includegraphics[width = \linewidth]{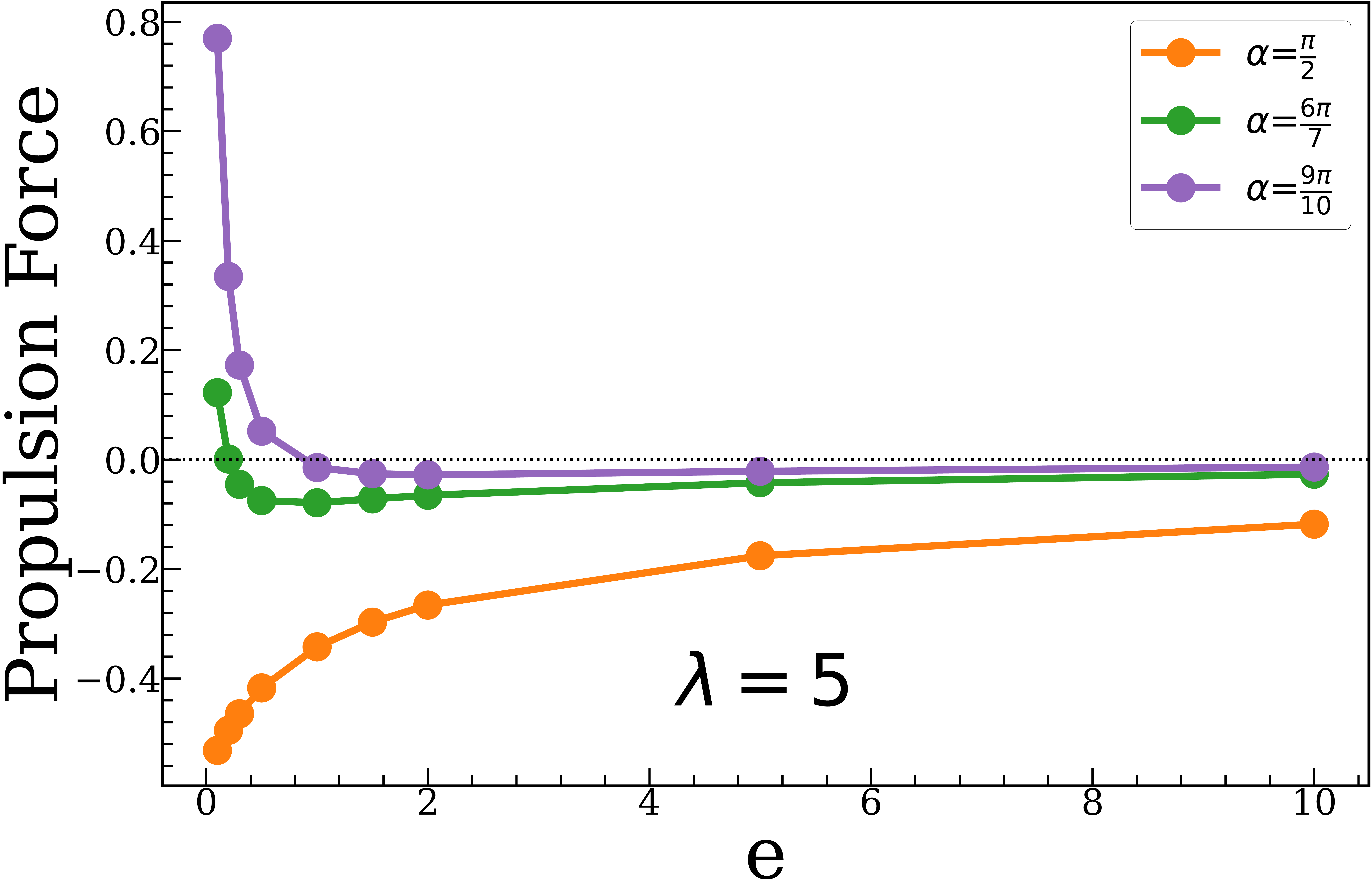}
    \caption{}
    \label{fig:plot_15_1_4_Cargo}
  \end{subfigure}%
  \hspace{2em}
 

    \caption{ Propulsion forces for the JM (a,b,c) and the cargo (d,e,f) with $\alpha =\frac{\pi}{2}, \frac{6 \pi}{7}, \frac{9 \pi}{10} $  and $\lambda$ = (a,d) 1, (b,e) 2 and (c,f) 5.}
  \label{fig:lambda2}
\end{figure}

\begin{figure} 
\begin{subfigure}{.9\linewidth}
  \begin{subfigure}{.3\linewidth}
    \centering
    \includegraphics[width = \linewidth]{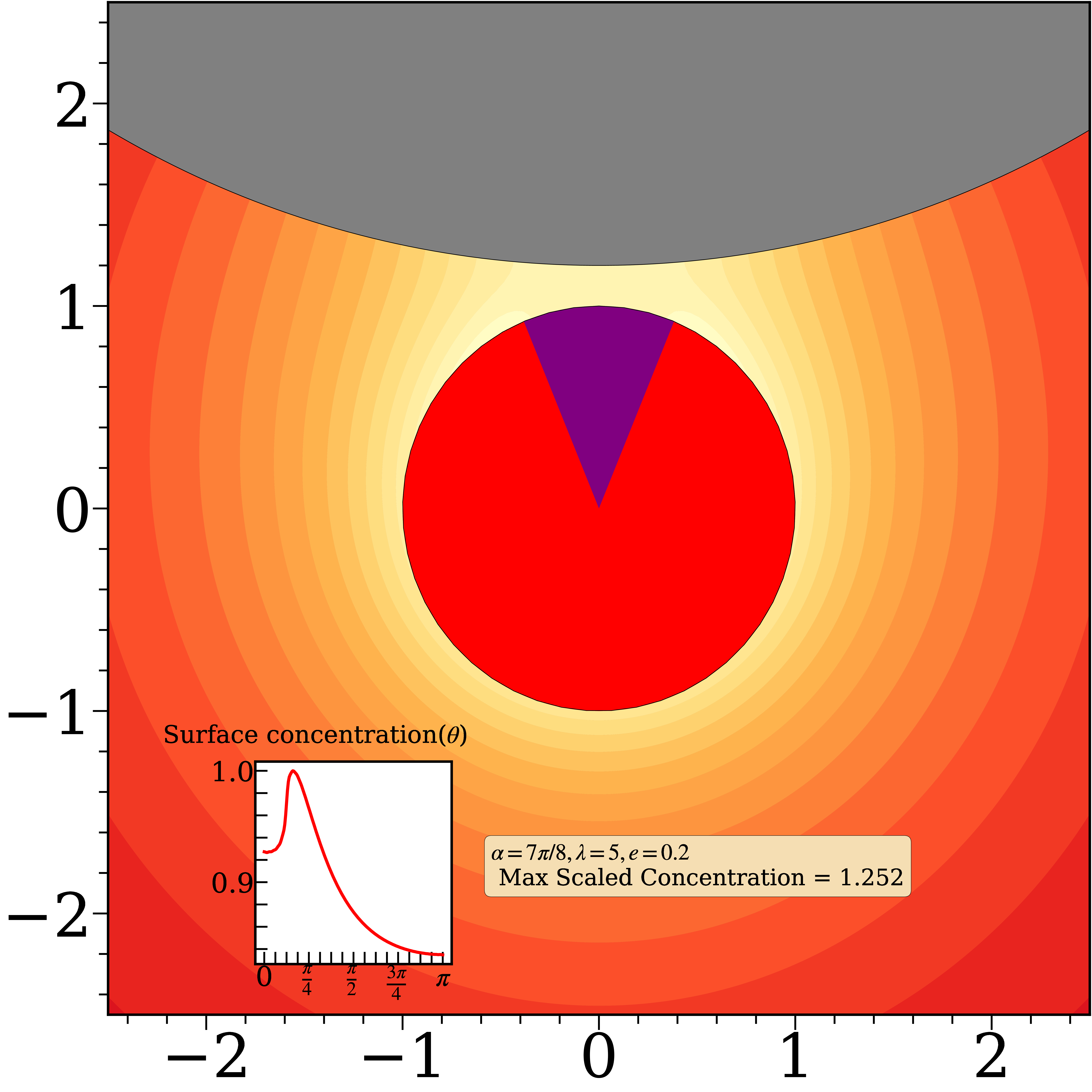}
    \caption{}
    \label{fig:9conc}
  \end{subfigure}%
  \hspace{0.5em}
    \begin{subfigure}{.3\linewidth}
    \centering
    \includegraphics[width = \linewidth]{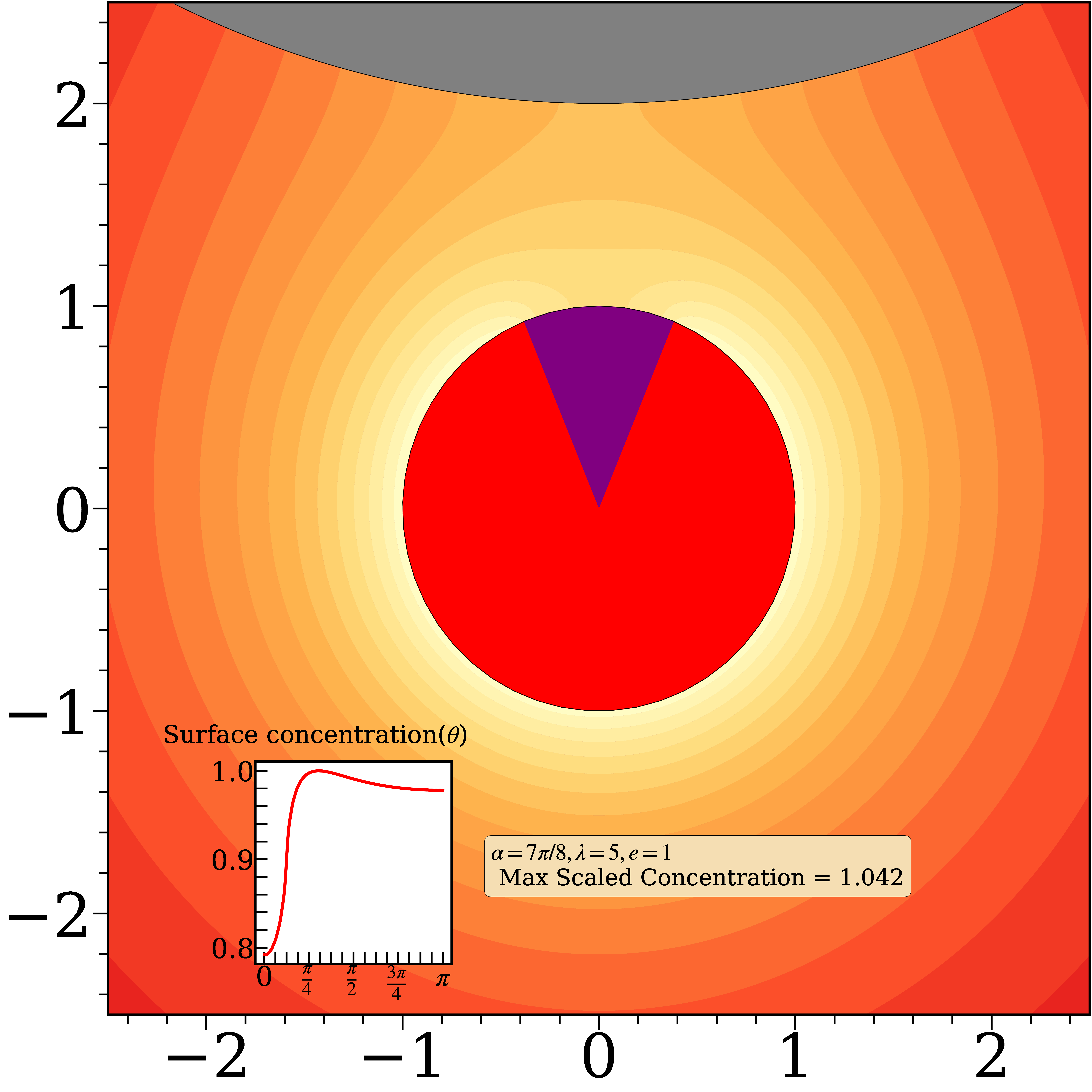}
    \caption{}
    \label{fig:10conc}
  \end{subfigure}
  \hspace{0.5em}
    \begin{subfigure}{.3\linewidth}
    \centering
    \includegraphics[width = \linewidth]{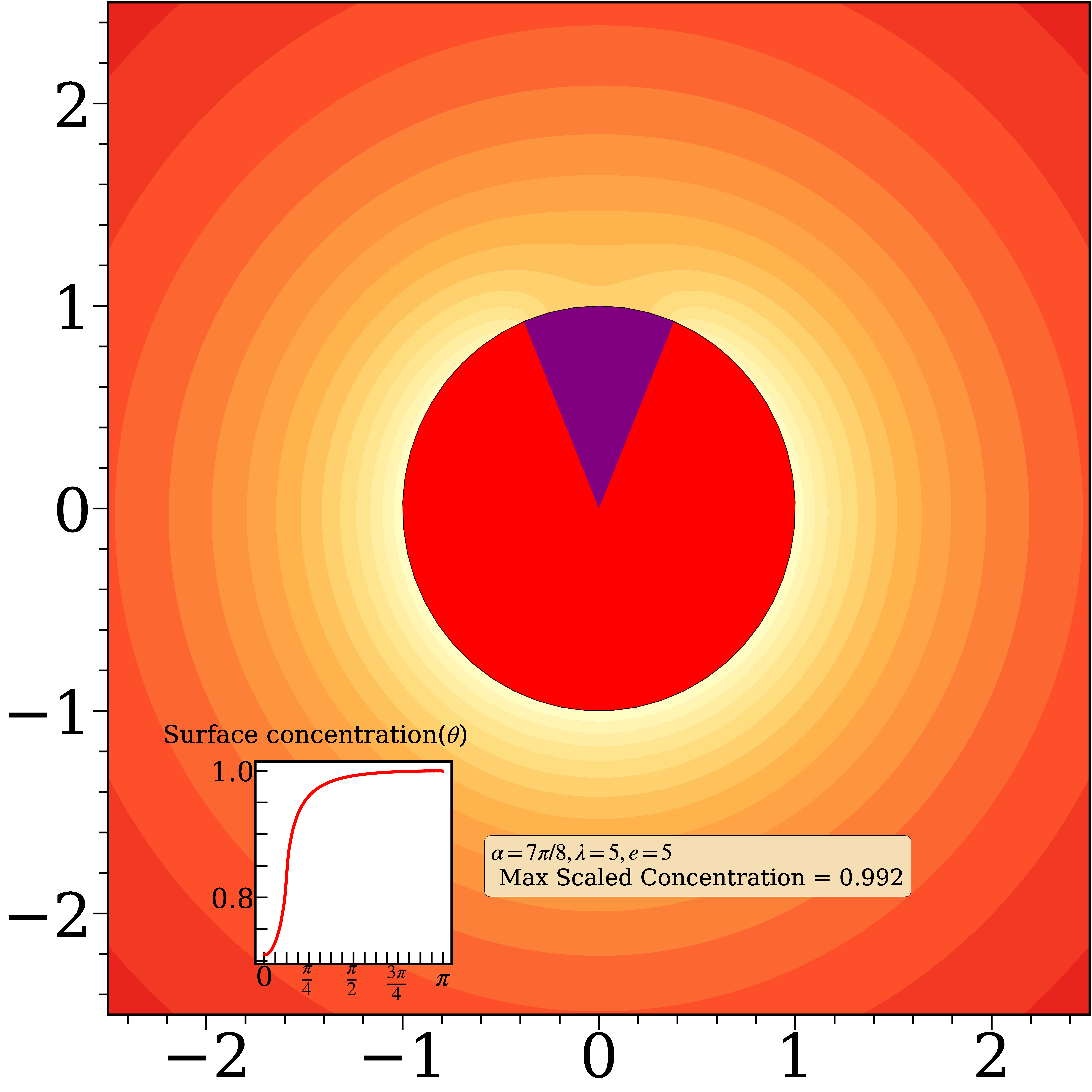}
    \caption{}
    \label{fig:11conc}
  \end{subfigure}%
 

  \begin{subfigure}{.3\linewidth}
    \centering
    \includegraphics[width = \linewidth]{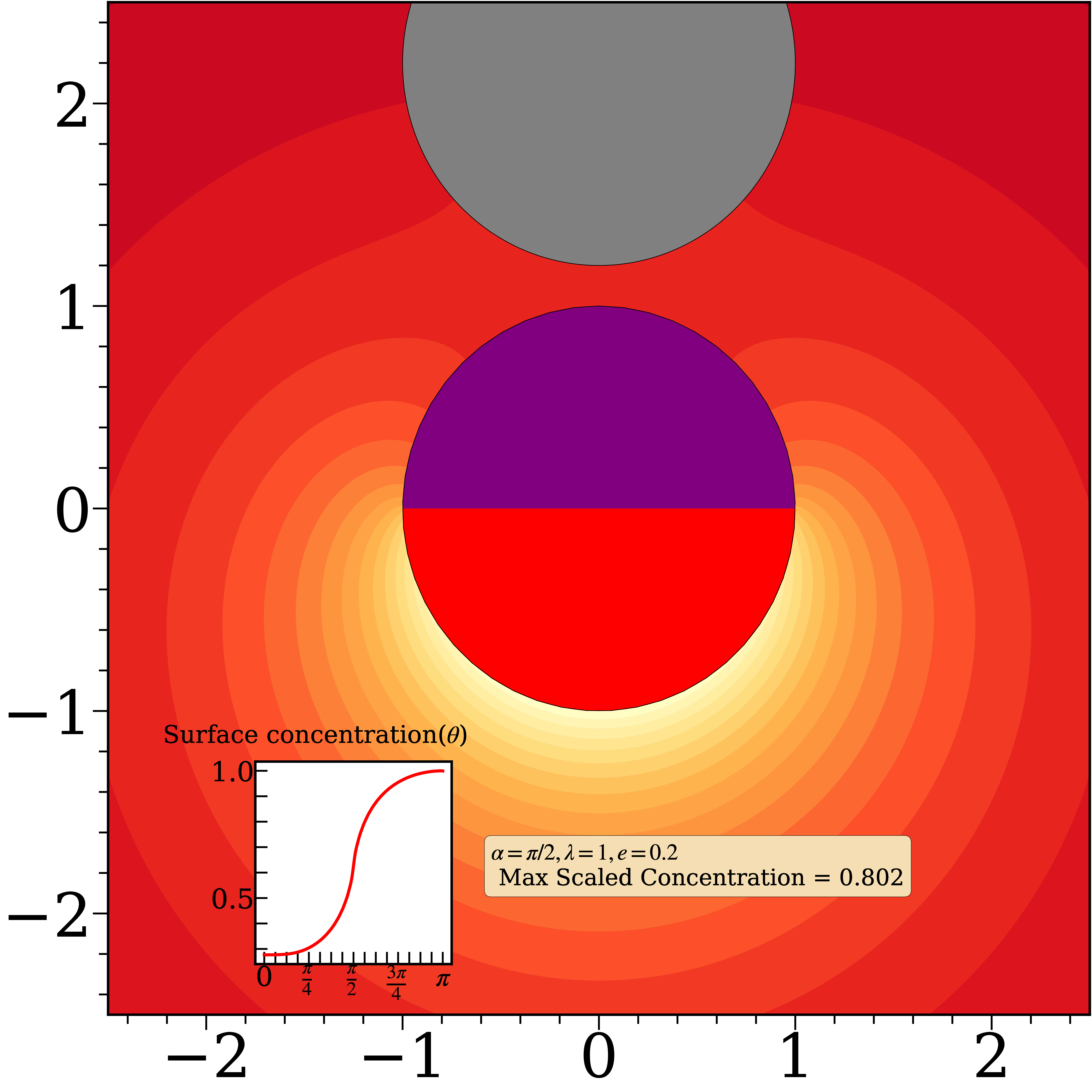}
    \caption{}
    \label{fig:14conc}
  \end{subfigure}%
  \hspace{0.5em}
    \begin{subfigure}{.3\linewidth}
    \centering
    \includegraphics[width = \linewidth]{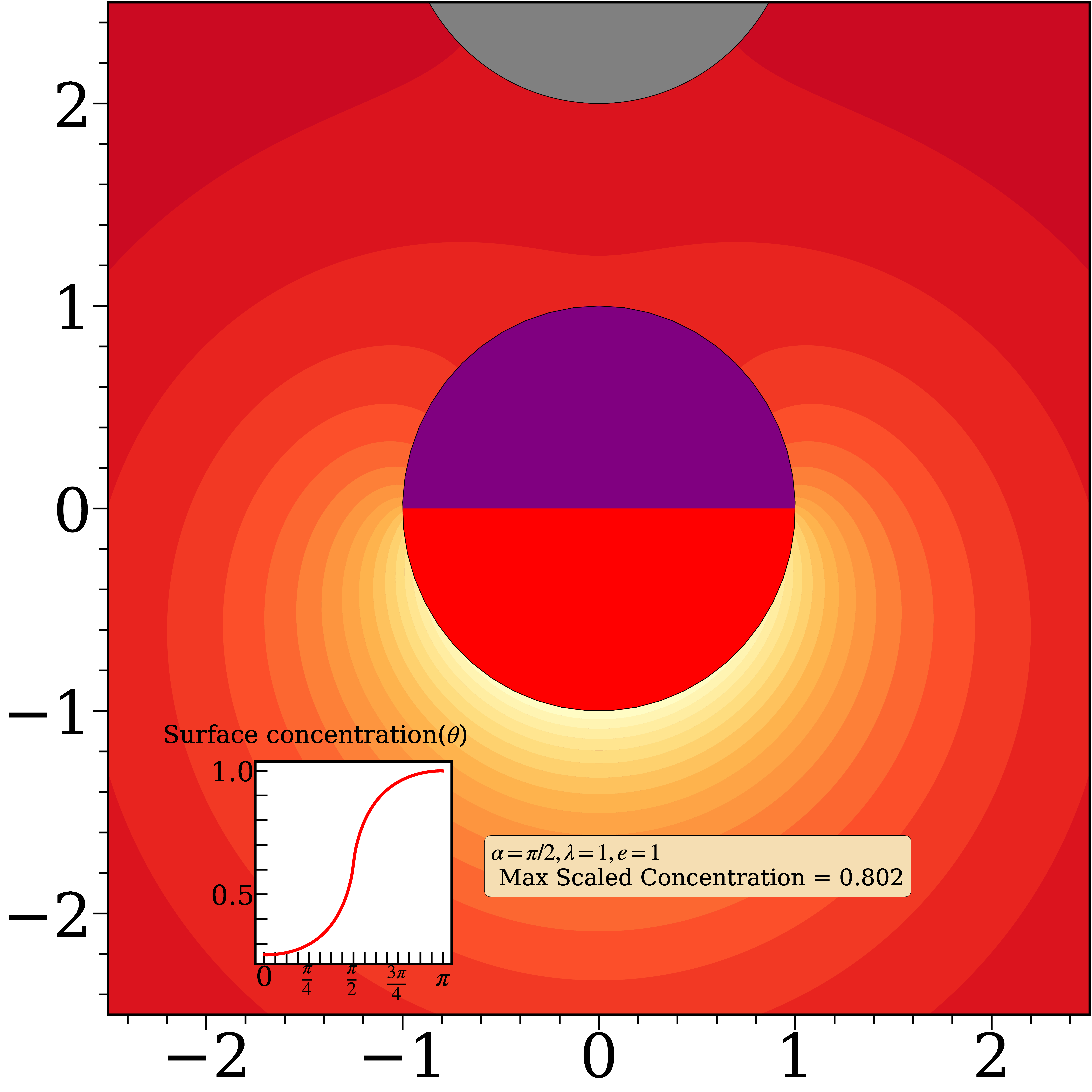}
    \caption{}
    \label{fig:12conc}
  \end{subfigure}%
  \hspace{0.5em}
    \begin{subfigure}{.3\linewidth}
    \centering
    \includegraphics[width = \linewidth]{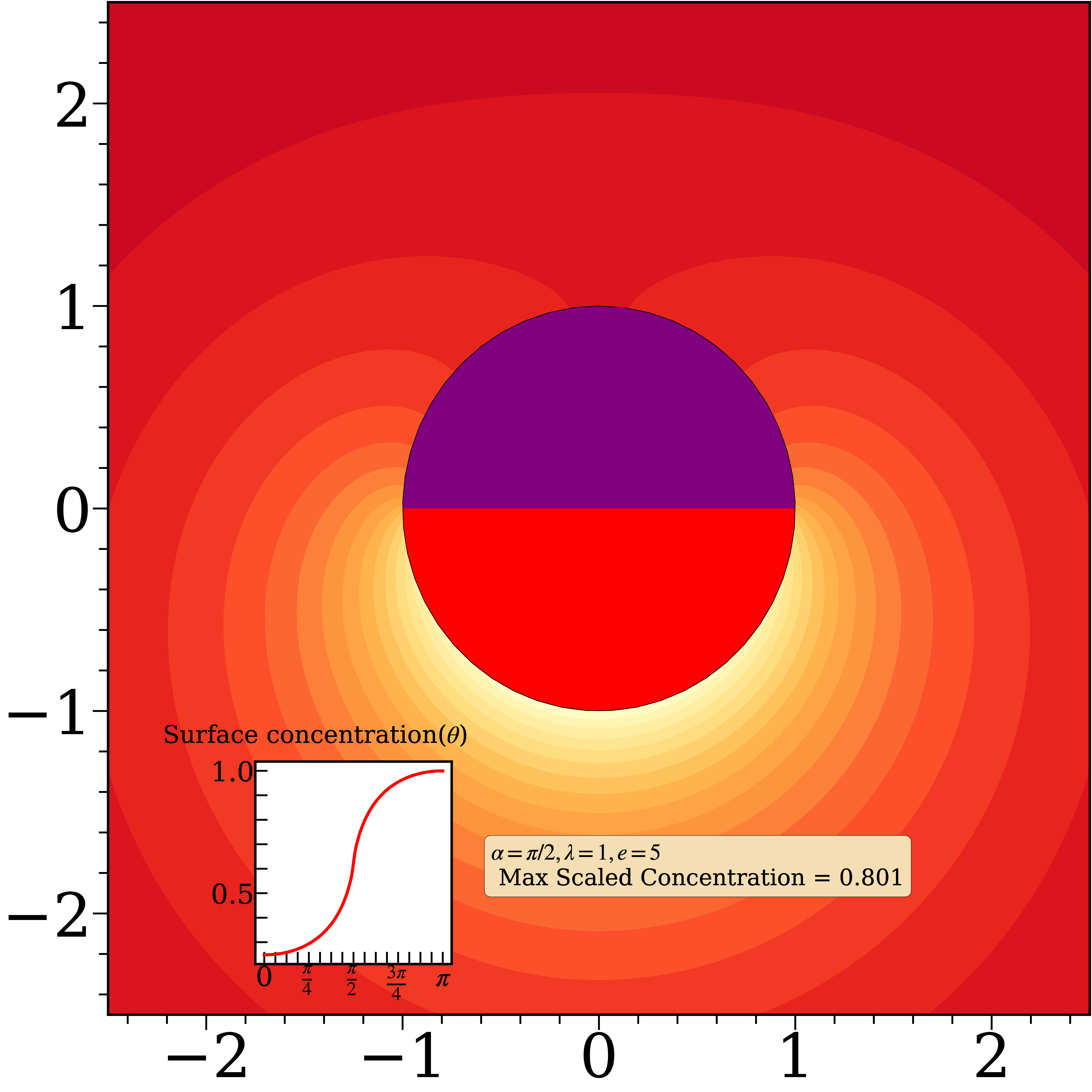}
    \caption{}
    \label{fig:13conc}
  \end{subfigure}%
  \end{subfigure}
\begin{subfigure}{0.07\linewidth}
\includegraphics[width = \linewidth,valign=c]{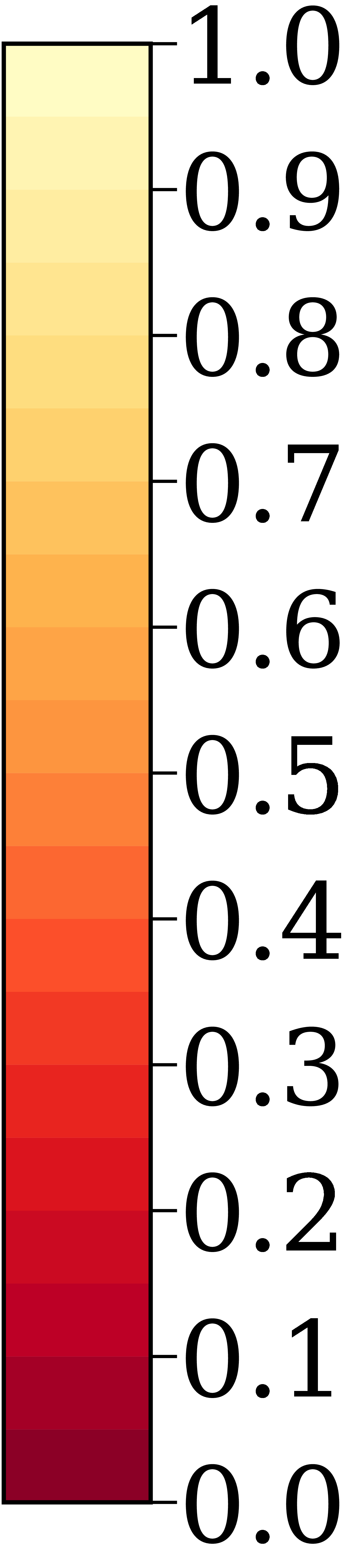}
\caption{}
\label{fig:conclegend}
\end{subfigure}%

    \caption{ Normalized concentration fields (with an inset of surface concentration along the JM's surface) for the JM and the cargo with $(\lambda=5, \alpha=\frac{7\pi}{8})$ at $e = $ (a) 0.2, (b) 1, (c) 5 and for $(\lambda=1, \alpha=\frac{\pi}{2})$ at $e = $  (d) 0.2, (e) 1, (f) 5.}
  \label{fig:concentration}
\end{figure}

\begin{figure} 
\begin{subfigure}{0.9\linewidth}
  \begin{subfigure}{.3\linewidth}
    \centering
    \includegraphics[width = \linewidth]{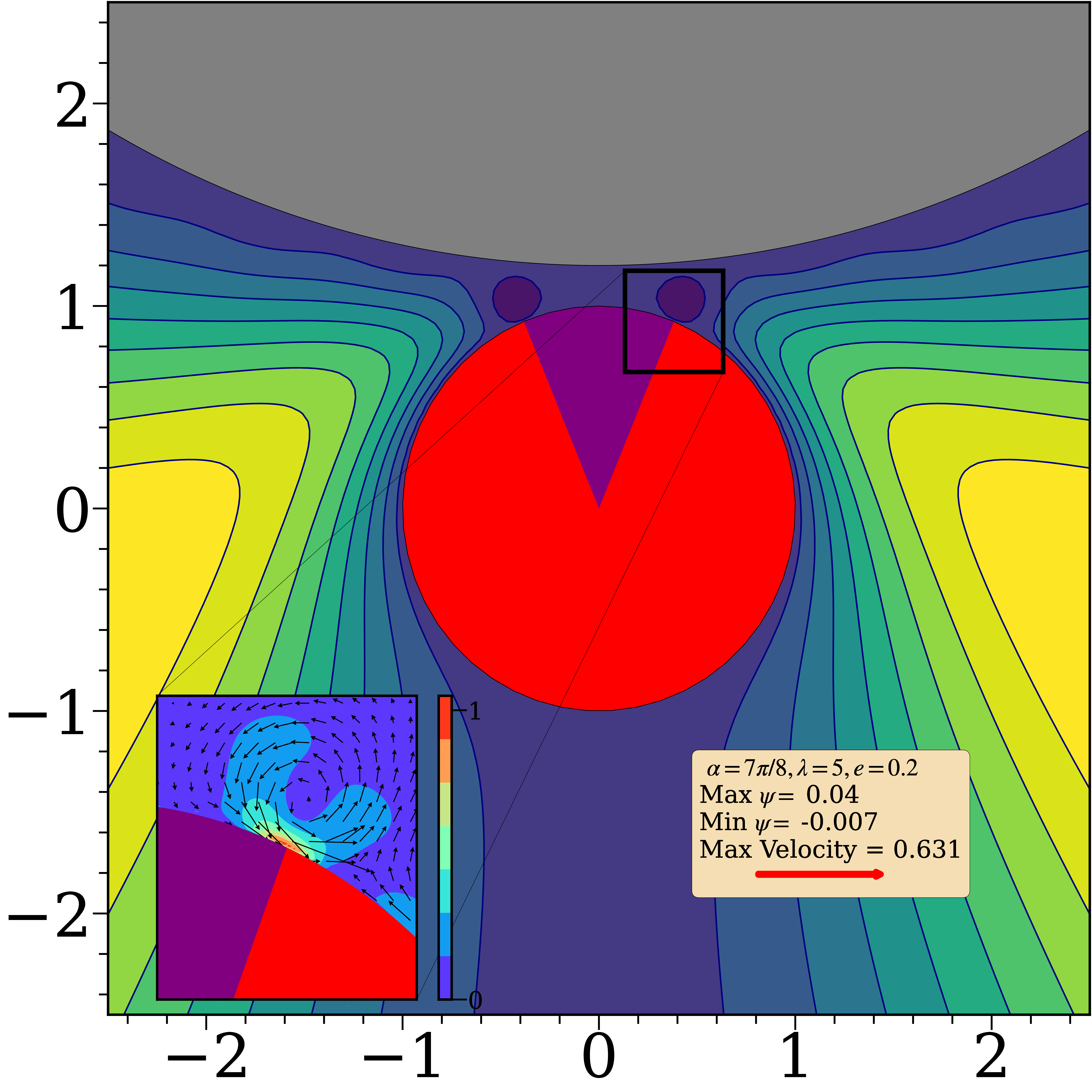}
    \caption{}
    \label{fig:9}
  \end{subfigure}%
  \hspace{1em}
    \begin{subfigure}{.3\linewidth}
    \centering
    \includegraphics[width = \linewidth]{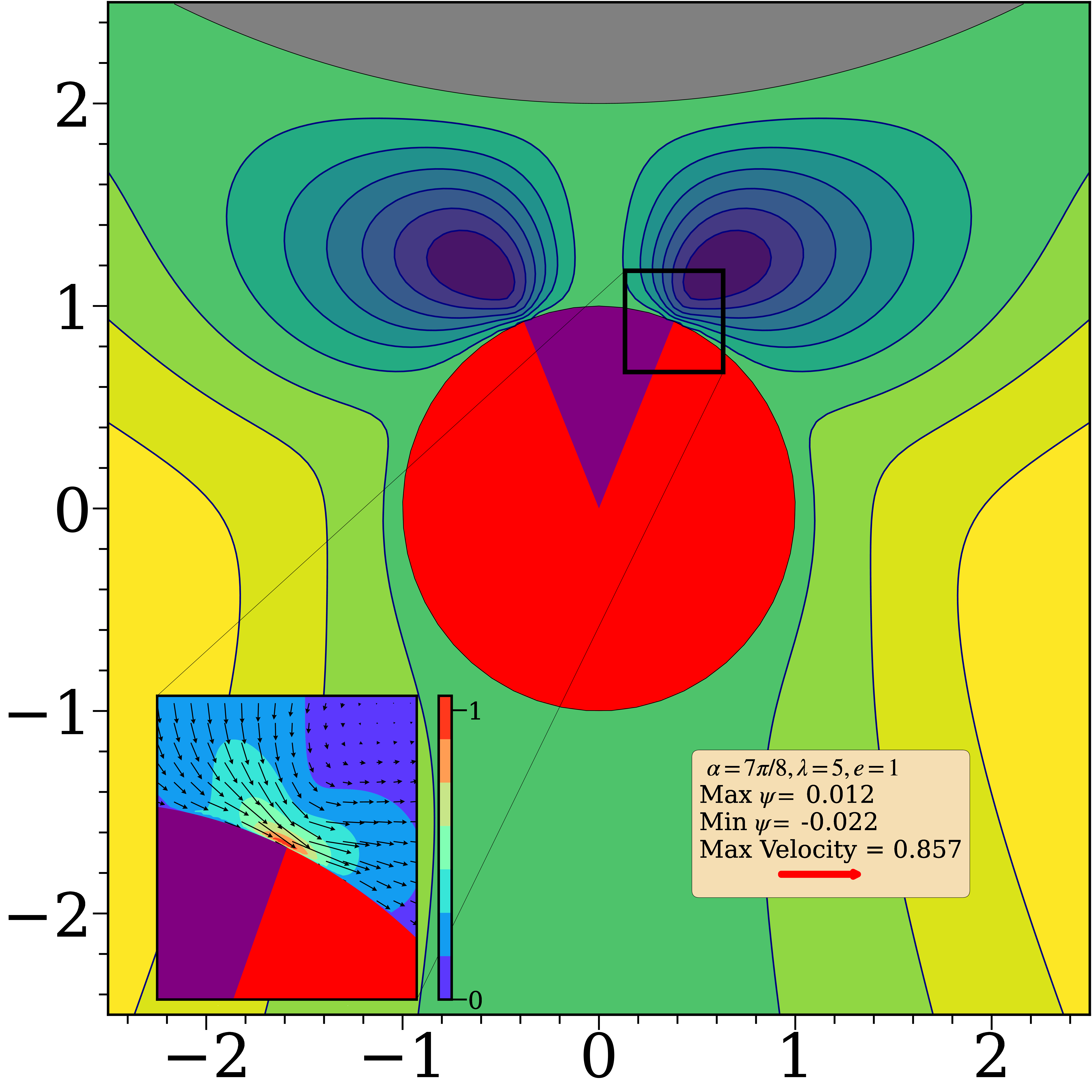}
    \caption{}
    \label{fig:10}
  \end{subfigure}%
  \hspace{1em}
    \begin{subfigure}{.3\linewidth}
    \centering
    \includegraphics[width = \linewidth]{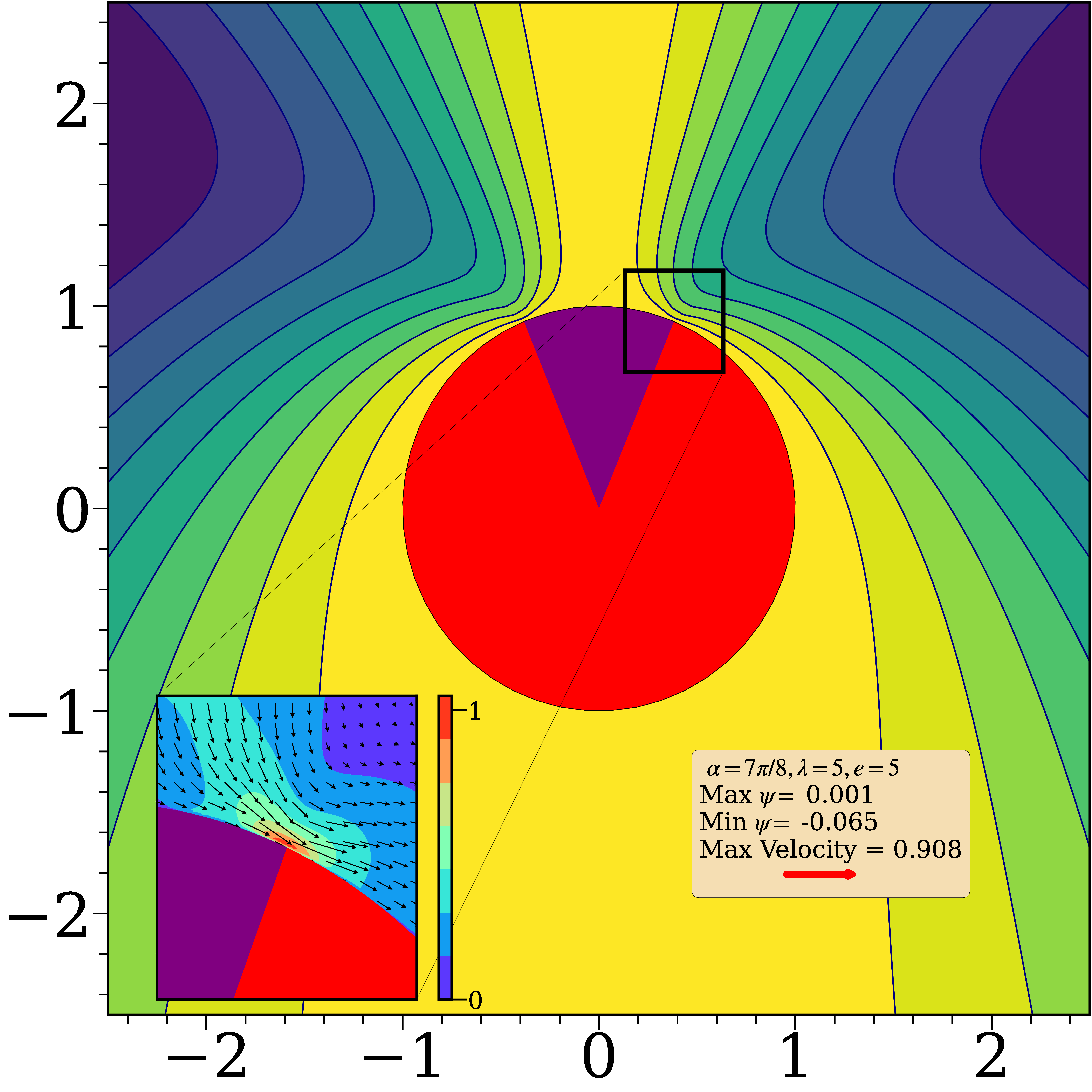}
    \caption{}
    \label{fig:11}
  \end{subfigure}%
  \hspace{1em}

  \begin{subfigure}{.3\linewidth}
    \centering
    \includegraphics[width = \linewidth]{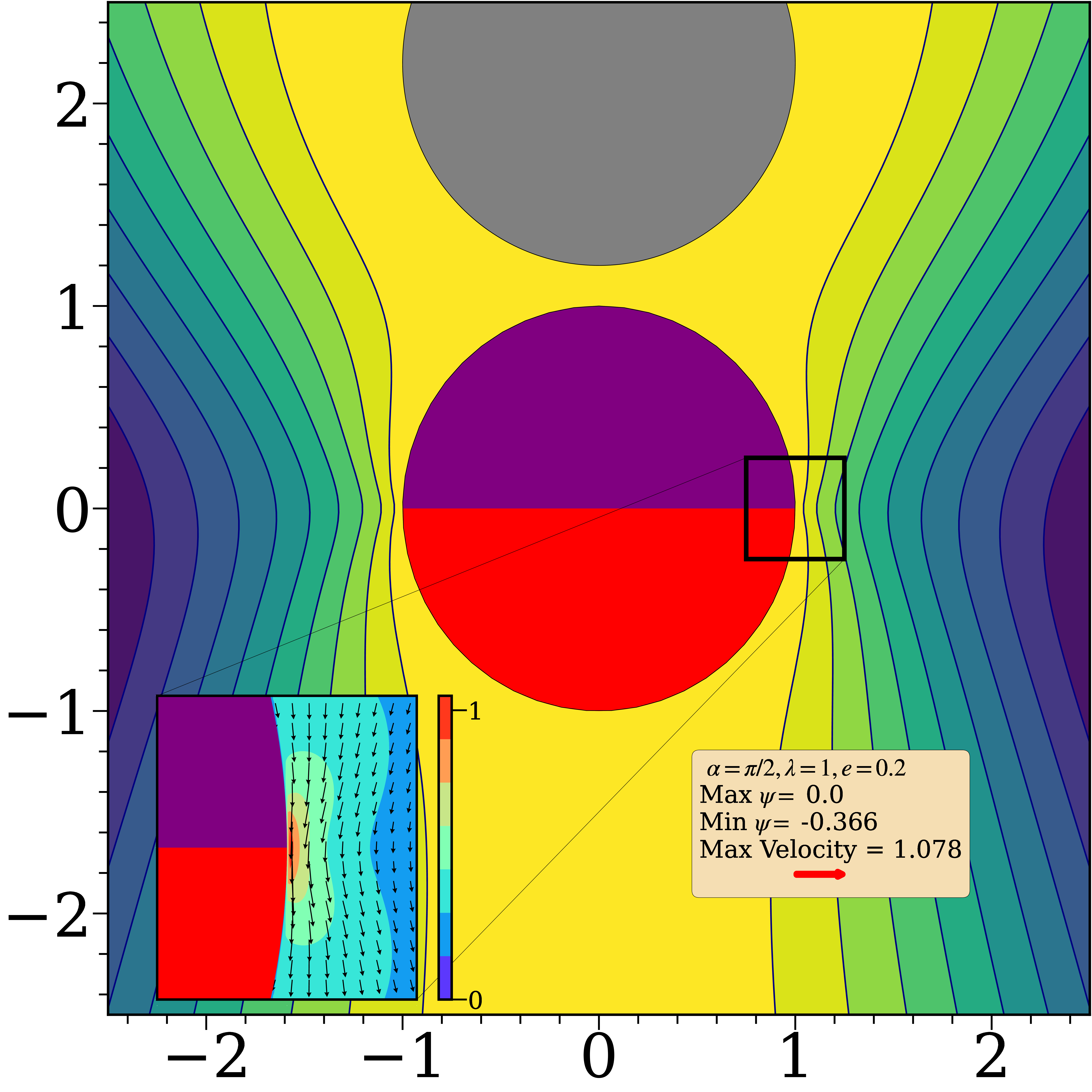}
    \caption{}
    \label{fig:14}
  \end{subfigure}%
  \hspace{1em}
    \begin{subfigure}{.3\linewidth}
    \centering
    \includegraphics[width = \linewidth]{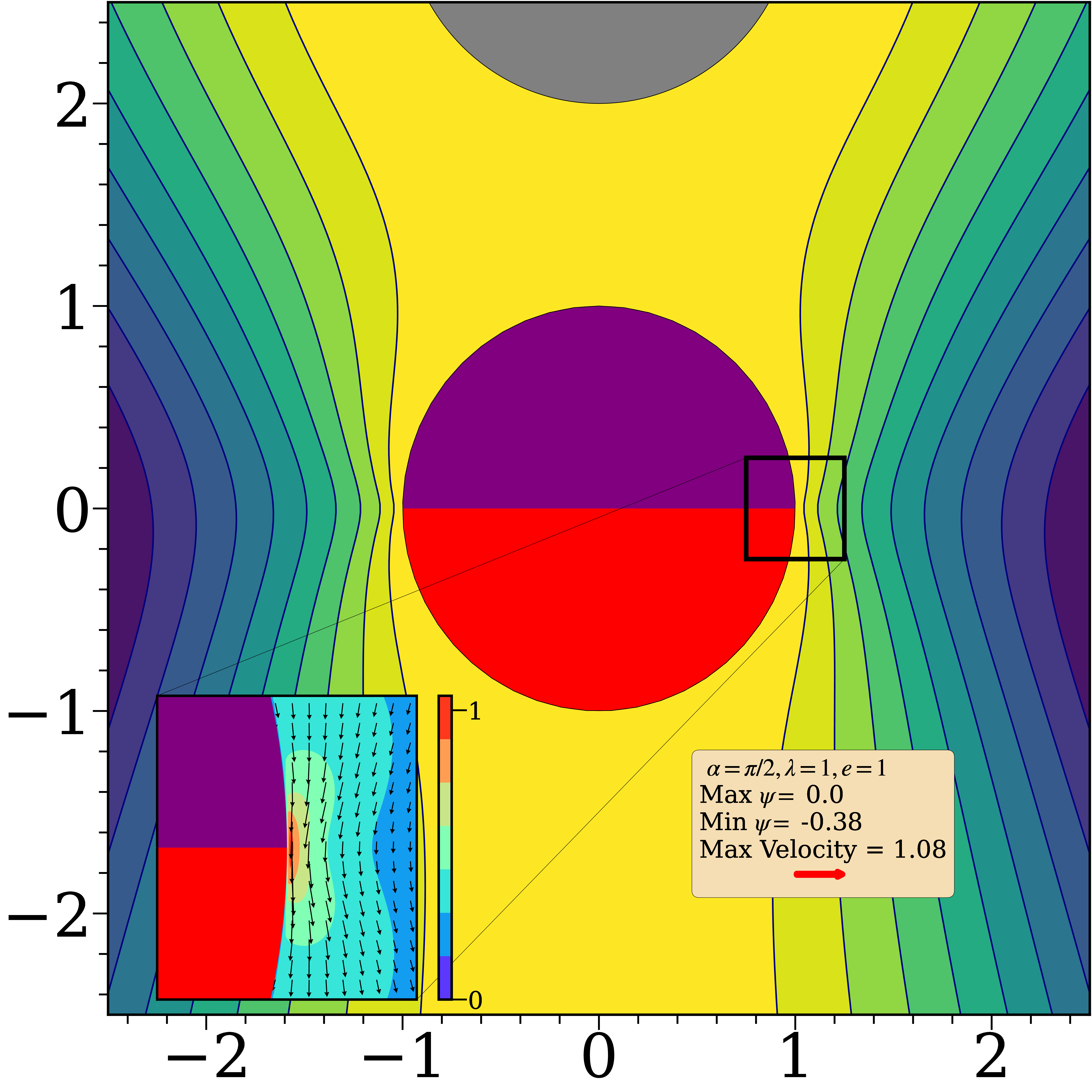}
    \caption{}
    \label{fig:12}
  \end{subfigure}%
  \hspace{1em}
    \begin{subfigure}{.3\linewidth}
    \centering
    \includegraphics[width = \linewidth]{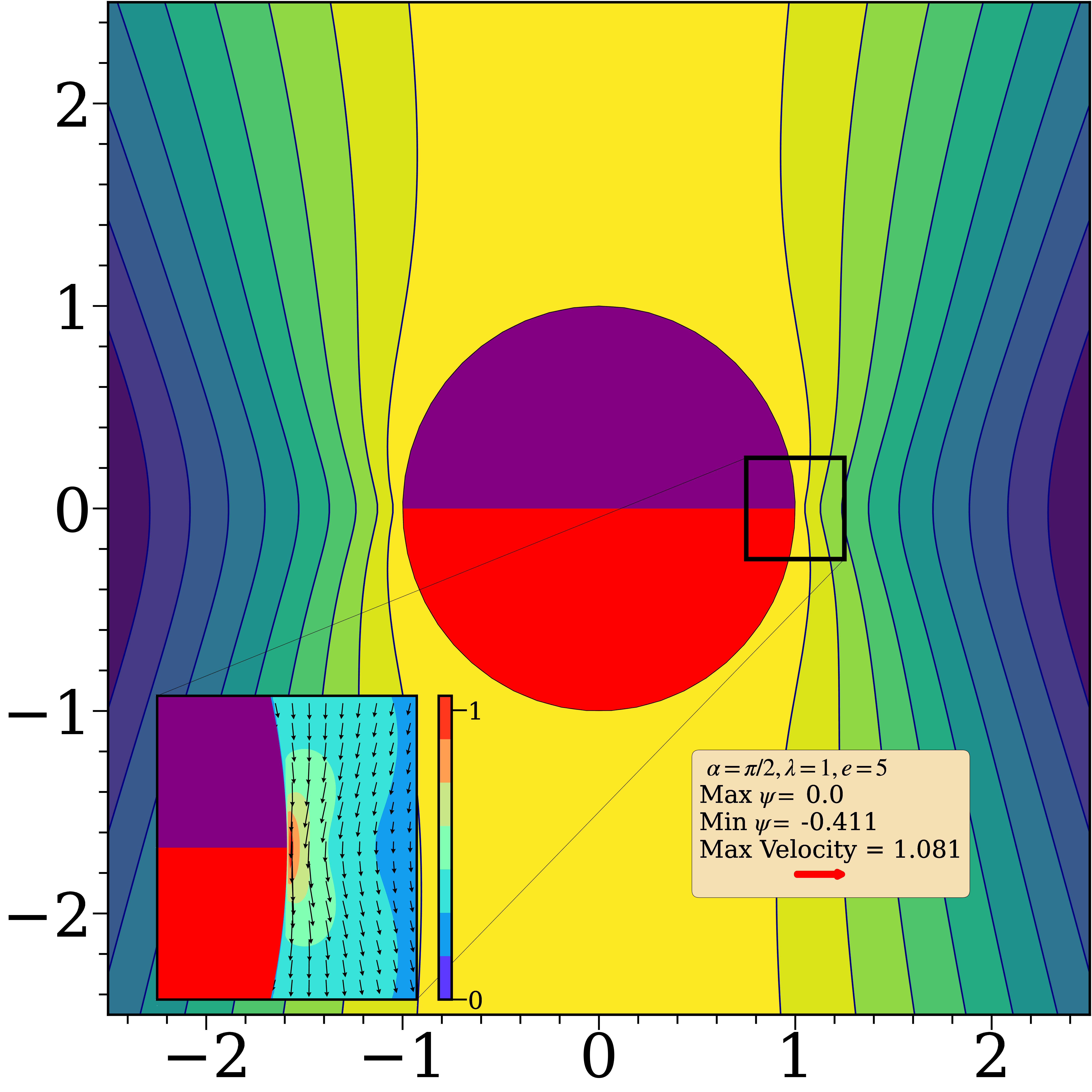}
    \caption{}
    \label{fig:13}
  \end{subfigure}%
\end{subfigure}
\begin{subfigure}{0.07\linewidth}
\includegraphics[width = \linewidth,valign=c]{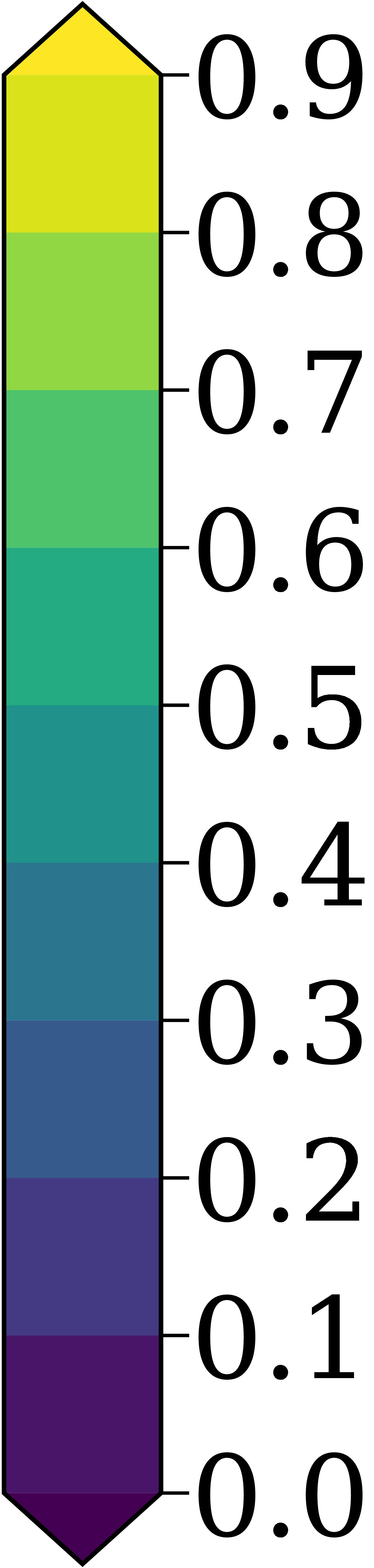}
\caption{}
\label{fig:streamlegend}
\end{subfigure}
    \caption{  Normalized streamlines and velocity fields for the JM and the cargo with $(\lambda=5, \alpha=\frac{7\pi}{8})$ at $e = $ (a) 0.2, (b) 1, (c) 5 and for $(\lambda=1, \alpha=\frac{\pi}{2})$ at $e = $ (d) 0.2, (e) 1, (f) 5.}
  \label{fig:streamplots}
\end{figure}

In Figure \ref{fig:lambda2} we consider the propulsion forces for $\lambda=1,2$ and $5$ for 3 different cap sizes $\alpha=\frac{\pi}{2},\frac{6\pi}{7}$ and $\frac{9\pi}{10}$. For sufficiently small values of $\alpha$, the force on the JM and the cargo are towards each other and above a certain stagnant cap size, the JM and the cargo exhibit a reversal of the direction of propulsion. To understand the reason, it is necessary to take a look at the concentration field of the produced solute. Figure \ref{fig:concentration} shows the concentration field and the direction of concentration gradients on the surface of the JM for select separation distances and Figure \ref{fig:streamplots} shows the corresponding flow field and streamlines. It is evident that under select circumstances that depend on ($\alpha,\lambda,e$), the concentration of solutes exhibits a non monotonic behavior along the surface of the JM. The accumulation of solutes at closer separations in between the JM and the cargo lead to a reversal of the direction of the tangential concentration gradient on it's surface. This leads to opposing slip velocities along the JM surface from \ref{eq:BCv2} creating competing contributions to the propulsion forces as the corresponding flow from sub-problem 3 (\ref{fig:sub3}) is purely driven by slip velocity.  

\begin{figure}     
    \begin{subfigure}[b]{0.48\textwidth}
             \centering
             \includegraphics[width=0.985\textwidth]{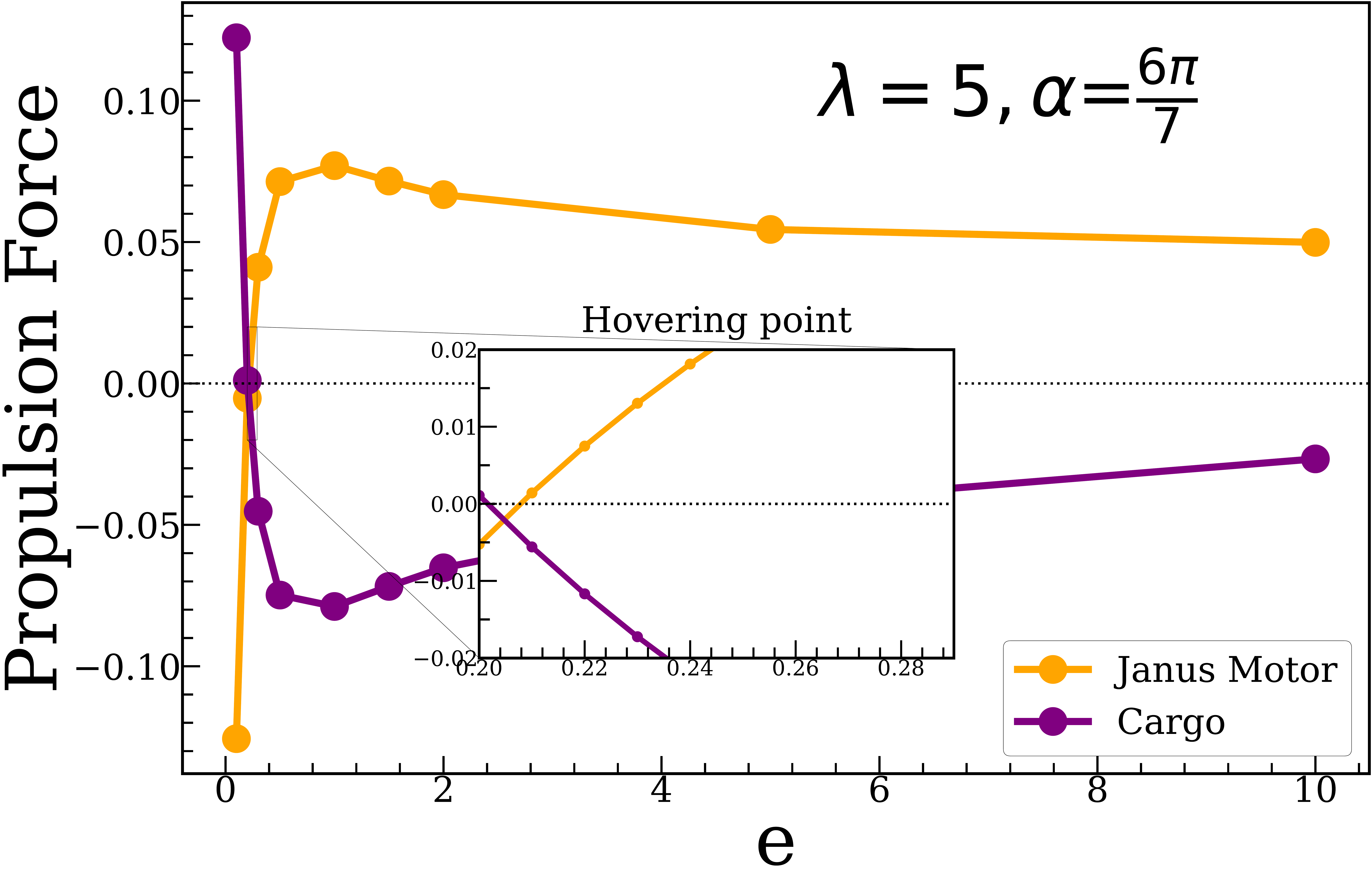}
             \caption{}
             \label{fig:Prop_6pi_7,5}
    \end{subfigure}
    \begin{subfigure}[b]{0.48\textwidth}
         \centering
         \includegraphics[width=0.985\textwidth]{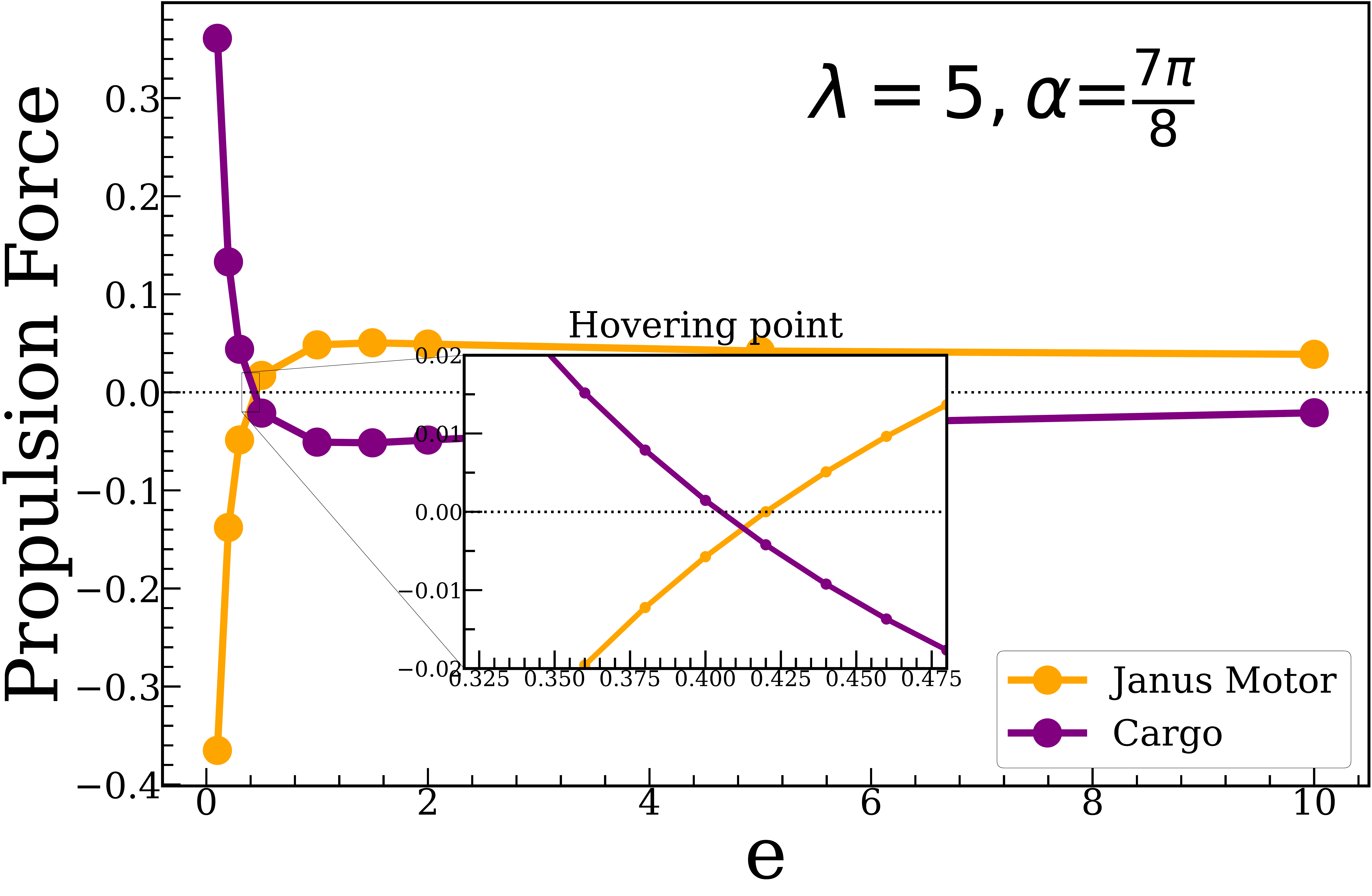}
         \caption{}
         \label{fig:Prop_7pi_8,5}
    \end{subfigure}
    
    \begin{subfigure}[b]{0.48\textwidth}
         \centering
         \includegraphics[width=0.985\textwidth]{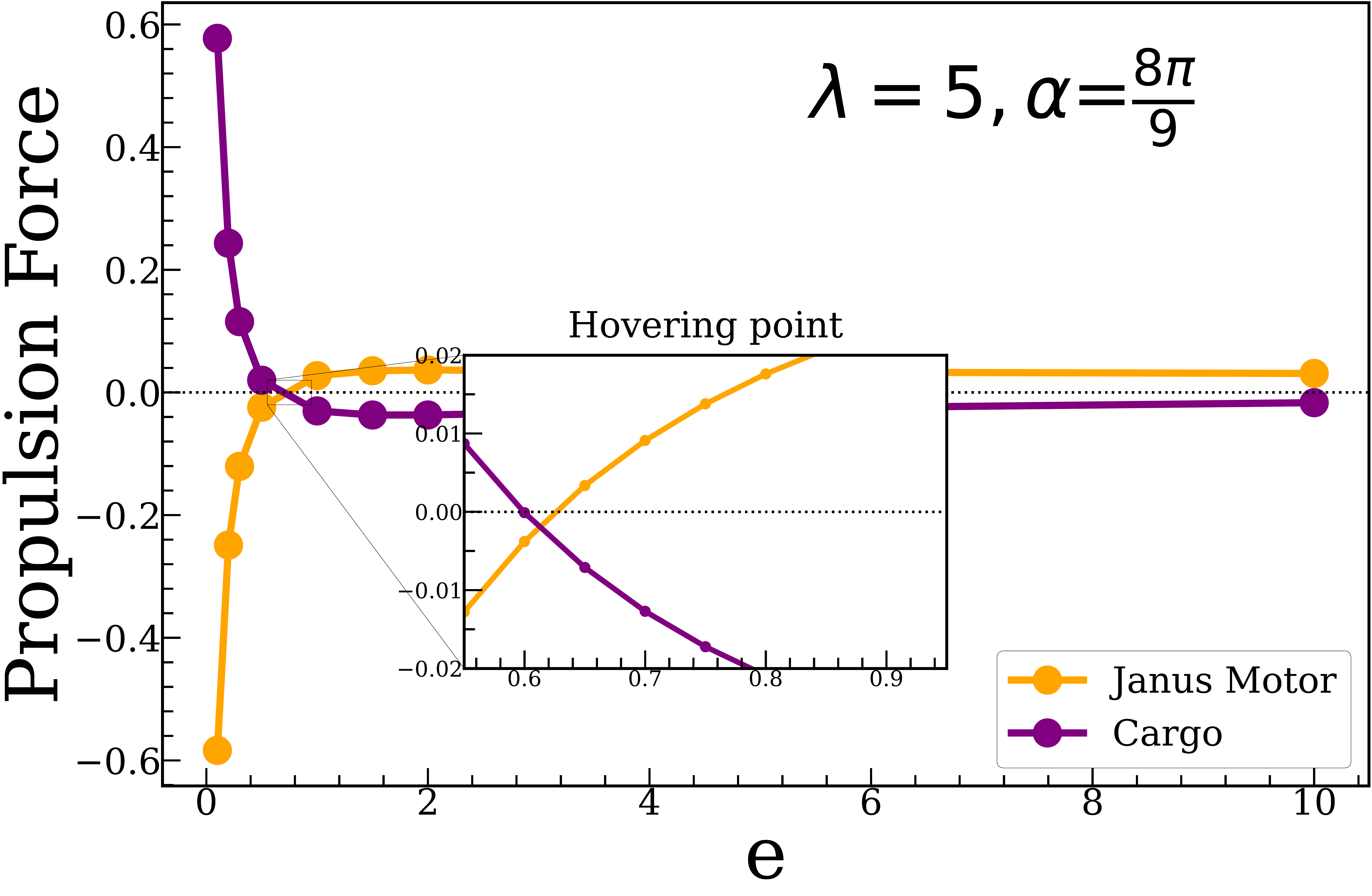}
         \caption{}
         \label{fig:Prop_8pi_9,5}
    \end{subfigure}
    \begin{subfigure}[b]{0.48\textwidth}
         \centering
         \includegraphics[width=0.985\textwidth]{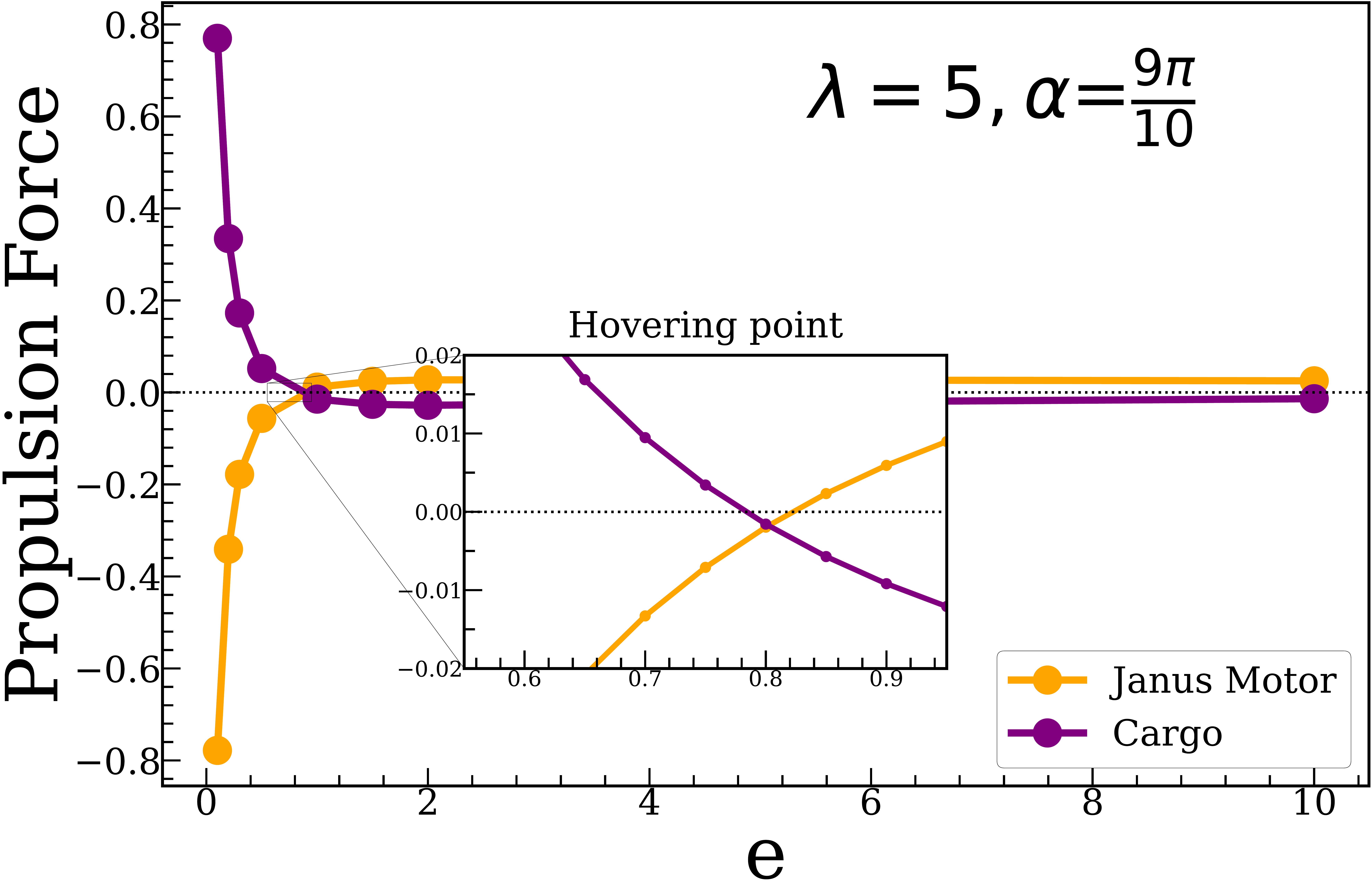}
         \caption{}
         \label{fig:Prop_9pi_10,5}
    \end{subfigure}%

\caption{Plots of the propulsion forces on the JM and the cargo for $\lambda =5$ depicting the monotonic increase of hovering separation distance $(e_h)$  with increasing stagnant cap angles ($\alpha$).} \label{fig:lambda5}
\end{figure}

If the cargo is held stationary (I.e. $V=0$), then the resulting velocity of the Janus particle can be directly found from the force free condition \ref{eq:FJnet} which simplifies to $U=-\frac{F^c_{J}}{F^a_{J}}$. The JM velocity $U=0 $, if $F^c_{J}=0$. Therefore,  if the propulsion force on the JM vanishes for a given configuration of $(\lambda, \alpha)$ at a fixed separation distance $e=e_h$, the JM is considered to be "hovering" at this fixed position with respect to the cargo. The concentration field still generates a flow due to the slip on the surface of the JM creating a flow field around the two spheres. Note that since the cargo is held stationary, it can still experience a force from this flow (I.e. $F^c_{C}\neq0$) as seen from the different x intercepts of the two forces in Figure \ref{fig:lambda5}. Upon inspecting the direction of the forces in the hovering configurations, it is evident that the hovering state is stable to translational perturbations of the JM along the line of centers. Further study is necessary to confirm the stability of the hovering configuration to non-axisymmetric perturbations.

Figure \ref{fig:lambda5} shows that when increasing $\alpha$ keeping $\lambda$ fixed, the hovering separation distance $e_h$ also increases as expected given that the larger stagnant cap size reduces the asymmetry in the distribution of the solute around the JM. Therefore, a lesser extent of accumulation of the solute between the JM and the cargo can cause opposing tangential concentration gradients on the surface of the JM leading to opposing slip velocities that allow the JM to hover at further distances from the cargo. 

by the corollary, the hovering separation decreases with decreasing $\alpha$ and there exists a minimum cap size ($\alpha_{min}$) for a given ($\lambda$) where $e_h\to0$. For $\alpha<\alpha_{min}$, the JM will tend to crash into the cargo if not for lubrication effects.

When we increase $\lambda$, the minimum cap angle required to allow hovering $\alpha_{min}$ reduces, as can be seen by looking at Figure \ref{fig:lambda2}. Hence, there exists a global minimum, $\alpha_{min}$ that corresponds to the case where $\lambda \rightarrow \infty$ that is physically equivalent to the case of a JM normally approaching a planar wall. This problem has been studied in detail by \cite{mozaffari2016self} where the authors showed that the JM can hover over a wall at a minimum stagnant cap angle of about  $\alpha_{min}\approx\ang{145}$

 In the limit of $e \rightarrow \infty$, the propulsion force on the cargo goes to zero. In this limit the JM experiences a propulsion force which approaches the value of a single JM at infinite separation here.   For the JM to hover at infinite separation, the active cap must cover the entire motor so that the resulting asymmetry in the concentration field vanishes. In other words, as $e \rightarrow \infty, $ $\alpha_{min} \rightarrow \pi$.

\begin{figure}
    \begin{subfigure}[b]{0.48\textwidth}
    \includegraphics[width=1\textwidth]{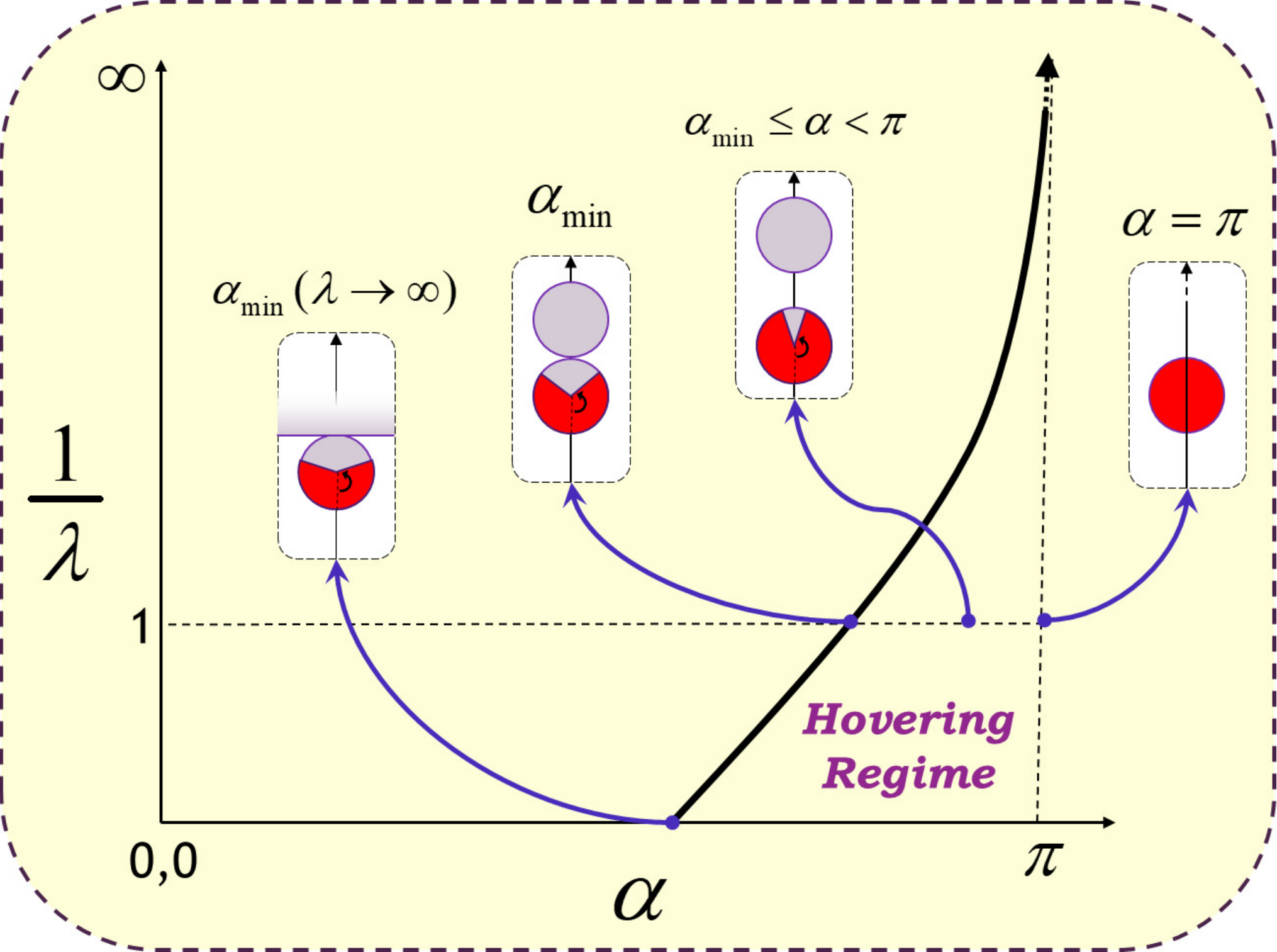} 
    \caption{}
\label{fig:phasediagram_a}
\end{subfigure}
\begin{subfigure}[b]{0.48\textwidth}
    \includegraphics[width=1\textwidth]{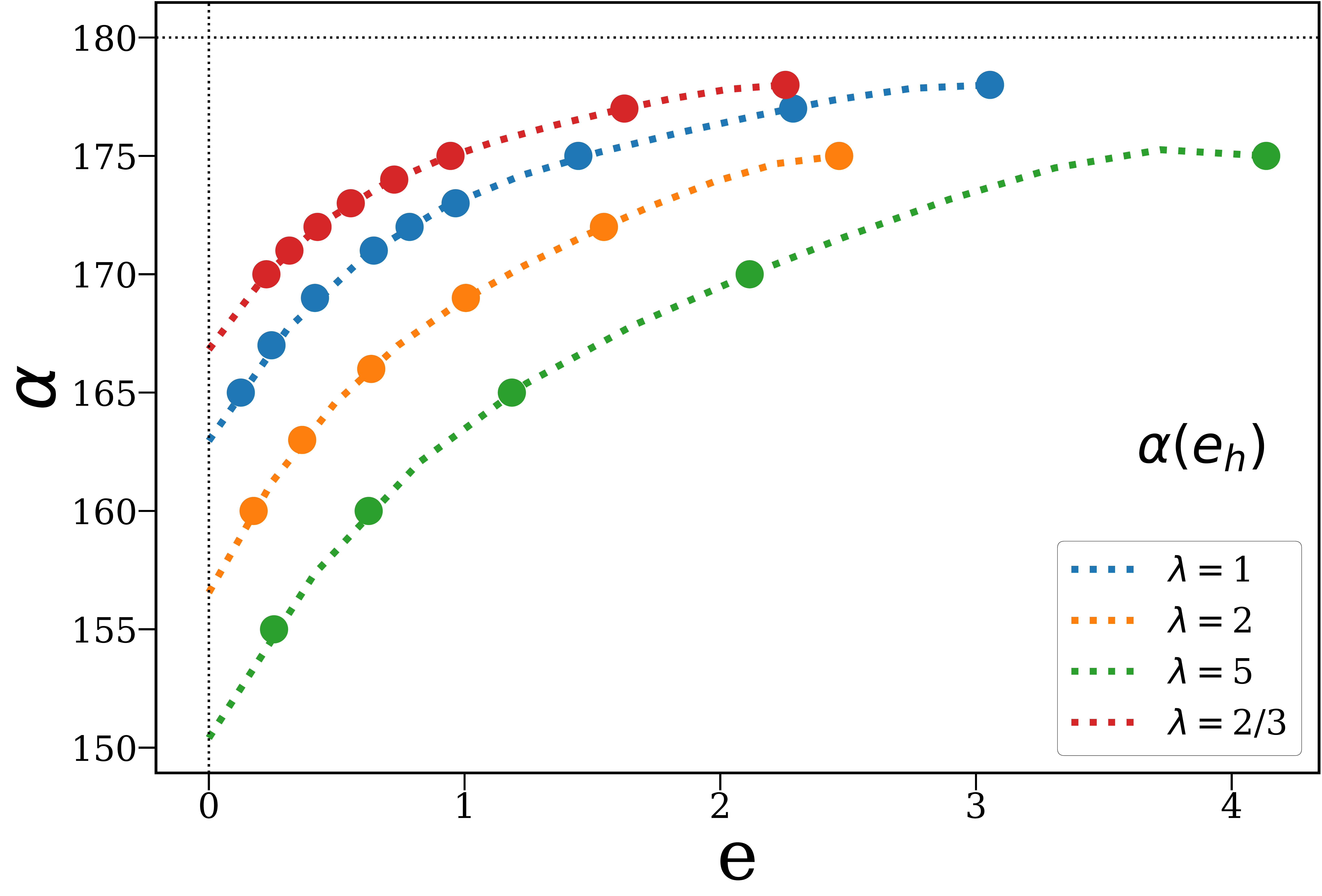}
    \caption{}    
    \label{fig:phasediagram_b}
\end{subfigure}
 \begin{subfigure}[b]{0.48\textwidth}
    \includegraphics[width=1\textwidth]{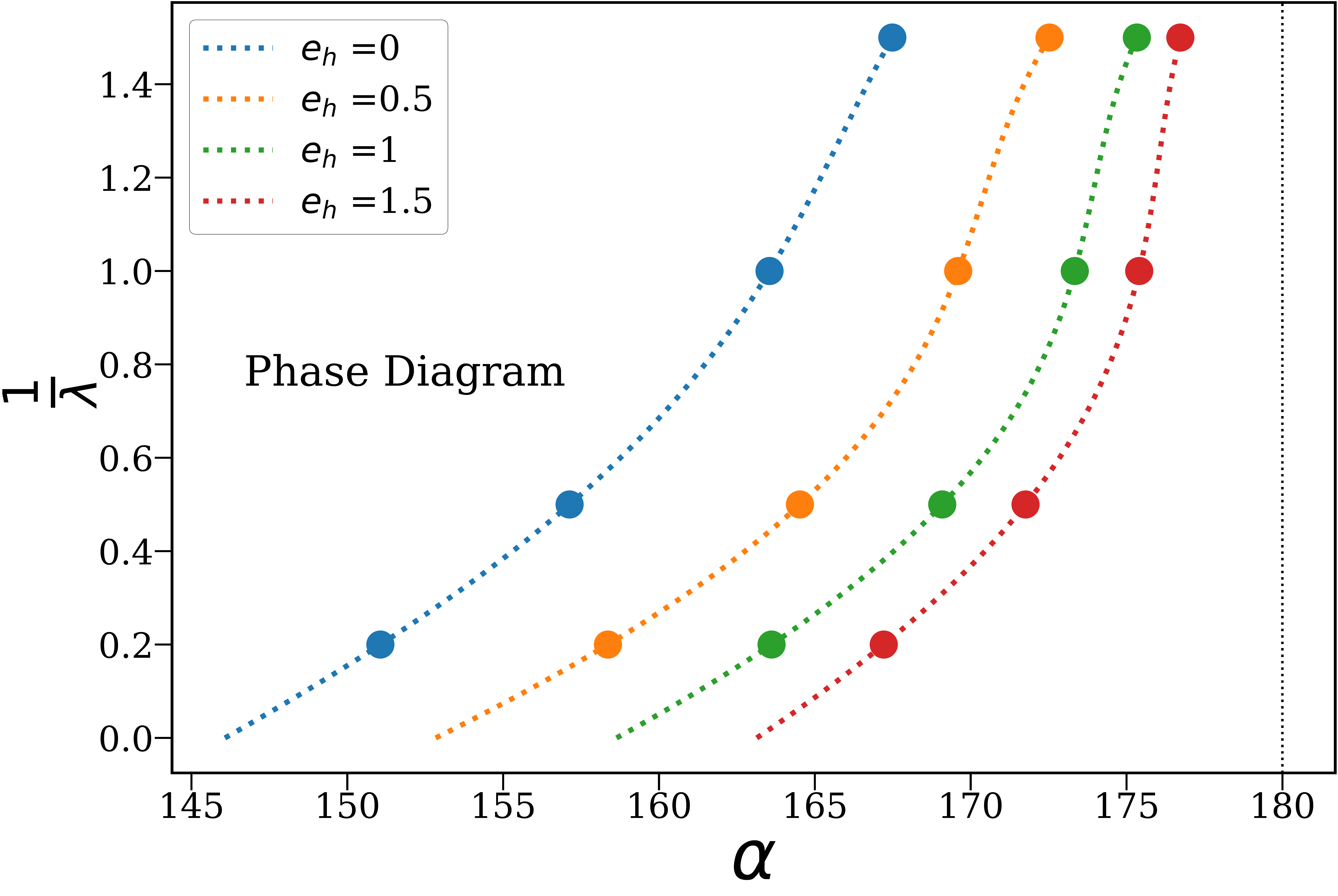}
    \caption{}    
    \label{fig:phasediagram_c}
\end{subfigure}
 \begin{subfigure}[b]{0.48\textwidth}
    \includegraphics[width=1\textwidth]{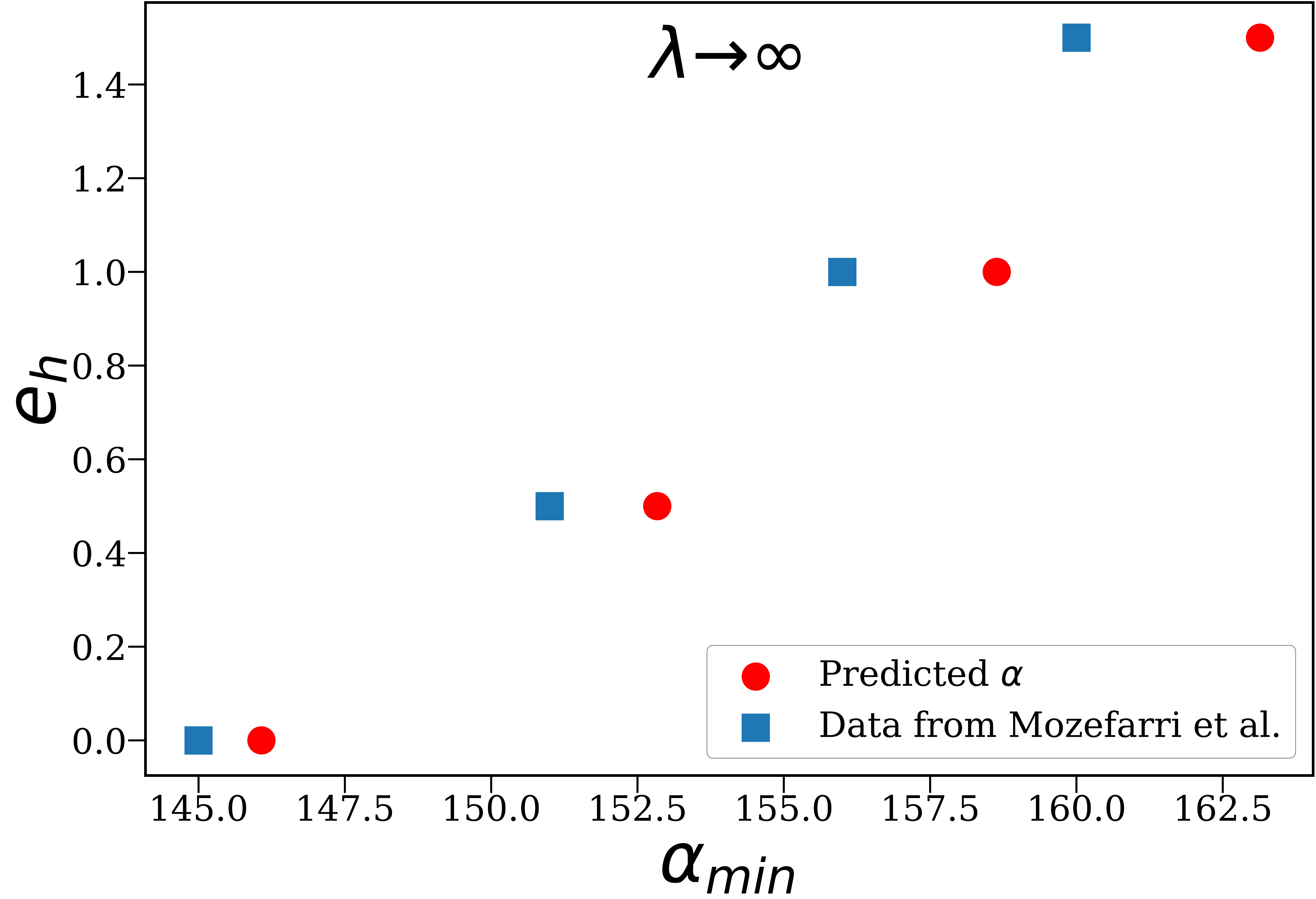}
    \caption{}    
    \label{fig:validation_ali}
\end{subfigure}
\caption{$(a)$ Illustration of the phase boundary of configurations ($\lambda,\alpha$) with hovering states. $(b)$ Polynomial fit of $\alpha_{min}$ as function of $e_h$ for different sizes of the JM and the cargo. $(c)$ Phase diagram estimated from extrapolation data in Figure \ref{fig:phasediagram_b}. $(d)$ Comparison of the extrapolation of Figure \ref{fig:phasediagram_c} to the limit of  $\lambda\to\infty$ with the literature \cite{mozaffari2016self}. }
\label{fig:phasediagram}
\end{figure}

 Based on these observations, we create a phase diagram that predicts the existence of the hovering point in the trajectory of approach of the JM to the cargo. Figure \ref{fig:phasediagram_a} depicts the proposed phase diagram with illustrations of the various configurations that the two spheres sustain at steady state. Since, we do not have an analytical solution to the minimum cap size for a given size ratio (I.e. $\alpha_{min}(\lambda)$), we can use an extrapolation approach to approximate the same. For a given $\lambda$, we estimate the hovering separation distance ($e_h$) to within $\pm 0.005$ by seeking a change in sign of the force on the JM with systematic increments of $e$. We then use a bi quadratic polynomial to generate a least squares fit of $e_h$ as a function of $\alpha$. We can then predict  $\alpha_{min}$ by extrapolating the fitted polynomial to $e_h=0$. Plots of the fitted polynomial for $\frac{1}{\lambda}=0.2,0.5,1$ and $1.5$ are shown in Figure \ref{fig:phasediagram_b}. Plotting the predicted  $\alpha_{min}$ with respect to $\frac{1}{\lambda}$, we can approximate the phase boundary of the possible configurations ($\alpha,\lambda$) where the JM and the stationary cargo can experience a hovering state (Figure \ref{fig:phasediagram_c}). We can use a similar approach to interpolate the required $\alpha$ for any $e_h$. Figure \ref{fig:phasediagram_c} also shows lines of constant hovering separation distances $e_h = 0.5, 1, 1.5$. We can further use a similar polynomial fitting to extrapolate the generated iso-separation lines and the phase boundary to the limiting case of $\lambda\to\infty$ (I.e., the x intercepts in Figure \ref{fig:phasediagram_c}) to compare with the existing literature on hovering of JM near a planar wall from \cite{mozaffari2016self} (Figure \ref{fig:validation_ali}. where in spite of the double extrapolation, there is a reasonable agreement. Note, in Figure \ref{fig:phasediagram}, the data points are plotted along with their corresponding polynomial fits (dotted lines) whereas in the plots depicting the velocities or the propulsion forces of the cargo and the JM, the lines are only view guides connecting the data points. 

\subsection{\label{sec:movingcargo}Moving Cargo}

\begin{figure} 
  \centering
  \begin{subfigure}{.4\linewidth}
    \centering
    \includegraphics[width = \linewidth]{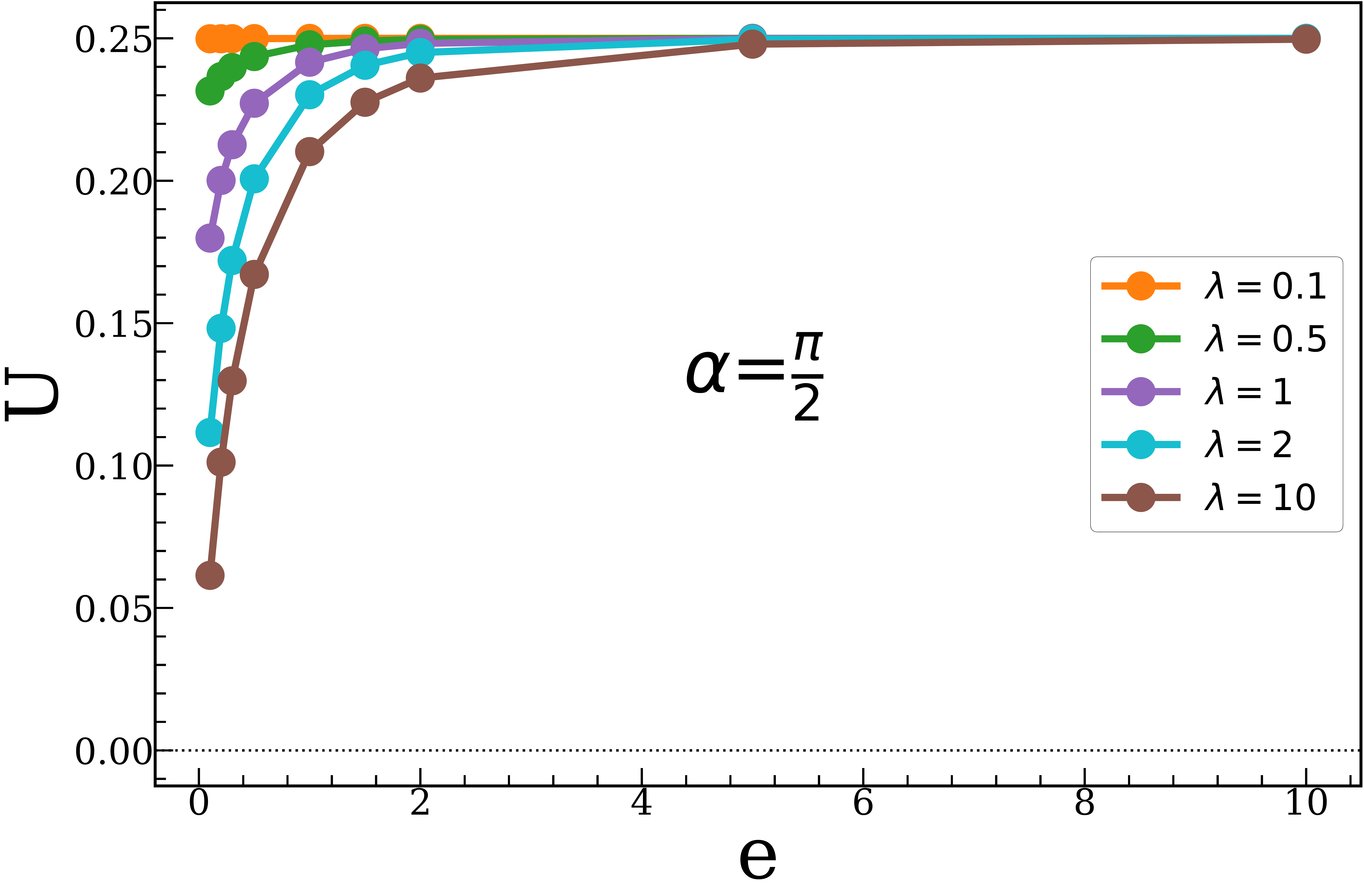}
    \caption{}
    \label{fig:velplot_JM}
  \end{subfigure}%
  \hspace{2em}
   \begin{subfigure}{.4\linewidth}
    \centering
    \includegraphics[width = \linewidth]{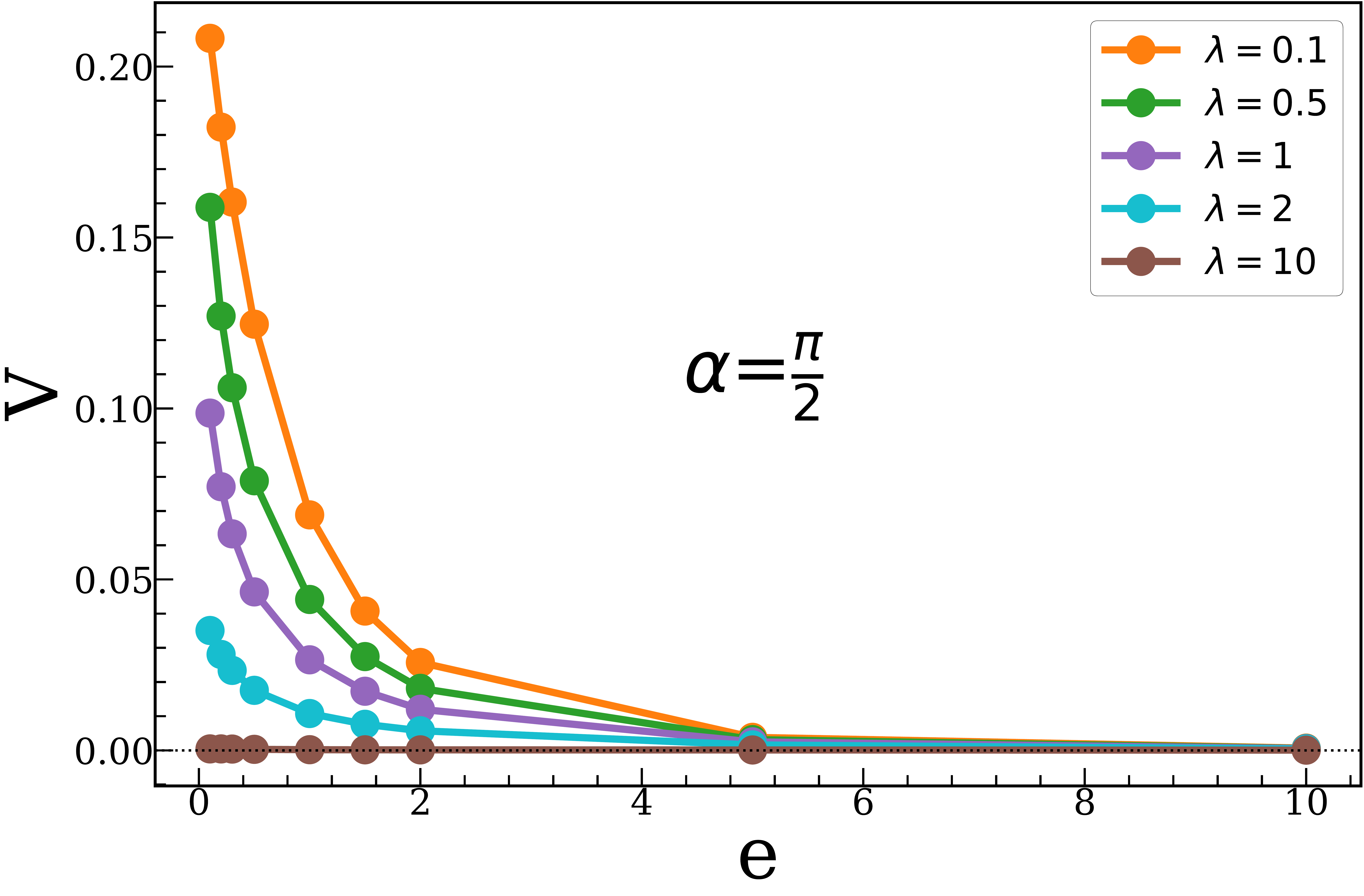}
    \caption{}
    \label{fig:velplot_Cargo}
  \end{subfigure}%
\caption{Velocities for the JM and the cargo for $\alpha=\frac{\pi}{2}$ and a range of values for $\lambda$.}
\label{fig:velPi_2}
\end{figure}

In the earlier subsection \ref{sec:stationarycargo}, the cargo was held stationary thereby allowing us to determine the motion of the JM by looking only at the propulsion forces. In this subsection, the cargo is allowed to freely move due to the hydrodynamic forces generated by sub-problems \ref{fig:sub1} and \ref{fig:sub2} in Figure \ref{fig:threesubprobs}. For the case of $\alpha=\frac{\pi}{2}$, the velocities of the JM ($U$) and the cargo ($V$) are presented for a range of values of $\lambda$ in Figure \ref{fig:velPi_2}. Here, we see that $U\rightarrow0.25$ as $e\rightarrow\infty$ as expected with the cargo becoming stationary and the terminal velocity approaching $\frac{1-cos^{2}\alpha}{4}$ (JM in an unbounded fluid \cite{mozaffari2016self}). As shown in this scenario, for sufficiently small values of $\alpha$ for a given $\lambda$, We note that the relative velocity between the two spheres reduces ($U\to V$) as $e\rightarrow0$. However, note that the propulsion component of the forces (i.e. $F^{c}$ from the third sub-problem \ref{fig:sub3}) does not include the lubrication effects from the hydrodynamic drag. Thus, the relative velocity (however small) between the two spheres provides the lubrication forces to counter the increasing propulsion forces on the two spheres towards each other. In such a scenario, the JM and the cargo approach a constant "towing" velocity in the limit of making contact with each other.

\begin{figure}     
    \begin{subfigure}[b]{0.48\textwidth}
             \centering
             \includegraphics[width=0.985\textwidth]{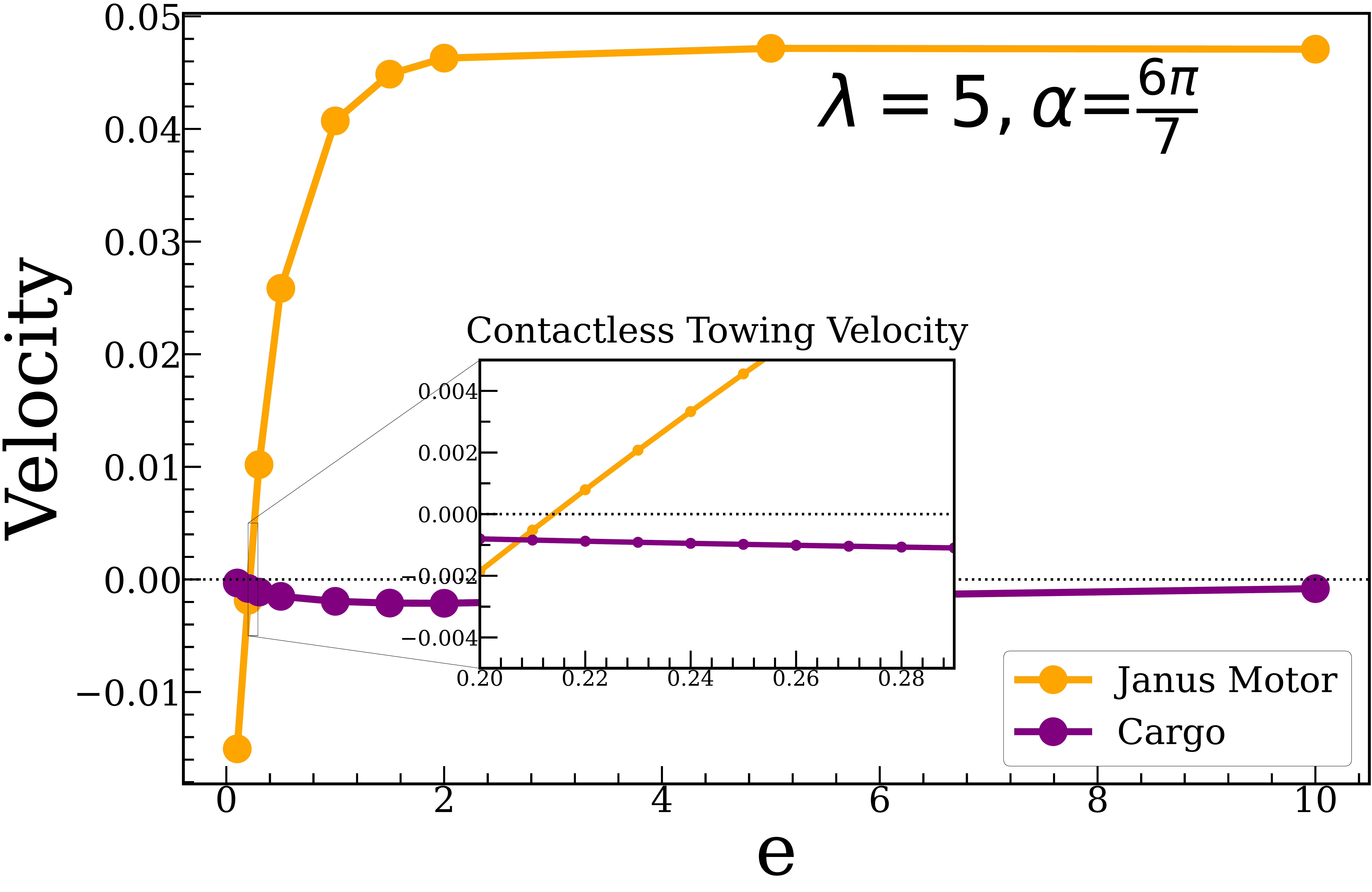}
             \caption{}
             \label{fig:Prop_6pi_7,5,vel}
    \end{subfigure}
    \begin{subfigure}[b]{0.48\textwidth}
         \centering
         \includegraphics[width=0.985\textwidth]{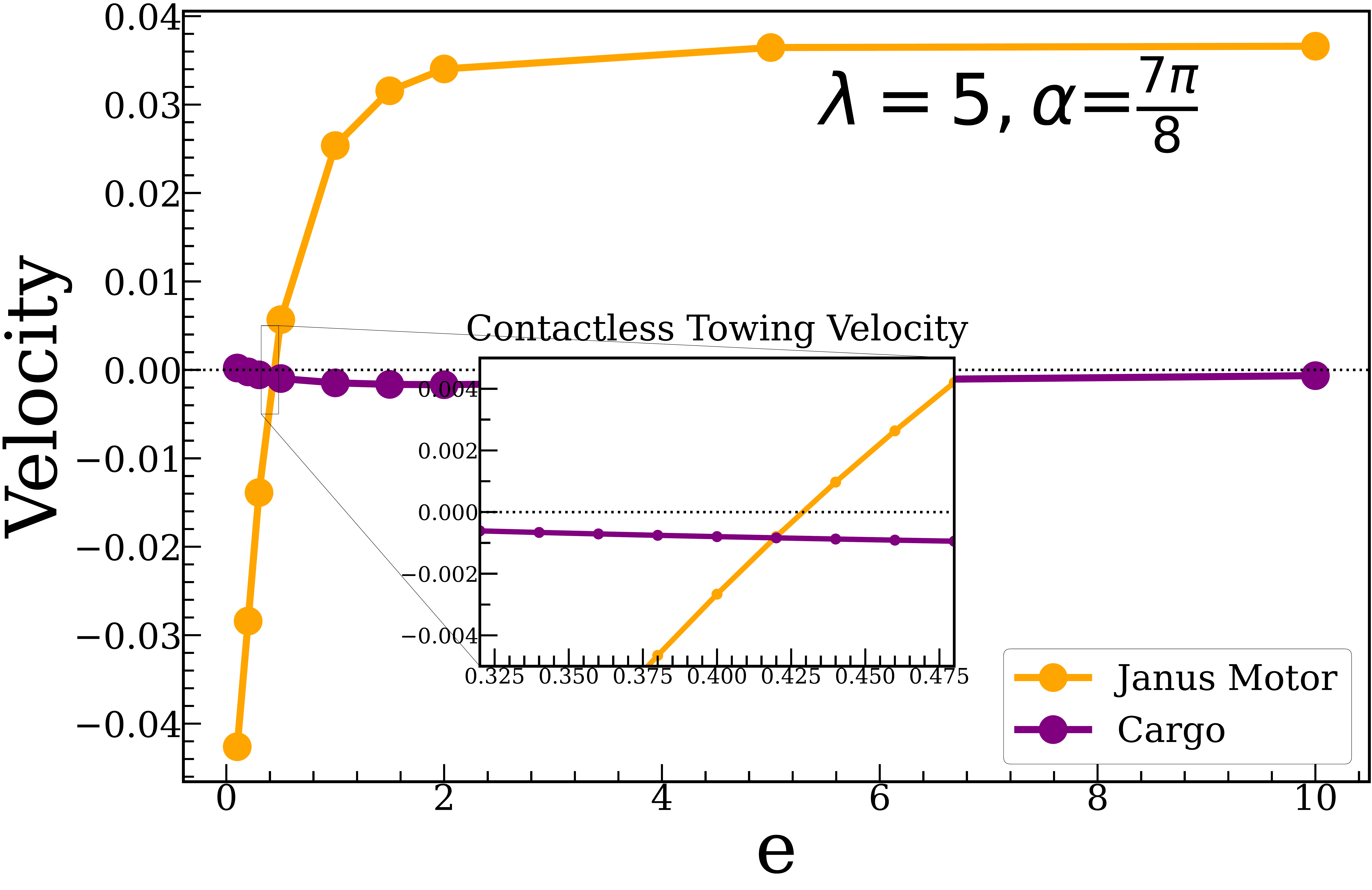}
         \caption{}
         \label{fig:Prop_7pi_8,5,vel}
    \end{subfigure}
    
    \begin{subfigure}[b]{0.48\textwidth}
         \centering
         \includegraphics[width=0.985\textwidth]{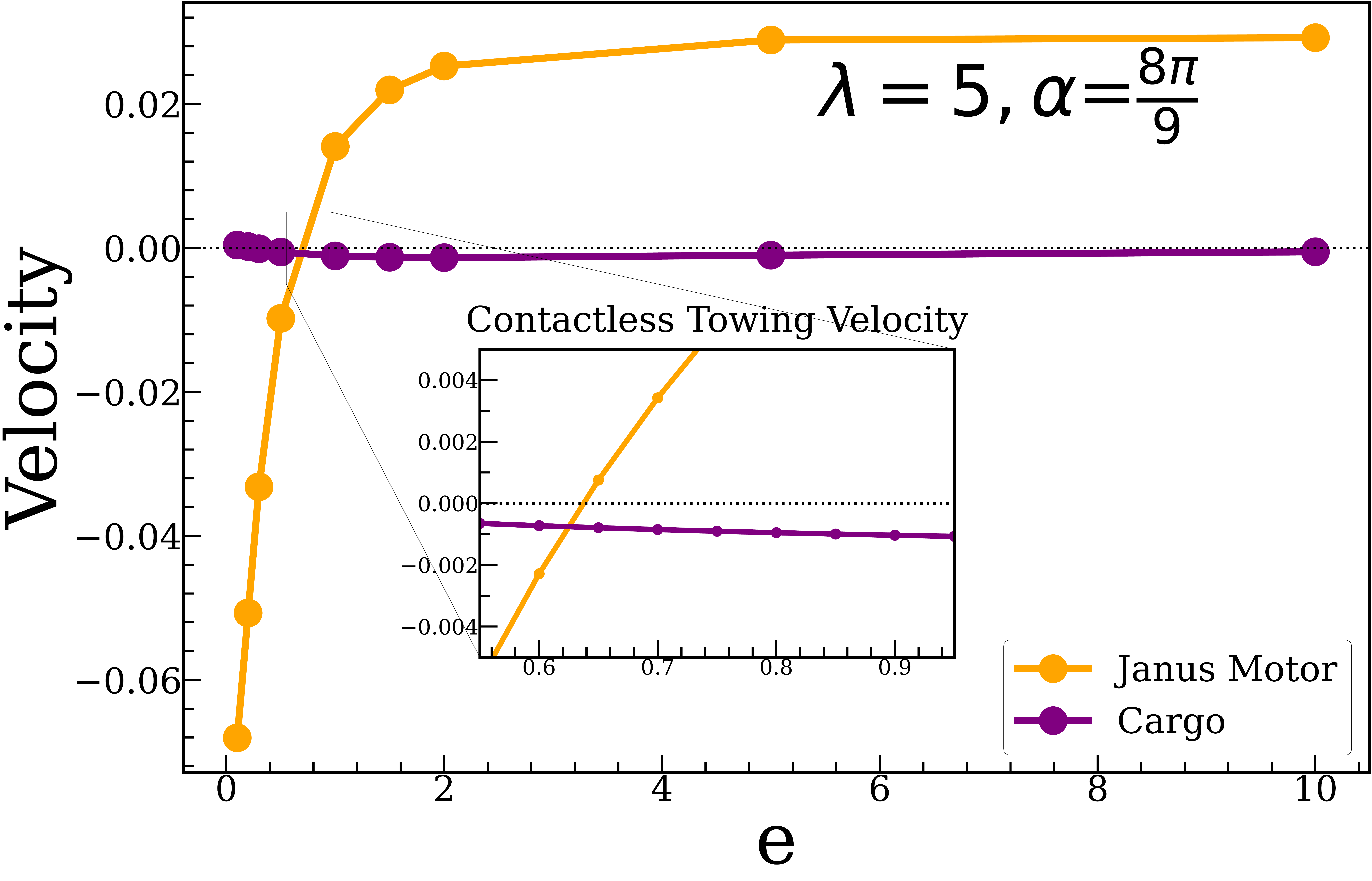}
         \caption{}
         \label{fig:Prop_9pi_10,5,vel}
    \end{subfigure}
    \begin{subfigure}[b]{0.48\textwidth}
         \centering
         \includegraphics[width=0.985\textwidth]{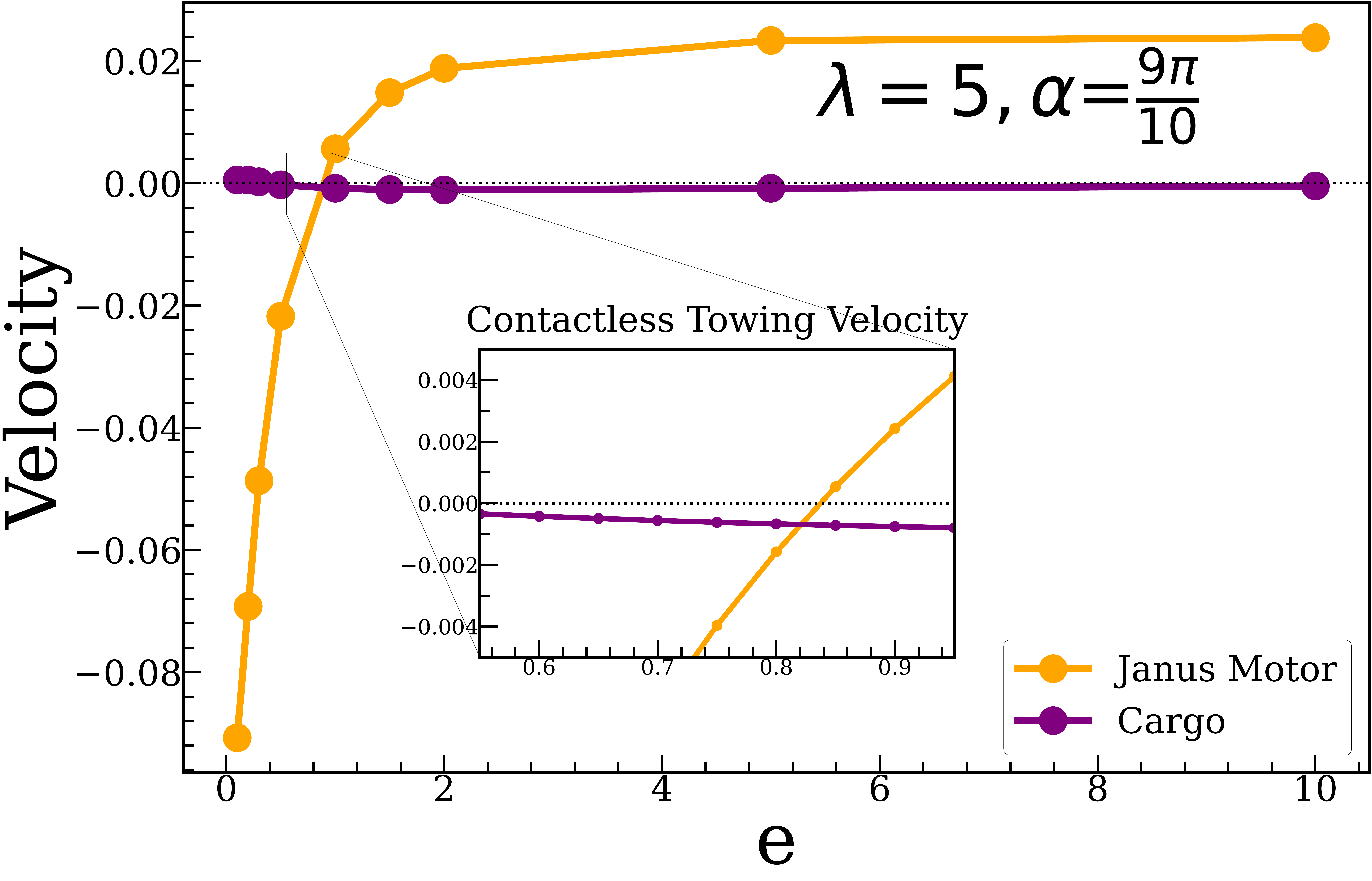}
         \caption{}
         \label{fig:Prop_8pi_9,5,vel}
    \end{subfigure}%

\caption{Plots of the velocities of the JM and a freely moving cargo corresponding to the propulsion forces depicted in Figure \ref{fig:lambda5} for $\lambda =5$ and different stagnant cap sizes.} \label{fig:lambda5vel}
\end{figure}

The solutions to the velocities of the JM and the cargo for different stagnant cap sizes of the JM for a size ratio $\lambda=5$ are presented in Figure \ref{fig:lambda5vel}. Note, the corresponding propulsion forces were shown in Figure \ref{fig:lambda5}. In these cases, the values of $\alpha$ are sufficiently large to allow enough solute accumulation such that the Jm and the cargo exhibit a reversal in direction of motion. The insets also show that for each value of $\alpha$ there exists a corresponding separation distance $e_{ct}$ at which the JM and the cargo have the same velocity. At $e_{ct}$, the two spheres are able to move in tandem and the motion can be described as a "contactless towing". From these results, this state appears to be stable to axisymmetric perturbations like the state of hovering described in section \ref{sec:stationarycargo}. In general, $e_{ct} \neq e_h$. Furthermore, $e_{ct}$ is also in general different from the separation distance at which the propulsion forces are equal (I.e., $F_J^c=F_C^c$) except for the case of $\lambda=1$, since the hydrodynamic drags for equal sized spheres moving in tandem are equal \cite{stimson1926motion}. Further investigation will be necessary to develop a phase diagram showing configurations where contactless towing is feasible.

\section{\label{sec:Conclusion}Conclusion}
We investigated the motion of a JM towards a spherical cargo for different relative ratios $\lambda$ and for different sizes of the active region on the JM (active cap angle $\alpha$).  We also developed a twin multipole approach for the axisymmetric Stokes stream function using the translation functions developed for the eigenfunctions in terms of Gegenbauer polynomials. 
 When the cargo is held stationary, we discussed the existence of two possible stable steady states along the approach trajectory of the motor. Under select conditions under which the JM comes into a stationary hover at some fixed distance away from the cargo. This state is referred to as the "hovering state" . Outside the hovering conditions, the JM continually approaches the cargo to the limit of making contact. When the cargo and the JM are free to move, they can achieve a similar limit of contact and translate together at a steady terminal velocity. This state is referred to as the "towing state". However, under select conditions, the two spheres can translate at a steady terminal velocity at a finite separation distance. We call the latter state "contact less towing state".  Finally, we developed a phase diagram based on the qualitative observations of the properties of the hovering regime for the stationary cargo case.

%
%

\appendix
\section{Translational Theorems modified}
Using the relationship ${g_n}(\gamma ) = {{\left( {{p_{n + 1}}(\gamma ) - {p_{n - 1}}(\gamma )} \right)} \mathord{\left/
 {\vphantom {{\left( {{p_{n + 1}}(\gamma ) - {p_{n - 1}}(\gamma )} \right)} {\left( {2n + 1} \right)}}} \right.
 \kern-\nulldelimiterspace} {\left( {2n + 1} \right)}}$, the eigen functions in \ref{eq:eigenexpansion} can be rewritten as 
 \begin{equation}
 \label{eq:eigen1restated}
\frac{{{g_n}({\gamma _2})}}{{{r_2}^n}} = \left( {\frac{1}{{2n + 1}}} \right)\left( {\frac{{{p_{n + 1}}({\gamma _2})}}{{{r_2}^n}} - \frac{{{p_{n - 1}}({\gamma _2})}}{{{r_2}^n}}} \right)
\end{equation}
\begin{equation}
\label{eq:eigen2restated}
\frac{{{g_n}({\gamma _2})}}{{{r_2}^{n - 2}}} = \left( {\frac{{{r_2}^2}}{{2n + 1}}} \right)\left( {\frac{{{p_{n + 1}}({\gamma _2})}}{{{r_2}^n}} - \frac{{{p_{n - 1}}({\gamma _2})}}{{{r_2}^n}}} \right)
 \end{equation}
Using the translation theorem for the legendre polynomial functions \ref{eq:translation}, the expressions in \ref{eq:eigen1restated} and \ref{eq:eigen2restated} can further be simplified to
\begin{equation}
\label{eq:beforesacks1}
\frac{{{g_n}({\gamma _2})}}{{{r_2}^n}} = \left( {\frac{1}{{2n + 1}}} \right)\left( {\frac{{{r_2}^2}}{{{R^{n + 2}}}}\sum\limits_{m = 0}^\infty  {\left( {\begin{array}{*{20}{c}}
{m + n + 1}\\
{n + 1}
\end{array}} \right){{\left( {\frac{{{r_1}}}{R}} \right)}^m}} {p_m}({\gamma _1}) - \frac{1}{{{R^n}}}\sum\limits_{m = 0}^\infty  {\left( {\begin{array}{*{20}{c}}
{m + n - 1}\\
{n - 1}
\end{array}} \right){{\left( {\frac{{{r_1}}}{R}} \right)}^m}} {p_m}({\gamma _1})} \right)
\end{equation}
\begin{equation}
    \label{eq:beforesacks2}
    \frac{{{g_n}({\gamma _2})}}{{{r_2}^{n - 2}}} = \left( {\frac{1}{{2n + 1}}} \right)\left( {\frac{{{r_2}^4}}{{{R^{n + 2}}}}\sum\limits_{m = 0}^\infty  {\left( {\begin{array}{*{20}{c}}
{m + n + 1}\\
{n + 1}
\end{array}} \right){{\left( {\frac{{{r_1}}}{R}} \right)}^m}} {p_m}({\gamma _1}) - \frac{{{r_2}^2}}{{{R^n}}}\sum\limits_{m = 0}^\infty  {\left( {\begin{array}{*{20}{c}}
{m + n - 1}\\
{n - 1}
\end{array}} \right){{\left( {\frac{{{r_1}}}{R}} \right)}^m}} {p_m}({\gamma _1})} \right)
\end{equation}
We can expand ${r_2}^n = {\left( {{R^2} + {r_1}^2 - 2R{r_1}{\gamma _1}} \right)^{\frac{n}{2}}}$ as follows \cite{sack1964generalization}
\begin{equation}
    \label{eq:powersofr}
    {r_2}^n = \sum\limits_{l = 0} {{R_{nl}}({r_1},R){p_l}({\gamma _1})} 
\end{equation}
where ${R_{nl}}({r_1},R)$ are the radial functions as defined in \cite{sack1964generalization}. The summation in \ref{eq:powersofr} is shown to be finite and to truncate at $l \le n/2$  for positive, even values of $n$. Equations \ref{eq:beforesacks1} and \ref{eq:beforesacks2} reduce to
\begin{equation}
    \label{eq:eigen1withproductlaplace}
    \frac{{{g_n}({\gamma _2})}}{{{r_2}^n}} = \left( {\frac{1}{{2n + 1}}} \right)\left( {\sum\limits_{l = 0}^1 {\sum\limits_{m = 0}^\infty  {\frac{{{R_{2l}}({r_1},R)}}{{{R^{n + 2}}}}\left( {\begin{array}{*{20}{c}}
{m + n + 1}\\
{n + 1}
\end{array}} \right){{\left( {\frac{{{r_1}}}{R}} \right)}^m}{p_m}({\gamma _1}){p_l}({\gamma _1})} }  - \sum\limits_{m = 0}^\infty  {\frac{1}{{{R^n}}}\left( {\begin{array}{*{20}{c}}
{m + n - 1}\\
{n - 1}
\end{array}} \right){{\left( {\frac{{{r_1}}}{R}} \right)}^m}} {p_m}({\gamma _1})} \right)
\end{equation}
\begin{equation}
	\begin{split}
		\label{eq:eigen2withproductlaplace}
		\frac{{{g_n}({\gamma _2})}}{{{r_2}^{n - 2}}} &= \left( {\frac{1}{{2n + 1}}} \right)\left( {\sum\limits_{l = 0}^2 {\sum\limits_{m = 0}^\infty  {\frac{{{R_{4l}}({r_1},R)}}{{{R^{n + 2}}}}\left( {\begin{array}{*{20}{c}}
							{m + n + 1}\\
							{n + 1}
					\end{array}} \right){{\left( {\frac{{{r_1}}}{R}} \right)}^m}} {p_m}({\gamma _1}){p_l}({\gamma _1})}} \right)
			\\ &-\left( {\frac{1}{{2n + 1}}} \right)\left({ \sum\limits_{l = 0}^1 {\sum\limits_{m = 0}^\infty  {\frac{{{R_{2l}}({r_1},R)}}{{{R^n}}}\left( {\begin{array}{*{20}{c}}
							{m + n - 1}\\
							{n - 1}
					\end{array}} \right){{\left( {\frac{{{r_1}}}{R}} \right)}^m}} {p_m}({\gamma _1})} {p_l}({\gamma _1})} \right).
	\end{split}
\end{equation}
noting that ${p_0}(\gamma ) = 1,{p_1}(\gamma ) = \gamma $  and  ${p_2}(\gamma ) = \left( {3{\gamma ^2} - 1} \right)/2$ and using Bonnets recursion theorem, we can linearize the product of the Legendre polynomials in \ref{eq:eigen1withproductlaplace} and \ref{eq:eigen2withproductlaplace}. Finally, grouping coefficients for different orders of the Legendre polynomials, we derive the translation theorem for the Eigen functions given in  \ref{eq:transtheorem1} and \ref{eq:transtheorem2} with the following definitions for $\Omega_1-\Omega_8$

\begin{flalign}
&\Omega_1 = {m + n + 1 \choose n + 1 } \Biggl( \frac{R_{20}}{R^{m+n+2}} \Biggr) - {m + n - 1 \choose n - 1 } \Biggl( \frac{1}{R^{m+n}}   \Biggr), \\[10pt]
&\Omega_2 =  {m + n + 1 \choose n + 1 }  \Biggl( \frac{m}{2m+1}  \Biggr) \Biggl( \frac{R_{21}}{R^{m+n+2}} \Biggr), \\[10pt]
&\Omega_3 =  {m + n + 1 \choose n + 1 }  \Biggl( \frac{m + 1}{2m+1}  \Biggr) \Biggl( \frac{R_{21}}{R^{m+n+2}} \Biggr), \\[10pt]
\begin{split}
&\Omega_4 = {m + n + 1 \choose n + 1} \Biggl( \frac{1}{R^{m+n+2}}   \Biggr) \Biggl\{R_{40} + R_{42} \Biggl[1 - \frac{3}{2}   \Biggl(  \frac{(m+1)(m+2)  }{(2m+1)(2m+3)}   + \frac{m(m-1)}{(2m+1)(2m-1)}           \Biggr) \Biggr] \Biggr\} \\
& - {m + n - 1 \choose n - 1}  \Biggl( \frac{R_{20}}{R^{m+n}} \Biggr),
\end{split}
\\[10pt]
&\Omega_5 = \frac{m}{2m+1} \Biggl[ {m + n + 1 \choose n + 1 } \Biggl( \frac{R_{41}}{R^{m+n+2}} \Biggr) - { m + n - 1 \choose n - 1} \Biggl( \frac{R_{21}}{R^{m+n}}  \Biggr)  \Biggl], \\[10pt]
&\Omega_6 = \frac{m + 1}{2m+1} \Biggl[ {m + n + 1 \choose n + 1 } \Biggl( \frac{R_{41}}{R^{m+n+2}} \Biggr) - { m + n - 1 \choose n - 1} \Biggl( \frac{R_{21}}{R^{m+n}}  \Biggr)  \Biggl], \\[10pt]
&\Omega_7 = \frac{3}{2} \Biggl( \frac{(m+1)(m+2) }{(2m+1)(2m+3)}    \Biggr)  { m + n + 1 \choose n + 1}  \Biggl( \frac{R_{42}}{R^{m+n+2}}  \Biggr)  ,\\[10pt]
&\Omega_8 = \frac{3}{2} \Biggl( \frac{m(m-1) }{(2m+1)(2m-1)}    \Biggr)  { m + n + 1 \choose n + 1}  \Biggl( \frac{R_{42}}{R^{m+n+2}}  \Biggr),  \\[10pt]
\end{flalign}

We can interchange the indices 1, 2 to translate the Eigenfunctions to the second sphere. We note that the final form of the translation theorems \ref{eq:transtheorem1},\ref{eq:transtheorem2} are valid only along the surfaces of the two spheres and the corresponding values of ${R_{ij}} \Leftrightarrow {R_{ij}}({r_1},R)$  for $(i,j) = (2,0),(2,1),(4,0),(4,1),(4,2)$.are defined as
\begin{flalign}
{R_{20}}({r_1},R) = {R^2}F( - 1, - 3/2;3/2;{\left( {{r_1}/R} \right)^2})\\
{R_{21}}({r_1},R) =  - 2{R^2}\left( {\frac{{{r_1}}}{R}} \right)F(0, - 3/2;5/2;{\left( {{r_1}/R} \right)^2})\\
{R_{40}}({r_1},R) = {R^4}F( - 2, - 5/2;3/2;{\left( {{r_1}/R} \right)^2})\\
{R_{41}}({r_1},R) =  - 4{R^4}\left( {\frac{{{r_1}}}{R}} \right)F( - 1, - 5/2;5/2;{\left( {{r_1}/R} \right)^2})\\
{R_{42}}({r_1},R) = \left( {8/3} \right){R^4}{\left( {\frac{{{r_1}}}{R}} \right)^2}F(0, - 5/2;7/2;{\left( {{r_1}/R} \right)^2})
\end{flalign}
where the function $F$ is Gauss' hypergeometric function defined as
\begin{equation}
    F(a,b;c;x)=1 + \sum\limits_{k = 1}^\infty  {\frac{{{{(a)}_k}{{(b)}_k}}}{{k!{{(c)}_k}}}} {\left( x \right)^k}
\end{equation}
with 
\begin{equation}
    {(a)_k} = \Gamma (a + k)/\Gamma (a)
\end{equation}
\section{COMSOL}\label{sec:apB}
We used COMSOL Multiphysics to collaborate the validity of our diffusiophoretic results. We place our two particles in a large cube $(250R_1 \times 250R_1 \times 250R_1)$ to create an infinite medium (more on these dimensions later), with the Janus motor’s center corresponding to the cube’s center/origin.  We define the concentration flux on the surface of the Janus motor (with radius $R_1$), and its velocity in terms the concentration gradient.  We also assign the necessary boundary conditions, including perfect slip on the outer box (not including the inlet or outlet),

\begin{flalign}\label{eq:perfslip}
&{\bf{u}} \cdot {\bm{n}}\big|_{wall} = 0, \\[5pt]
&\bm{n} \cdot{\bm{\sigma}} \cdot {\bm{t}}\big|_{wall}  = 0.
\end{flalign}

We assign an inlet, the bottom of the box, with no velocity (${\bm{V}} = \bm{0}$) since there is nothing flowing inward.  Our outlet, the top of the box, conditions are

\begin{flalign}\label{eq:outlet}
&{\bm{n \cdot\bm{\sigma}}} \cdot {\bm{t}}\big|_{outlet}  = 0, \\[5pt]
&{\bm{n} \cdot \bm{\sigma} \cdot \bm{n}} \big|_{outlet}   = 1\hspace{0.2cm} atm.
\end{flalign}
                       
For the cargo, we set boundary conditions of no slip (${\bm{u}} = {\bm{0}}$) and of no flux ($J = 0$).  Finally, we set the sides of the box to have no concentration ($C = 0$).   

COMSOL then discretizes the volume by tetrahedral mesh and the surfaces by triangles.  COMSOL employs the finite element method to solve the concentration and the Stokes equations, by discretizing those equations and solving them using the weak formulation.  We choose to have the concentration and velocity solved using quadratic shape functions, while the pressure is solved using linear ones. 

To ensure our COMSOL results were themselves converged, we tested the mesh, relative tolerance, and box size. We use the test case of $\lambda = 1$, $\alpha = \pi/2$, and $R = 2.1$ Starting with the box size, we decreased the box on all sides to prove that the box size we used for all of our models $(250R_1 \times 250R_1 \times 250R_1)$ returned a converged result.  The results of these models can be seen in Table \ref{tab:BoxSize} and show that even for $25\%$ closer wall our forces are within $1\%$  of the original box size.    Next we turn to the mesh to ensure that we have small enough elements, given that mesh must be fine when dealing with steep gradients.  We prove that our mesh is fine enough, by showing that going both coarser and finer does not make a difference (Table \ref{tab:MeshSize}).   Lastly we looked at the relative tolerance.  Our original tolerance is $0.0004$ (a number decided a priori considered both with the mesh and computation time).  We are able to show that our relative tolerance returns a converged result (Table \ref{tab:Tolerance}), and finally prove that our model decisions are all returning accurate and precise results.  We see these results of our test case (only using original values for mesh, box size, and tolerance) in our plots when comparing to the twin multipole approach in Figure \ref{fig:COMS_pi_2,1}.  We also see the problem where we have hovering in Figure \ref{fig:COMS_7pi_8,5} and find that COMSOL is able to capture this hovering behavior as well.    

\begin{table}[]
\begin{tabular}{|l|l|l|}
\hline
                  &        &         \\
                  & $F^{JP}$ & $F^C$     \\ \hline
Original Size     & 0.3935 & -0.2089 \\ \hline
10{\%} Closer Box & 0.3936 & -0.2087 \\ \hline
25{\%} Closer Box & 0.3940 & -0.2084 \\ \hline
\end{tabular}
\caption{We use the case of $\lambda = 1$, $\alpha = \pi/2$, and $R = 2.1$. We show how the box size does not effect convergence. Our forces are all nondimensionalized. }
\label{tab:BoxSize}
\end{table}

\begin{figure} [htb!]
  \centering
  \begin{subfigure}{.3\linewidth}
    \centering
    \includegraphics[width = \linewidth]{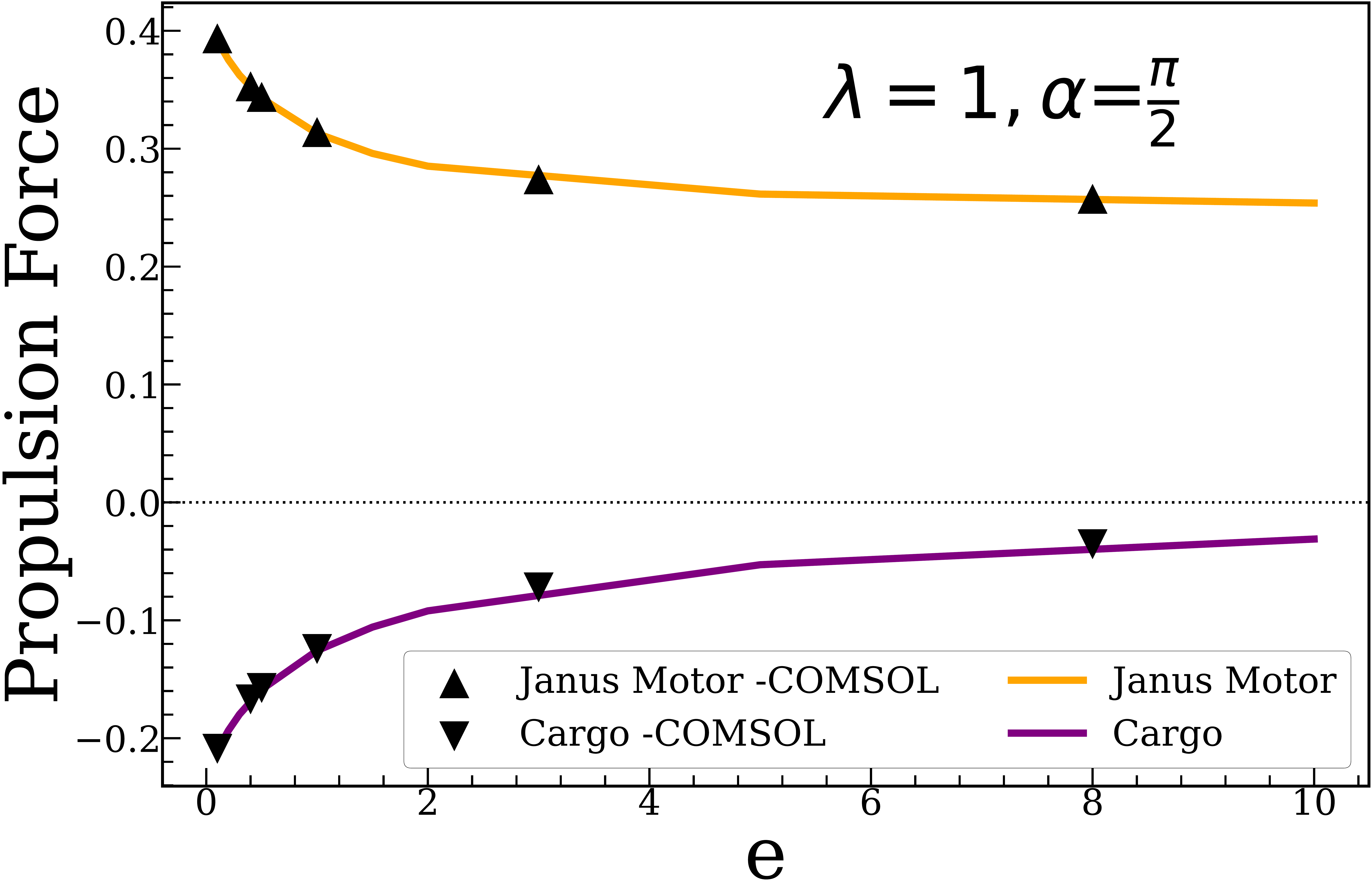}
    \caption{}
    \label{fig:COMS_pi_2,1}
  \end{subfigure}%
  \hspace{2em}
  \begin{subfigure}{.3\linewidth}
    \centering
    \includegraphics[width = \linewidth]{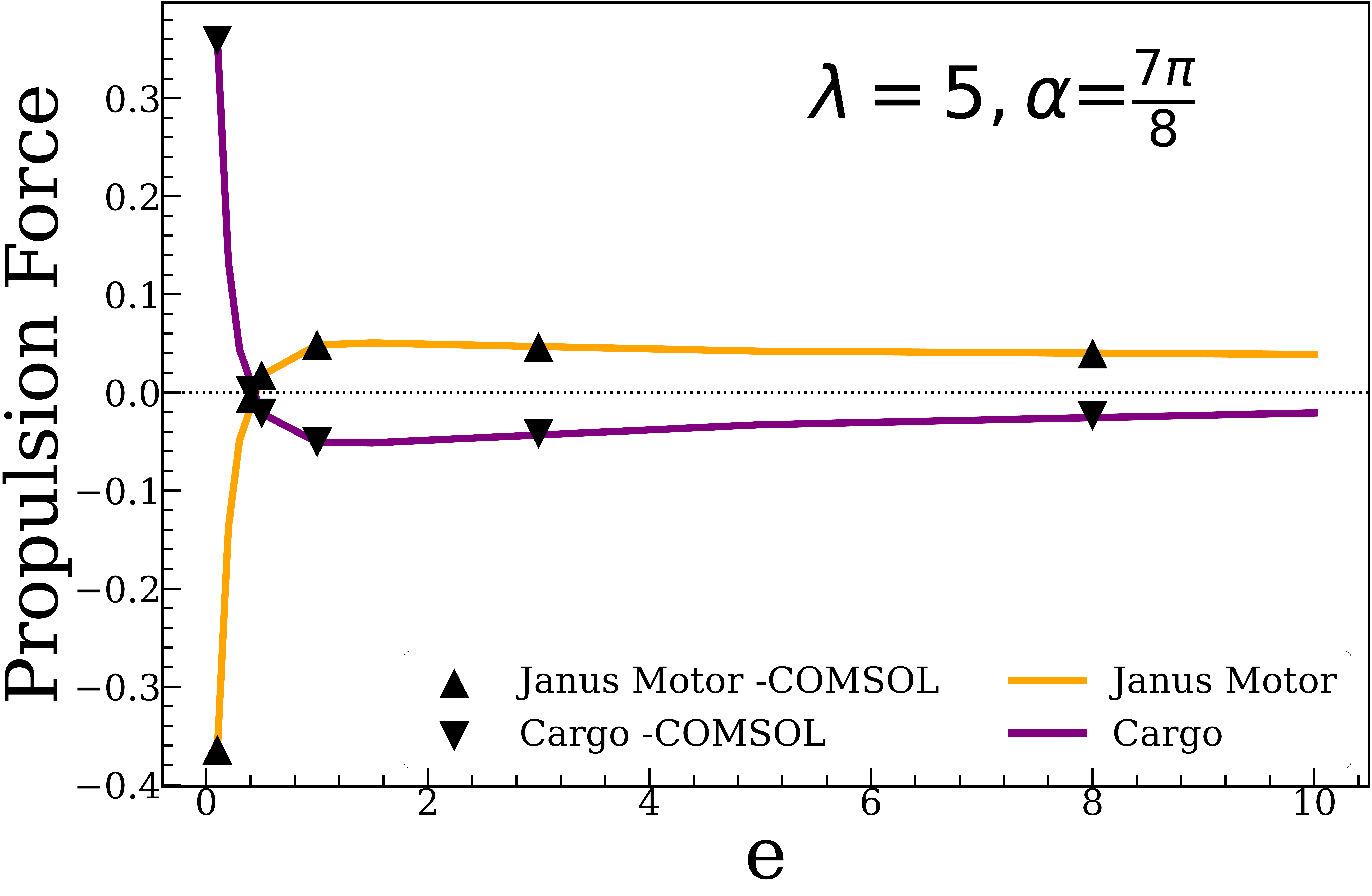}
    \caption{}
    \label{fig:COMS_7pi_8,5}
  \end{subfigure}%
    \caption{Two examples, one of hovering and one of towing, of how well the COMSOL results match the twin multipole approach results.}
  \label{}
\end{figure}

\begin{table}[]
\begin{tabular}{|l|l|l|}
\hline
                  &        &         \\
                  & $F^{JP}$ & $F^C$     \\ \hline
5{\%} Finer Mesh & 0.3933 & -0.2089\\ \hline                  
Original Mesh     & 0.3935 & -0.2089 \\ \hline
5{\%} Coarser Mesh &0.3927  & -0.2089 \\ \hline
\end{tabular}
\caption{We use the case of $\lambda = 1$, $\alpha = \pi/2$, and $R = 2.1$. We show how our mesh is already converged. Our forces are all nondimensionalized. }
\label{tab:MeshSize}
\end{table}

\begin{table}[]
\begin{tabular}{|l|l|l|}
\hline
                  &        &         \\
                  & $F^{JP}$ & $F^C$     \\ \hline
Smaller Tolerance (0.00009) & 0.3933 & -0.2089\\ \hline                  
Original Tolerance (0.0004)    & 0.3933 & -0.2089 \\ \hline
Larger Tolerance (0.0005)  &0.3933  & -0.2089 \\ \hline
\end{tabular}
\caption{We use the case of $\lambda = 1$, $\alpha = \pi/2$, and $R = 2.1$. We show how our relative tolerance already returns a converged result.  Our forces are all nondimensionalized. }
\label{tab:Tolerance}
\end{table}

We can also validate the twin sphere solutions by comparing the results for the forces $F_{_J}^{^{a}}+F_{_J}^{^{b}},F_{_C}^{^{a}}+F_{_C}^{^{b}}$ to the hydrodynamic drag values for two spheres moving in tandem from tabulated results of the Stimson and Jeffrey solutions (S\&J) \cite{stimson1926motion} taken from \cite{cooley1969slow}. 
The forces (scaled by $-6\pi\mu R_{1}$) on the Janus motor and the cargo are given by $f_1$ and $f_2$ respectively.  
\begin{figure}
  \centering
  \begin{subfigure}{.3\linewidth}
    \centering
    \includegraphics[width = \linewidth]{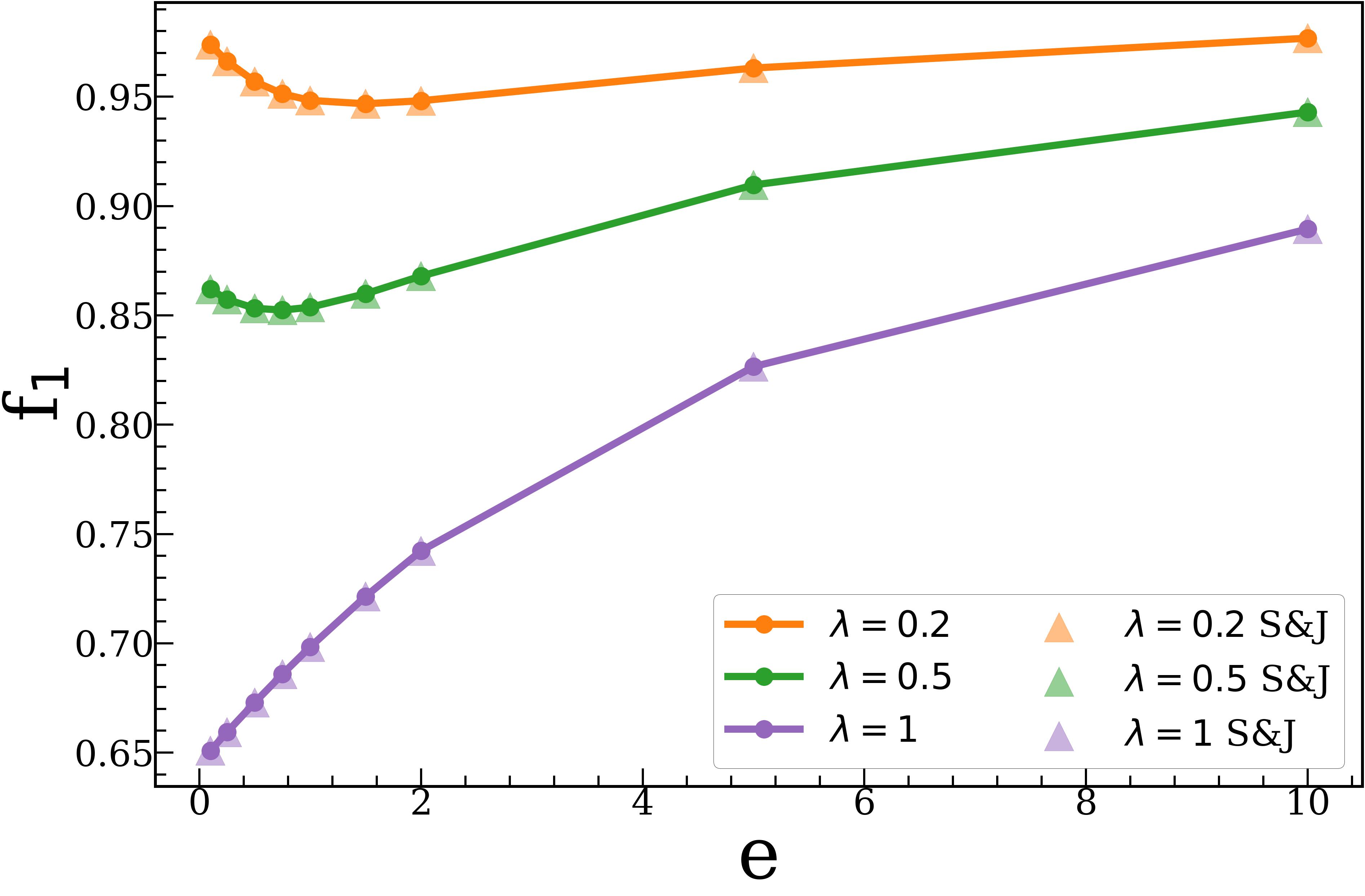}
    \caption{}
    \label{fig:JP_validation}
  \end{subfigure}%
  \hspace{2em}
  \begin{subfigure}{.3\linewidth}
    \centering
    \includegraphics[width = \linewidth]{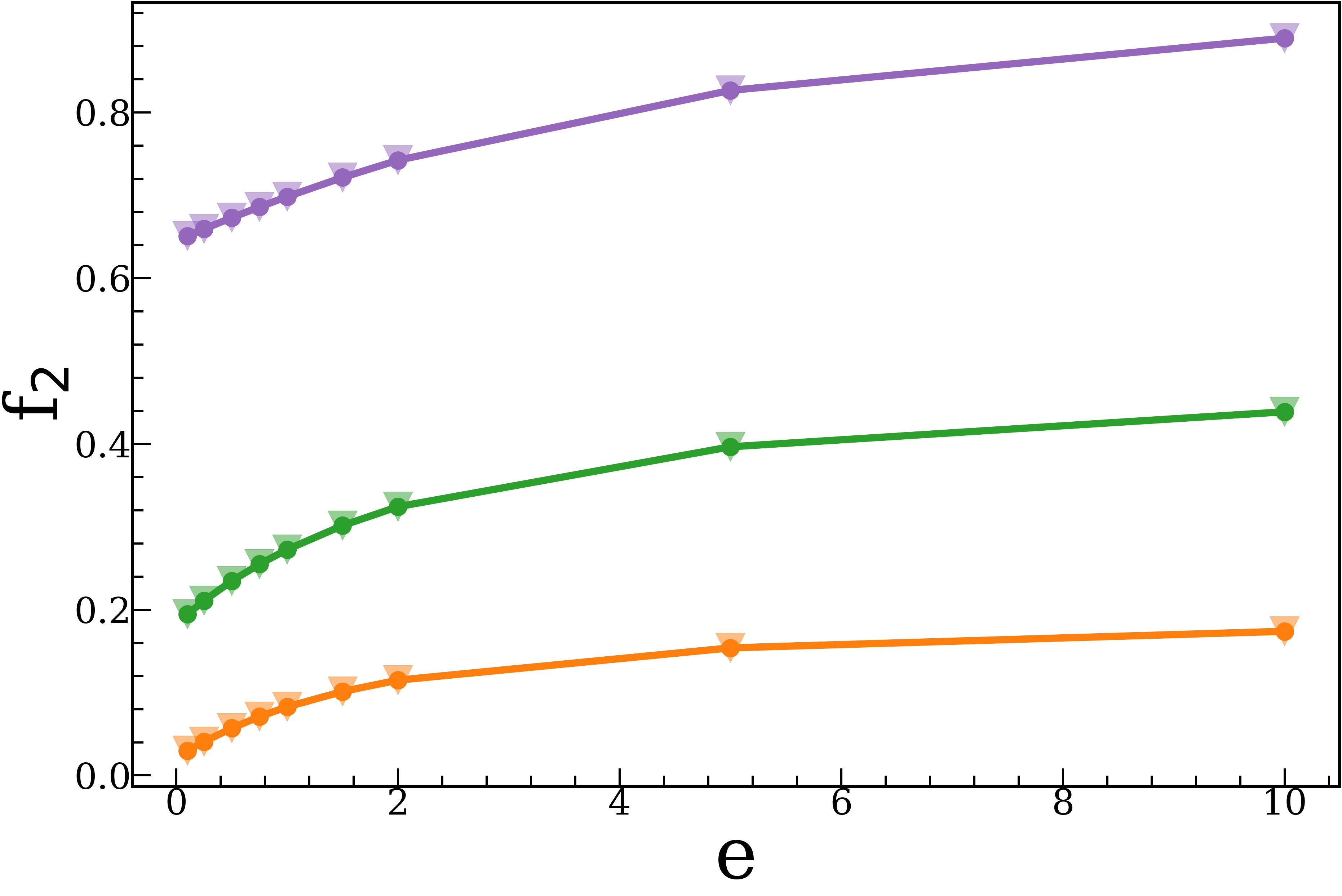}
    \caption{}
    \label{fig:cargo_validation}
  \end{subfigure}%
    \caption{Validation of the twin multipole approach using the hydrodynamic forces for the Janus motor and the cargo from \cite{cooley1969slow} (S\&J).}
  \label{validationcurveshydrodynamic}
\end{figure}

\clearpage
\bibliography{AxisymmetricPaper}

\providecommand{\noopsort}[1]{}\providecommand{\singleletter}[1]{#1}%
\begin{thebibliography}{39}%
\makeatletter
\providecommand \@ifxundefined [1]{%
 \@ifx{#1\undefined}
}%
\providecommand \@ifnum [1]{%
 \ifnum #1\expandafter \@firstoftwo
 \else \expandafter \@secondoftwo
 \fi
}%
\providecommand \@ifx [1]{%
 \ifx #1\expandafter \@firstoftwo
 \else \expandafter \@secondoftwo
 \fi
}%
\providecommand \natexlab [1]{#1}%
\providecommand \enquote  [1]{``#1''}%
\providecommand \bibnamefont  [1]{#1}%
\providecommand \bibfnamefont [1]{#1}%
\providecommand \citenamefont [1]{#1}%
\providecommand \href@noop [0]{\@secondoftwo}%
\providecommand \href [0]{\begingroup \@sanitize@url \@href}%
\providecommand \@href[1]{\@@startlink{#1}\@@href}%
\providecommand \@@href[1]{\endgroup#1\@@endlink}%
\providecommand \@sanitize@url [0]{\catcode `\\12\catcode `\$12\catcode
  `\&12\catcode `\#12\catcode `\^12\catcode `\_12\catcode `\%12\relax}%
\providecommand \@@startlink[1]{}%
\providecommand \@@endlink[0]{}%
\providecommand \url  [0]{\begingroup\@sanitize@url \@url }%
\providecommand \@url [1]{\endgroup\@href {#1}{\urlprefix }}%
\providecommand \urlprefix  [0]{URL }%
\providecommand \Eprint [0]{\href }%
\providecommand \doibase [0]{https://doi.org/}%
\providecommand \selectlanguage [0]{\@gobble}%
\providecommand \bibinfo  [0]{\@secondoftwo}%
\providecommand \bibfield  [0]{\@secondoftwo}%
\providecommand \translation [1]{[#1]}%
\providecommand \BibitemOpen [0]{}%
\providecommand \bibitemStop [0]{}%
\providecommand \bibitemNoStop [0]{.\EOS\space}%
\providecommand \EOS [0]{\spacefactor3000\relax}%
\providecommand \BibitemShut  [1]{\csname bibitem#1\endcsname}%
\let\auto@bib@innerbib\@empty
\bibitem [{\citenamefont {Lauga}\ and\ \citenamefont
  {Powers}(2009)}]{Lauga2009hydrodynamics}%
  \BibitemOpen
  \bibfield  {author} {\bibinfo {author} {\bibfnamefont {E.}~\bibnamefont
  {Lauga}}\ and\ \bibinfo {author} {\bibfnamefont {T.~R.}\ \bibnamefont
  {Powers}},\ }\bibfield  {title} {\bibinfo {title} {The hydrodynamics of
  swimming microorganisms},\ }\href@noop {} {\bibfield  {journal} {\bibinfo
  {journal} {Reports on progress in physics}\ }\textbf {\bibinfo {volume}
  {72}},\ \bibinfo {pages} {096601} (\bibinfo {year} {2009})}\BibitemShut
  {NoStop}%
\bibitem [{\citenamefont {Su}\ \emph {et~al.}(2019)\citenamefont {Su},
  \citenamefont {Price}, \citenamefont {Jing}, \citenamefont {Tian},
  \citenamefont {Liu},\ and\ \citenamefont {Qian}}]{su2019janus}%
  \BibitemOpen
  \bibfield  {author} {\bibinfo {author} {\bibfnamefont {H.}~\bibnamefont
  {Su}}, \bibinfo {author} {\bibfnamefont {C.-A.~H.}\ \bibnamefont {Price}},
  \bibinfo {author} {\bibfnamefont {L.}~\bibnamefont {Jing}}, \bibinfo {author}
  {\bibfnamefont {Q.}~\bibnamefont {Tian}}, \bibinfo {author} {\bibfnamefont
  {J.}~\bibnamefont {Liu}},\ and\ \bibinfo {author} {\bibfnamefont
  {K.}~\bibnamefont {Qian}},\ }\bibfield  {title} {\bibinfo {title} {Janus
  particles: design, preparation, and biomedical applications},\ }\href@noop {}
  {\bibfield  {journal} {\bibinfo  {journal} {Materials today bio}\ }\textbf
  {\bibinfo {volume} {4}},\ \bibinfo {pages} {100033} (\bibinfo {year}
  {2019})}\BibitemShut {NoStop}%
\bibitem [{\citenamefont {Tran}\ \emph {et~al.}(2014)\citenamefont {Tran},
  \citenamefont {Lesieur},\ and\ \citenamefont {Faivre}}]{tran2014janus}%
  \BibitemOpen
  \bibfield  {author} {\bibinfo {author} {\bibfnamefont {L.-T.-C.}\
  \bibnamefont {Tran}}, \bibinfo {author} {\bibfnamefont {S.}~\bibnamefont
  {Lesieur}},\ and\ \bibinfo {author} {\bibfnamefont {V.}~\bibnamefont
  {Faivre}},\ }\bibfield  {title} {\bibinfo {title} {Janus nanoparticles:
  materials, preparation and recent advances in drug delivery},\ }\href@noop {}
  {\bibfield  {journal} {\bibinfo  {journal} {Expert opinion on drug delivery}\
  }\textbf {\bibinfo {volume} {11}},\ \bibinfo {pages} {1061} (\bibinfo {year}
  {2014})}\BibitemShut {NoStop}%
\bibitem [{\citenamefont {Ho}\ \emph {et~al.}(2011)\citenamefont {Ho},
  \citenamefont {Sun},\ and\ \citenamefont {Sun}}]{ho2011monodisperse}%
  \BibitemOpen
  \bibfield  {author} {\bibinfo {author} {\bibfnamefont {D.}~\bibnamefont
  {Ho}}, \bibinfo {author} {\bibfnamefont {X.}~\bibnamefont {Sun}},\ and\
  \bibinfo {author} {\bibfnamefont {S.}~\bibnamefont {Sun}},\ }\bibfield
  {title} {\bibinfo {title} {Monodisperse magnetic nanoparticles for
  theranostic applications},\ }\href@noop {} {\bibfield  {journal} {\bibinfo
  {journal} {Accounts of chemical research}\ }\textbf {\bibinfo {volume}
  {44}},\ \bibinfo {pages} {875} (\bibinfo {year} {2011})}\BibitemShut
  {NoStop}%
\bibitem [{\citenamefont {Walther}\ and\ \citenamefont
  {Muller}(2013)}]{walther2013janus}%
  \BibitemOpen
  \bibfield  {author} {\bibinfo {author} {\bibfnamefont {A.}~\bibnamefont
  {Walther}}\ and\ \bibinfo {author} {\bibfnamefont {A.~H.}\ \bibnamefont
  {Muller}},\ }\bibfield  {title} {\bibinfo {title} {Janus particles:
  synthesis, self-assembly, physical properties, and applications},\
  }\href@noop {} {\bibfield  {journal} {\bibinfo  {journal} {Chemical reviews}\
  }\textbf {\bibinfo {volume} {113}},\ \bibinfo {pages} {5194} (\bibinfo {year}
  {2013})}\BibitemShut {NoStop}%
\bibitem [{\citenamefont {Purcell}(1977)}]{purcell1977life}%
  \BibitemOpen
  \bibfield  {author} {\bibinfo {author} {\bibfnamefont {E.~M.}\ \bibnamefont
  {Purcell}},\ }\bibfield  {title} {\bibinfo {title} {Life at low reynolds
  number},\ }\href@noop {} {\bibfield  {journal} {\bibinfo  {journal} {American
  journal of physics}\ }\textbf {\bibinfo {volume} {45}},\ \bibinfo {pages} {3}
  (\bibinfo {year} {1977})}\BibitemShut {NoStop}%
\bibitem [{\citenamefont {Anderson}\ and\ \citenamefont
  {Prieve}(1984)}]{anderson1984diffusiophoresis}%
  \BibitemOpen
  \bibfield  {author} {\bibinfo {author} {\bibfnamefont {J.~L.}\ \bibnamefont
  {Anderson}}\ and\ \bibinfo {author} {\bibfnamefont {D.~C.}\ \bibnamefont
  {Prieve}},\ }\bibfield  {title} {\bibinfo {title} {Diffusiophoresis:
  Migration of colloidal particles in gradients of solute concentration},\
  }\href@noop {} {\bibfield  {journal} {\bibinfo  {journal} {Separation and
  Purification Methods}\ }\textbf {\bibinfo {volume} {13}},\ \bibinfo {pages}
  {67} (\bibinfo {year} {1984})}\BibitemShut {NoStop}%
\bibitem [{\citenamefont {Anderson}(1989)}]{anderson1989colloid}%
  \BibitemOpen
  \bibfield  {author} {\bibinfo {author} {\bibfnamefont {J.~L.}\ \bibnamefont
  {Anderson}},\ }\bibfield  {title} {\bibinfo {title} {Colloid transport by
  interfacial forces},\ }\href@noop {} {\bibfield  {journal} {\bibinfo
  {journal} {Annual review of fluid mechanics}\ }\textbf {\bibinfo {volume}
  {21}},\ \bibinfo {pages} {61} (\bibinfo {year} {1989})}\BibitemShut {NoStop}%
\bibitem [{\citenamefont {Golestanian}\ \emph {et~al.}(2005)\citenamefont
  {Golestanian}, \citenamefont {Liverpool},\ and\ \citenamefont
  {Ajdari}}]{golestanian2005propulsion}%
  \BibitemOpen
  \bibfield  {author} {\bibinfo {author} {\bibfnamefont {R.}~\bibnamefont
  {Golestanian}}, \bibinfo {author} {\bibfnamefont {T.~B.}\ \bibnamefont
  {Liverpool}},\ and\ \bibinfo {author} {\bibfnamefont {A.}~\bibnamefont
  {Ajdari}},\ }\bibfield  {title} {\bibinfo {title} {Propulsion of a molecular
  machine by asymmetric distribution of reaction products},\ }\href@noop {}
  {\bibfield  {journal} {\bibinfo  {journal} {Physical review letters}\
  }\textbf {\bibinfo {volume} {94}},\ \bibinfo {pages} {220801} (\bibinfo
  {year} {2005})}\BibitemShut {NoStop}%
\bibitem [{\citenamefont {Howse}\ \emph {et~al.}(2007)\citenamefont {Howse},
  \citenamefont {Jones}, \citenamefont {Ryan}, \citenamefont {Gough},
  \citenamefont {Vafabakhsh},\ and\ \citenamefont
  {Golestanian}}]{howse2007self}%
  \BibitemOpen
  \bibfield  {author} {\bibinfo {author} {\bibfnamefont {J.~R.}\ \bibnamefont
  {Howse}}, \bibinfo {author} {\bibfnamefont {R.~A.}\ \bibnamefont {Jones}},
  \bibinfo {author} {\bibfnamefont {A.~J.}\ \bibnamefont {Ryan}}, \bibinfo
  {author} {\bibfnamefont {T.}~\bibnamefont {Gough}}, \bibinfo {author}
  {\bibfnamefont {R.}~\bibnamefont {Vafabakhsh}},\ and\ \bibinfo {author}
  {\bibfnamefont {R.}~\bibnamefont {Golestanian}},\ }\bibfield  {title}
  {\bibinfo {title} {Self-motile colloidal particles: from directed propulsion
  to random walk},\ }\href@noop {} {\bibfield  {journal} {\bibinfo  {journal}
  {Physical review letters}\ }\textbf {\bibinfo {volume} {99}},\ \bibinfo
  {pages} {048102} (\bibinfo {year} {2007})}\BibitemShut {NoStop}%
\bibitem [{\citenamefont {Ebbens}\ and\ \citenamefont
  {Howse}(2011)}]{ebbens2011direct}%
  \BibitemOpen
  \bibfield  {author} {\bibinfo {author} {\bibfnamefont {S.~J.}\ \bibnamefont
  {Ebbens}}\ and\ \bibinfo {author} {\bibfnamefont {J.~R.}\ \bibnamefont
  {Howse}},\ }\bibfield  {title} {\bibinfo {title} {Direct observation of the
  direction of motion for spherical catalytic swimmers},\ }\href@noop {}
  {\bibfield  {journal} {\bibinfo  {journal} {Langmuir}\ }\textbf {\bibinfo
  {volume} {27}},\ \bibinfo {pages} {12293} (\bibinfo {year}
  {2011})}\BibitemShut {NoStop}%
\bibitem [{\citenamefont {Ebbens}\ \emph {et~al.}(2012)\citenamefont {Ebbens},
  \citenamefont {Tu}, \citenamefont {Howse},\ and\ \citenamefont
  {Golestanian}}]{ebbens2012size}%
  \BibitemOpen
  \bibfield  {author} {\bibinfo {author} {\bibfnamefont {S.}~\bibnamefont
  {Ebbens}}, \bibinfo {author} {\bibfnamefont {M.-H.}\ \bibnamefont {Tu}},
  \bibinfo {author} {\bibfnamefont {J.~R.}\ \bibnamefont {Howse}},\ and\
  \bibinfo {author} {\bibfnamefont {R.}~\bibnamefont {Golestanian}},\
  }\bibfield  {title} {\bibinfo {title} {Size dependence of the propulsion
  velocity for catalytic janus-sphere swimmers},\ }\href@noop {} {\bibfield
  {journal} {\bibinfo  {journal} {Physical Review E}\ }\textbf {\bibinfo
  {volume} {85}},\ \bibinfo {pages} {020401} (\bibinfo {year}
  {2012})}\BibitemShut {NoStop}%
\bibitem [{\citenamefont {Zheng}\ \emph {et~al.}(2013)\citenamefont {Zheng},
  \citenamefont {Ten~Hagen}, \citenamefont {Kaiser}, \citenamefont {Wu},
  \citenamefont {Cui}, \citenamefont {Silber-Li},\ and\ \citenamefont
  {L{\"o}wen}}]{zheng2013non}%
  \BibitemOpen
  \bibfield  {author} {\bibinfo {author} {\bibfnamefont {X.}~\bibnamefont
  {Zheng}}, \bibinfo {author} {\bibfnamefont {B.}~\bibnamefont {Ten~Hagen}},
  \bibinfo {author} {\bibfnamefont {A.}~\bibnamefont {Kaiser}}, \bibinfo
  {author} {\bibfnamefont {M.}~\bibnamefont {Wu}}, \bibinfo {author}
  {\bibfnamefont {H.}~\bibnamefont {Cui}}, \bibinfo {author} {\bibfnamefont
  {Z.}~\bibnamefont {Silber-Li}},\ and\ \bibinfo {author} {\bibfnamefont
  {H.}~\bibnamefont {L{\"o}wen}},\ }\bibfield  {title} {\bibinfo {title}
  {Non-gaussian statistics for the motion of self-propelled janus particles:
  Experiment versus theory},\ }\href@noop {} {\bibfield  {journal} {\bibinfo
  {journal} {Physical Review E}\ }\textbf {\bibinfo {volume} {88}},\ \bibinfo
  {pages} {032304} (\bibinfo {year} {2013})}\BibitemShut {NoStop}%
\bibitem [{\citenamefont {Paxton}\ \emph {et~al.}(2004)\citenamefont {Paxton},
  \citenamefont {Kistler}, \citenamefont {Olmeda}, \citenamefont {Sen},
  \citenamefont {St.~Angelo}, \citenamefont {Cao}, \citenamefont {Mallouk},
  \citenamefont {Lammert},\ and\ \citenamefont {Crespi}}]{paxton2004catalytic}%
  \BibitemOpen
  \bibfield  {author} {\bibinfo {author} {\bibfnamefont {W.~F.}\ \bibnamefont
  {Paxton}}, \bibinfo {author} {\bibfnamefont {K.~C.}\ \bibnamefont {Kistler}},
  \bibinfo {author} {\bibfnamefont {C.~C.}\ \bibnamefont {Olmeda}}, \bibinfo
  {author} {\bibfnamefont {A.}~\bibnamefont {Sen}}, \bibinfo {author}
  {\bibfnamefont {S.~K.}\ \bibnamefont {St.~Angelo}}, \bibinfo {author}
  {\bibfnamefont {Y.}~\bibnamefont {Cao}}, \bibinfo {author} {\bibfnamefont
  {T.~E.}\ \bibnamefont {Mallouk}}, \bibinfo {author} {\bibfnamefont {P.~E.}\
  \bibnamefont {Lammert}},\ and\ \bibinfo {author} {\bibfnamefont {V.~H.}\
  \bibnamefont {Crespi}},\ }\bibfield  {title} {\bibinfo {title} {Catalytic
  nanomotors: autonomous movement of striped nanorods},\ }\href@noop {}
  {\bibfield  {journal} {\bibinfo  {journal} {Journal of the American Chemical
  Society}\ }\textbf {\bibinfo {volume} {126}},\ \bibinfo {pages} {13424}
  (\bibinfo {year} {2004})}\BibitemShut {NoStop}%
\bibitem [{\citenamefont {Sharifi-Mood}\ \emph {et~al.}(2013)\citenamefont
  {Sharifi-Mood}, \citenamefont {Koplik},\ and\ \citenamefont
  {Maldarelli}}]{sharifi2013diffusiophoretic}%
  \BibitemOpen
  \bibfield  {author} {\bibinfo {author} {\bibfnamefont {N.}~\bibnamefont
  {Sharifi-Mood}}, \bibinfo {author} {\bibfnamefont {J.}~\bibnamefont
  {Koplik}},\ and\ \bibinfo {author} {\bibfnamefont {C.}~\bibnamefont
  {Maldarelli}},\ }\bibfield  {title} {\bibinfo {title} {Diffusiophoretic
  self-propulsion of colloids driven by a surface reaction: the sub-micron
  particle regime for exponential and van der waals interactions},\ }\href@noop
  {} {\bibfield  {journal} {\bibinfo  {journal} {Physics of Fluids}\ }\textbf
  {\bibinfo {volume} {25}},\ \bibinfo {pages} {012001} (\bibinfo {year}
  {2013})}\BibitemShut {NoStop}%
\bibitem [{\citenamefont {Popescu}\ \emph {et~al.}(2016)\citenamefont
  {Popescu}, \citenamefont {Uspal},\ and\ \citenamefont
  {Dietrich}}]{popescu2016self}%
  \BibitemOpen
  \bibfield  {author} {\bibinfo {author} {\bibfnamefont {M.~N.}\ \bibnamefont
  {Popescu}}, \bibinfo {author} {\bibfnamefont {W.~E.}\ \bibnamefont {Uspal}},\
  and\ \bibinfo {author} {\bibfnamefont {S.}~\bibnamefont {Dietrich}},\
  }\bibfield  {title} {\bibinfo {title} {Self-diffusiophoresis of chemically
  active colloids},\ }\href@noop {} {\bibfield  {journal} {\bibinfo  {journal}
  {The European Physical Journal Special Topics}\ }\textbf {\bibinfo {volume}
  {225}},\ \bibinfo {pages} {2189} (\bibinfo {year} {2016})}\BibitemShut
  {NoStop}%
\bibitem [{\citenamefont {Brady}(2011)}]{brady2011particle}%
  \BibitemOpen
  \bibfield  {author} {\bibinfo {author} {\bibfnamefont {J.~F.}\ \bibnamefont
  {Brady}},\ }\bibfield  {title} {\bibinfo {title} {Particle motion driven by
  solute gradients with application to autonomous motion: continuum and
  colloidal perspectives},\ }\href@noop {} {\bibfield  {journal} {\bibinfo
  {journal} {Journal of Fluid Mechanics}\ }\textbf {\bibinfo {volume} {667}},\
  \bibinfo {pages} {216} (\bibinfo {year} {2011})}\BibitemShut {NoStop}%
\bibitem [{\citenamefont {Ebbens}\ and\ \citenamefont
  {Howse}(2010)}]{ebbens2010pursuit}%
  \BibitemOpen
  \bibfield  {author} {\bibinfo {author} {\bibfnamefont {S.~J.}\ \bibnamefont
  {Ebbens}}\ and\ \bibinfo {author} {\bibfnamefont {J.~R.}\ \bibnamefont
  {Howse}},\ }\bibfield  {title} {\bibinfo {title} {In pursuit of propulsion at
  the nanoscale},\ }\href@noop {} {\bibfield  {journal} {\bibinfo  {journal}
  {Soft Matter}\ }\textbf {\bibinfo {volume} {6}},\ \bibinfo {pages} {726}
  (\bibinfo {year} {2010})}\BibitemShut {NoStop}%
\bibitem [{\citenamefont {Wang}\ \emph {et~al.}(2013)\citenamefont {Wang},
  \citenamefont {Duan}, \citenamefont {Ahmed}, \citenamefont {Mallouk},\ and\
  \citenamefont {Sen}}]{wang2013small}%
  \BibitemOpen
  \bibfield  {author} {\bibinfo {author} {\bibfnamefont {W.}~\bibnamefont
  {Wang}}, \bibinfo {author} {\bibfnamefont {W.}~\bibnamefont {Duan}}, \bibinfo
  {author} {\bibfnamefont {S.}~\bibnamefont {Ahmed}}, \bibinfo {author}
  {\bibfnamefont {T.~E.}\ \bibnamefont {Mallouk}},\ and\ \bibinfo {author}
  {\bibfnamefont {A.}~\bibnamefont {Sen}},\ }\bibfield  {title} {\bibinfo
  {title} {Small power: Autonomous nano-and micromotors propelled by
  self-generated gradients},\ }\href@noop {} {\bibfield  {journal} {\bibinfo
  {journal} {Nano Today}\ }\textbf {\bibinfo {volume} {8}},\ \bibinfo {pages}
  {531} (\bibinfo {year} {2013})}\BibitemShut {NoStop}%
\bibitem [{\citenamefont {Kapral}(2013)}]{kapral2013perspective}%
  \BibitemOpen
  \bibfield  {author} {\bibinfo {author} {\bibfnamefont {R.}~\bibnamefont
  {Kapral}},\ }\bibfield  {title} {\bibinfo {title} {Perspective: Nanomotors
  without moving parts that propel themselves in solution},\ }\href@noop {}
  {\bibfield  {journal} {\bibinfo  {journal} {The Journal of Chemical Physics}\
  }\textbf {\bibinfo {volume} {138}},\ \bibinfo {pages} {020901} (\bibinfo
  {year} {2013})}\BibitemShut {NoStop}%
\bibitem [{\citenamefont {Moran}\ and\ \citenamefont
  {Posner}(2017)}]{moran2017phoretic}%
  \BibitemOpen
  \bibfield  {author} {\bibinfo {author} {\bibfnamefont {J.~L.}\ \bibnamefont
  {Moran}}\ and\ \bibinfo {author} {\bibfnamefont {J.~D.}\ \bibnamefont
  {Posner}},\ }\bibfield  {title} {\bibinfo {title} {Phoretic
  self-propulsion},\ }\href@noop {} {\bibfield  {journal} {\bibinfo  {journal}
  {Annual Review of Fluid Mechanics}\ }\textbf {\bibinfo {volume} {49}},\
  \bibinfo {pages} {511} (\bibinfo {year} {2017})}\BibitemShut {NoStop}%
\bibitem [{\citenamefont {Baraban}\ \emph
  {et~al.}(2012{\natexlab{a}})\citenamefont {Baraban}, \citenamefont {Makarov},
  \citenamefont {Streubel}, \citenamefont {Monch}, \citenamefont {Grimm},
  \citenamefont {Sanchez},\ and\ \citenamefont
  {Schmidt}}]{baraban2012catalytic}%
  \BibitemOpen
  \bibfield  {author} {\bibinfo {author} {\bibfnamefont {L.}~\bibnamefont
  {Baraban}}, \bibinfo {author} {\bibfnamefont {D.}~\bibnamefont {Makarov}},
  \bibinfo {author} {\bibfnamefont {R.}~\bibnamefont {Streubel}}, \bibinfo
  {author} {\bibfnamefont {I.}~\bibnamefont {Monch}}, \bibinfo {author}
  {\bibfnamefont {D.}~\bibnamefont {Grimm}}, \bibinfo {author} {\bibfnamefont
  {S.}~\bibnamefont {Sanchez}},\ and\ \bibinfo {author} {\bibfnamefont {O.~G.}\
  \bibnamefont {Schmidt}},\ }\bibfield  {title} {\bibinfo {title} {Catalytic
  janus motors on microfluidic chip: deterministic motion for targeted cargo
  delivery},\ }\href@noop {} {\bibfield  {journal} {\bibinfo  {journal} {ACS
  nano}\ }\textbf {\bibinfo {volume} {6}},\ \bibinfo {pages} {3383} (\bibinfo
  {year} {2012}{\natexlab{a}})}\BibitemShut {NoStop}%
\bibitem [{\citenamefont {Baraban}\ \emph
  {et~al.}(2012{\natexlab{b}})\citenamefont {Baraban}, \citenamefont
  {Tasinkevych}, \citenamefont {Popescu}, \citenamefont {Sanchez},
  \citenamefont {Dietrich},\ and\ \citenamefont
  {Schmidt}}]{baraban2012transport}%
  \BibitemOpen
  \bibfield  {author} {\bibinfo {author} {\bibfnamefont {L.}~\bibnamefont
  {Baraban}}, \bibinfo {author} {\bibfnamefont {M.}~\bibnamefont
  {Tasinkevych}}, \bibinfo {author} {\bibfnamefont {M.~N.}\ \bibnamefont
  {Popescu}}, \bibinfo {author} {\bibfnamefont {S.}~\bibnamefont {Sanchez}},
  \bibinfo {author} {\bibfnamefont {S.}~\bibnamefont {Dietrich}},\ and\
  \bibinfo {author} {\bibfnamefont {O.}~\bibnamefont {Schmidt}},\ }\bibfield
  {title} {\bibinfo {title} {Transport of cargo by catalytic janus
  micro-motors},\ }\href@noop {} {\bibfield  {journal} {\bibinfo  {journal}
  {Soft Matter}\ }\textbf {\bibinfo {volume} {8}},\ \bibinfo {pages} {48}
  (\bibinfo {year} {2012}{\natexlab{b}})}\BibitemShut {NoStop}%
\bibitem [{\citenamefont {Bayati}\ and\ \citenamefont
  {Najafi}(2016)}]{bayati2016dynamics}%
  \BibitemOpen
  \bibfield  {author} {\bibinfo {author} {\bibfnamefont {P.}~\bibnamefont
  {Bayati}}\ and\ \bibinfo {author} {\bibfnamefont {A.}~\bibnamefont
  {Najafi}},\ }\bibfield  {title} {\bibinfo {title} {Dynamics of two
  interacting active janus particles},\ }\href@noop {} {\bibfield  {journal}
  {\bibinfo  {journal} {The Journal of chemical physics}\ }\textbf {\bibinfo
  {volume} {144}},\ \bibinfo {pages} {134901} (\bibinfo {year}
  {2016})}\BibitemShut {NoStop}%
\bibitem [{\citenamefont {Nasouri}\ and\ \citenamefont
  {Golestanian}(2020)}]{nasouri2020exact}%
  \BibitemOpen
  \bibfield  {author} {\bibinfo {author} {\bibfnamefont {B.}~\bibnamefont
  {Nasouri}}\ and\ \bibinfo {author} {\bibfnamefont {R.}~\bibnamefont
  {Golestanian}},\ }\bibfield  {title} {\bibinfo {title} {Exact axisymmetric
  interaction of phoretically active janus particles},\ }\href@noop {}
  {\bibfield  {journal} {\bibinfo  {journal} {Journal of Fluid Mechanics}\
  }\textbf {\bibinfo {volume} {905}} (\bibinfo {year} {2020})}\BibitemShut
  {NoStop}%
\bibitem [{\citenamefont {Rojas-P{\'e}rez}\ \emph {et~al.}(2021)\citenamefont
  {Rojas-P{\'e}rez}, \citenamefont {Delmotte},\ and\ \citenamefont
  {Michelin}}]{rojas2021hydrochemical}%
  \BibitemOpen
  \bibfield  {author} {\bibinfo {author} {\bibfnamefont {F.}~\bibnamefont
  {Rojas-P{\'e}rez}}, \bibinfo {author} {\bibfnamefont {B.}~\bibnamefont
  {Delmotte}},\ and\ \bibinfo {author} {\bibfnamefont {S.}~\bibnamefont
  {Michelin}},\ }\bibfield  {title} {\bibinfo {title} {Hydrochemical
  interactions of phoretic particles: a regularized multipole framework},\
  }\href@noop {} {\bibfield  {journal} {\bibinfo  {journal} {Journal of Fluid
  Mechanics}\ }\textbf {\bibinfo {volume} {919}} (\bibinfo {year}
  {2021})}\BibitemShut {NoStop}%
\bibitem [{\citenamefont {Sharifi-Mood}\ \emph {et~al.}(2016)\citenamefont
  {Sharifi-Mood}, \citenamefont {Mozaffari},\ and\ \citenamefont
  {C{\'o}rdova-Figueroa}}]{sharifi2016pair}%
  \BibitemOpen
  \bibfield  {author} {\bibinfo {author} {\bibfnamefont {N.}~\bibnamefont
  {Sharifi-Mood}}, \bibinfo {author} {\bibfnamefont {A.}~\bibnamefont
  {Mozaffari}},\ and\ \bibinfo {author} {\bibfnamefont {U.~M.}\ \bibnamefont
  {C{\'o}rdova-Figueroa}},\ }\bibfield  {title} {\bibinfo {title} {Pair
  interaction of catalytically active colloids: from assembly to escape},\
  }\href@noop {} {\bibfield  {journal} {\bibinfo  {journal} {Journal of Fluid
  Mechanics}\ }\textbf {\bibinfo {volume} {798}},\ \bibinfo {pages} {910}
  (\bibinfo {year} {2016})}\BibitemShut {NoStop}%
\bibitem [{\citenamefont {Happel}\ and\ \citenamefont
  {Brenner}(2012)}]{happel2012low}%
  \BibitemOpen
  \bibfield  {author} {\bibinfo {author} {\bibfnamefont {J.}~\bibnamefont
  {Happel}}\ and\ \bibinfo {author} {\bibfnamefont {H.}~\bibnamefont
  {Brenner}},\ }\href@noop {} {\emph {\bibinfo {title} {Low Reynolds number
  hydrodynamics: with special applications to particulate media}}},\
  Vol.~\bibinfo {volume} {1}\ (\bibinfo  {publisher} {Springer Science \&
  Business Media},\ \bibinfo {year} {2012})\BibitemShut {NoStop}%
\bibitem [{\citenamefont {Teubner}(1982)}]{teubner1982motion}%
  \BibitemOpen
  \bibfield  {author} {\bibinfo {author} {\bibfnamefont {M.}~\bibnamefont
  {Teubner}},\ }\bibfield  {title} {\bibinfo {title} {The motion of charged
  colloidal particles in electric fields},\ }\href@noop {} {\bibfield
  {journal} {\bibinfo  {journal} {The Journal of Chemical Physics}\ }\textbf
  {\bibinfo {volume} {76}},\ \bibinfo {pages} {5564} (\bibinfo {year}
  {1982})}\BibitemShut {NoStop}%
\bibitem [{\citenamefont {Lee}\ and\ \citenamefont
  {Leal}(1980)}]{lee1980motion}%
  \BibitemOpen
  \bibfield  {author} {\bibinfo {author} {\bibfnamefont {S.}~\bibnamefont
  {Lee}}\ and\ \bibinfo {author} {\bibfnamefont {L.}~\bibnamefont {Leal}},\
  }\bibfield  {title} {\bibinfo {title} {Motion of a sphere in the presence of
  a plane interface. part 2. an exact solution in bipolar co-ordinates},\
  }\href@noop {} {\bibfield  {journal} {\bibinfo  {journal} {Journal of Fluid
  Mechanics}\ }\textbf {\bibinfo {volume} {98}},\ \bibinfo {pages} {193}
  (\bibinfo {year} {1980})}\BibitemShut {NoStop}%
\bibitem [{\citenamefont {Ross}(1968)}]{ross1968potential}%
  \BibitemOpen
  \bibfield  {author} {\bibinfo {author} {\bibfnamefont {D.}~\bibnamefont
  {Ross}},\ }\bibfield  {title} {\bibinfo {title} {The potential due to two
  point charges each at the centre of a spherical cavity and embedded in a
  dielectric medium},\ }\href@noop {} {\bibfield  {journal} {\bibinfo
  {journal} {Australian Journal of Physics}\ }\textbf {\bibinfo {volume}
  {21}},\ \bibinfo {pages} {817} (\bibinfo {year} {1968})}\BibitemShut
  {NoStop}%
\bibitem [{\citenamefont {Jeffrey}(1973)}]{jeffrey1973conduction}%
  \BibitemOpen
  \bibfield  {author} {\bibinfo {author} {\bibfnamefont {D.~J.}\ \bibnamefont
  {Jeffrey}},\ }\bibfield  {title} {\bibinfo {title} {Conduction through a
  random suspension of spheres},\ }\href@noop {} {\bibfield  {journal}
  {\bibinfo  {journal} {Proceedings of the Royal Society of London. A.
  Mathematical and Physical Sciences}\ }\textbf {\bibinfo {volume} {335}},\
  \bibinfo {pages} {355} (\bibinfo {year} {1973})}\BibitemShut {NoStop}%
\bibitem [{\citenamefont {Jeffrey}\ and\ \citenamefont
  {Onishi}(1984)}]{jeffrey1984calculation}%
  \BibitemOpen
  \bibfield  {author} {\bibinfo {author} {\bibfnamefont {D.}~\bibnamefont
  {Jeffrey}}\ and\ \bibinfo {author} {\bibfnamefont {Y.}~\bibnamefont
  {Onishi}},\ }\bibfield  {title} {\bibinfo {title} {Calculation of the
  resistance and mobility functions for two unequal rigid spheres in
  low-reynolds-number flow},\ }\href@noop {} {\bibfield  {journal} {\bibinfo
  {journal} {Journal of Fluid Mechanics}\ }\textbf {\bibinfo {volume} {139}},\
  \bibinfo {pages} {261} (\bibinfo {year} {1984})}\BibitemShut {NoStop}%
\bibitem [{\citenamefont {Leal}(2007)}]{leal2007advanced}%
  \BibitemOpen
  \bibfield  {author} {\bibinfo {author} {\bibfnamefont {L.~G.}\ \bibnamefont
  {Leal}},\ }\href@noop {} {\emph {\bibinfo {title} {Advanced transport
  phenomena: fluid mechanics and convective transport processes}}},\
  Vol.~\bibinfo {volume} {7}\ (\bibinfo  {publisher} {Cambridge university
  press},\ \bibinfo {year} {2007})\BibitemShut {NoStop}%
\bibitem [{\citenamefont {Kim}\ and\ \citenamefont
  {Karrila}(2013)}]{kim2013microhydrodynamics}%
  \BibitemOpen
  \bibfield  {author} {\bibinfo {author} {\bibfnamefont {S.}~\bibnamefont
  {Kim}}\ and\ \bibinfo {author} {\bibfnamefont {S.~J.}\ \bibnamefont
  {Karrila}},\ }\href@noop {} {\emph {\bibinfo {title} {Microhydrodynamics:
  principles and selected applications}}}\ (\bibinfo  {publisher} {Courier
  Corporation},\ \bibinfo {year} {2013})\BibitemShut {NoStop}%
\bibitem [{\citenamefont {Mozaffari}\ \emph {et~al.}(2016)\citenamefont
  {Mozaffari}, \citenamefont {Sharifi-Mood}, \citenamefont {Koplik},\ and\
  \citenamefont {Maldarelli}}]{mozaffari2016self}%
  \BibitemOpen
  \bibfield  {author} {\bibinfo {author} {\bibfnamefont {A.}~\bibnamefont
  {Mozaffari}}, \bibinfo {author} {\bibfnamefont {N.}~\bibnamefont
  {Sharifi-Mood}}, \bibinfo {author} {\bibfnamefont {J.}~\bibnamefont
  {Koplik}},\ and\ \bibinfo {author} {\bibfnamefont {C.}~\bibnamefont
  {Maldarelli}},\ }\bibfield  {title} {\bibinfo {title} {Self-diffusiophoretic
  colloidal propulsion near a solid boundary},\ }\href@noop {} {\bibfield
  {journal} {\bibinfo  {journal} {Physics of Fluids}\ }\textbf {\bibinfo
  {volume} {28}} (\bibinfo {year} {2016})}\BibitemShut {NoStop}%
\bibitem [{\citenamefont {Stimson}\ and\ \citenamefont
  {Jeffery}(1926)}]{stimson1926motion}%
  \BibitemOpen
  \bibfield  {author} {\bibinfo {author} {\bibfnamefont {M.}~\bibnamefont
  {Stimson}}\ and\ \bibinfo {author} {\bibfnamefont {G.~B.}\ \bibnamefont
  {Jeffery}},\ }\bibfield  {title} {\bibinfo {title} {The motion of two spheres
  in a viscous fluid},\ }\href@noop {} {\bibfield  {journal} {\bibinfo
  {journal} {Proceedings of the Royal Society of London. Series A, Containing
  Papers of a Mathematical and Physical Character}\ }\textbf {\bibinfo {volume}
  {111}},\ \bibinfo {pages} {110} (\bibinfo {year} {1926})}\BibitemShut
  {NoStop}%
\bibitem [{\citenamefont {Sack}(1964)}]{sack1964generalization}%
  \BibitemOpen
  \bibfield  {author} {\bibinfo {author} {\bibfnamefont {R.}~\bibnamefont
  {Sack}},\ }\bibfield  {title} {\bibinfo {title} {Generalization of laplace's
  expansion to arbitrary powers and functions of the distance between two
  points},\ }\href@noop {} {\bibfield  {journal} {\bibinfo  {journal} {Journal
  of Mathematical Physics}\ }\textbf {\bibinfo {volume} {5}},\ \bibinfo {pages}
  {245} (\bibinfo {year} {1964})}\BibitemShut {NoStop}%
\bibitem [{\citenamefont {Cooley}\ and\ \citenamefont
  {O'neill}(1969)}]{cooley1969slow}%
  \BibitemOpen
  \bibfield  {author} {\bibinfo {author} {\bibfnamefont {M.}~\bibnamefont
  {Cooley}}\ and\ \bibinfo {author} {\bibfnamefont {M.}~\bibnamefont
  {O'neill}},\ }\bibfield  {title} {\bibinfo {title} {On the slow motion of two
  spheres in contact along their line of centres through a viscous fluid},\
  }in\ \href@noop {} {\emph {\bibinfo {booktitle} {Mathematical Proceedings of
  the Cambridge Philosophical Society}}},\ Vol.~\bibinfo {volume} {66}\
  (\bibinfo {organization} {Cambridge University Press},\ \bibinfo {year}
  {1969})\ pp.\ \bibinfo {pages} {407--415}\BibitemShut {NoStop}%
\end{thebibliography}%

\end{document}